\documentclass{egpubl}
\usepackage{egsr2023}

\SpecialIssuePaper         

\CGFccby

\usepackage[T1]{fontenc}
\usepackage{dfadobe}  

\usepackage{cite}  
\BibtexOrBiblatex
\electronicVersion
\PrintedOrElectronic
\ifpdf \usepackage[pdftex]{graphicx} \pdfcompresslevel=3
\else \usepackage[dvips]{graphicx} \fi

\usepackage{egweblnk} 

\usepackage{booktabs} 
\usepackage[ruled]{algorithm2e} 
\usepackage[nolist]{acronym}
\usepackage{amsmath}
\usepackage{amssymb}
\usepackage{color}
\usepackage{xspace}
\usepackage{bm}

\usepackage{caption}
\usepackage[labelfont=rm,textfont=rm,font+=footnotesize]{subcaption}

\usepackage{mathtools}

\usepackage[binary-units=true]{siunitx}
\DeclareSIUnit\pixel{P}
\usepackage[super]{nth}

\makeatletter
\@namedef{ver@everyshi.sty}{}
\makeatother

\usepackage{listings}
\usepackage{comment}

\usepackage{tikz}
\usetikzlibrary{spy,arrows}
\usepackage{cleveref}
\usepackage[switch]{lineno}
\usepackage{csquotes}
\usepackage{multirow}

\usepackage[nolist]{acronym}
\usepackage{color,soul}
\usepackage[edges]{forest}
\usepackage{blindtext}
\usepackage{mwe}
\usepackage{nicefrac}
\usepackage{tablefootnote}
\usepackage{adjustbox}
\usepackage{enumitem} 
\usepackage{url}
\usepackage{placeins}

\usepackage{tabularx}
\usepackage{multirow}
\usepackage{array}
\usepackage{booktabs}

\usepackage{import}

\usepackage{layouts}
\usetikzlibrary{shadows.blur}
\usetikzlibrary{positioning}
\usetikzlibrary{arrows.meta}
\usetikzlibrary{positioning}
\usetikzlibrary{spy}
\usepackage[skins]{tcolorbox}
\usepackage{pgfplots}

\newcounter{GoalCNT}%
\Crefname{figure}{Fig.}{Figs.}
\Crefname{equation}{Eqn.}{Eqns.}
\Crefname{algorithm}{Algo.}{Algos.}
\Crefname{section}{Sec.}{Secs.}
\Crefname{table}{Tab.}{Tabs.}
\creflabelformat{equation}{#2\textup{#1}#3}

\SetAlFnt{\small}
\SetAlCapFnt{\small}
\SetAlCapNameFnt{\small}
\SetAlCapHSkip{0pt}

\makeatletter
\DeclareRobustCommand\onedot{\futurelet\@let@token\@onedot}
\def\@onedot{\ifx\@let@token.\else.\null\fi}
\def\etal{\textit{et~al\onedot}\xspace}
\def\eg{e.g\onedot, } 
\def\ie{i.e\onedot, } 
\def\cf{c.f\onedot} 
 
\def\vs{vs\onedot~}

\makeatother

\newcommand{\twoc}[1]{\multicolumn{2}{c}{#1}}

\newcommand{\shst}[1]{\shortstack{#1}}

\input{macros}

\definecolor{darkgreen}{rgb}{0.0,0.5,0.0}
\definecolor{pastelgreen}{rgb}{0.42,0.72,0.65}
\definecolor{aquamarine}{rgb}{0.13, 0.42, 0.53}
\definecolor{orange}{rgb}{0.8,0.4,0}
\definecolor{purple}{rgb}{0.5,0,0.5}

\begin{acronym}
	\acro{API}{Application Programming Interface}
	\acro{AR}{Augmented Reality}
	\acro{CG}{Computer Graphics}
	\acro{CNN}{Convolutional Neural Network}
	\acro{CPU}{Central Processing Unit}
	\acro{CV}{Computer Vision}
	\acro{DL}{Deep Learning}
	\acro{DNN}{Deep Neural Network}
	\acro{EPE}{Endpoint Error}
	\acro{FPS}{Frames per second}
	\acro{GPU}{Graphical Processing Unit}
	\acro{IB-AR}{Image-based Artistic Rendering}
	\acro{K}{Kilo, thousand}
	\acro{M}{Million}
 	\acro{MB}{Megabyte}
	\acro{NPR}{Non-photorealistic Rendering}
	\acro{NPU}{Neural Processing Unit}
	\acro{NST}{Neural Style Transfer}
	\acro{ReLU}{Rectified Linear Unit}
	\acro{SGD}{Stochastic Gradient Descent}
	\acro{SOTA}{State-of-the-art}
	\acro{TFLOPS}{$10^{12}$ Floating Point Operations Per Second}
	\acro{TSI}{Temporal Slice Image}
\end{acronym}

\title[Interactive Control over Temporal Consistency while Stylizing Video Streams]%
{Interactive Control over Temporal Consistency \\ while Stylizing Video Streams}

\author[S. Shekhar et al.]
{\parbox{\textwidth}{\centering Sumit Shekhar$^{1*}$\orcid{0000-0002-5683-2290},
		Max Reimann$^{1*}$\orcid{0000-0003-2146-4229}, 
		Moritz Hilscher$^1$,
		Amir Semmo$^{1,2}$\orcid{0000-0002-1553-4940}, \\
		J{\"u}rgen D{\"o}llner$^{1}$,
		and Matthias Trapp$^1$\orcid{0000-0003-3861-5759} \\ 
	}
	\\
	{\parbox{\textwidth}{\centering $^1$Hasso Plattner Institute for Digital Engineering, University of Potsdam, Germany \\
			$^2$Digital Masterpieces GmbH, Germany \\  
			(\enquote{*} denotes equal contribution)
		} 
	}
}

\begin{document}

\newcommand{\afterfloat}{\vspace{0.0cm}}
\newcommand{\beforepar}{\vspace{0.0cm}}
\newcommand{\beforesec}{\vspace{-0.1cm}}

\newcommand{\specFigSize}{0.155\textwidth}

\newlength{\tempdima}
\newcommand{\rowname}[1]
{\rotatebox{90}{\makebox[\tempdima][c]{\text{#1}}}}

\maketitle

\begin{abstract}
Image stylization has seen significant advancement and widespread interest over the years, leading to the development of a multitude of techniques.
Extending these stylization techniques, such as \ac{NST}, to videos is often achieved by applying them on a per-frame basis.
However, per-frame stylization usually lacks temporal consistency, expressed by undesirable flickering artifacts.
Most of the existing approaches for enforcing temporal consistency suffer from one or more of the following drawbacks: They (1) are only suitable for a limited range of techniques, (2) do not support online processing as they require the complete video as input, (3) cannot provide consistency for the task of stylization, or (4) do not provide interactive consistency control.
Domain-agnostic techniques for temporal consistency aim to eradicate flickering completely but typically disregard aesthetic aspects. For stylization tasks, however, consistency control is an essential requirement as a certain amount of flickering adds to the artistic look and feel.
Moreover, making this control interactive is paramount from a usability perspective.  
To achieve the above requirements, we propose an approach that stylizes video streams in real-time at full HD resolutions while providing interactive consistency control.
We develop a lite optical-flow network that operates at 80 \ac{FPS} on desktop systems with sufficient accuracy.
Further, we employ an adaptive combination of local and global consistency features and enable interactive selection between them.
Objective and subjective evaluations demonstrate that our method is superior to state-of-the-art video consistency approaches. \href{https://maxreimann.github.io/stream-consistency}{maxreimann.github.io/stream-consistency}


\end{abstract}  

\newcommand{\beforesection}{\vspace{0.0cm}}
\newcommand{\beforesubsection}{\vspace{0.0cm}}
\newcommand{\beforesubsubsection}{\vspace{0.0cm}}
\newcommand{\afterfigure}{\vspace{-0.2cm}}

\definecolor{dkgreen}       {rgb}{0.0,0.5,0.0}
\definecolor{dkalg}       {rgb}{0.0,0.6,0.2}
\definecolor{dkblue}       {rgb}{0.0,0.0,0.7}
\definecolor{dkred}       {rgb}{0.9,0.0,0.1}

\colorlet{agood}{dkgreen}
\colorlet{abad}{dkred}
\newcommand{\agood}[1]{\textcolor{agood}{#1}}
\newcommand{\abad}[1]{\textcolor{abad}{#1}}
\newcommand{\aequal}[0]{$\pm$ 0\%}
\newcommand{\abest}[1]{\underline{#1}}

\section{Introduction}
\label{Sec:Introduction}

\begin{table*}[tb]
\caption{Comparing existing consistent video filtering methods with ours with regards to consistency control. Here, the color \textcolor{darkgreen}{green} denotes the aspect which is favorable to interactive consistency-control while the color \textcolor{red}{red} denotes otherwise (\enquote{N/A} denotes Not-Applicable).}
\label{tab:consist_cntrl_compare}
\centering
\resizebox{\textwidth}{!}{
\begin{tabular}{|l|l|l|l|l|l|l|}
\hline
Aspects & Bonneel~\etal~\cite{Bonneel_Blind2015} & Yao~\etal~\cite{Yao_Occlusion2017} & Lai~\etal~\cite{Lai_Learning2018} & Shekhar~\etal~\cite{Shekhar_Consistent2019} & Thiomonier~\etal~\cite{Thimonier_learning2021} & Ours \\ \hline
Requires pre-processing?                                                                   & \textcolor{darkgreen}{No}      & \textcolor{red}{Yes} & \textcolor{darkgreen}{No}  & \textcolor{red}{Yes}     & \textcolor{darkgreen}{No}         & \textcolor{darkgreen}{No}   \\ 
Provides consistency control at inference time?& \textcolor{darkgreen}{Yes}     & \textcolor{red}{No}  & \textcolor{red}{No}  & \textcolor{darkgreen}{Yes}     & \textcolor{red}{No}         & \textcolor{darkgreen}{Yes}  \\
Provides interactive consistency control?                                                    & \textcolor{red}{No}      & N/A  & N/A  & \textcolor{darkgreen}{Yes}     & N/A         & \textcolor{darkgreen}{Yes}  \\ \hline
\end{tabular}%
}
\end{table*}

\begin{figure*}[tb]
	\begin{subfigure}{0.195\textwidth}%
		\includegraphics[trim=4cm 0 4cm 3cm, clip, width=\textwidth]{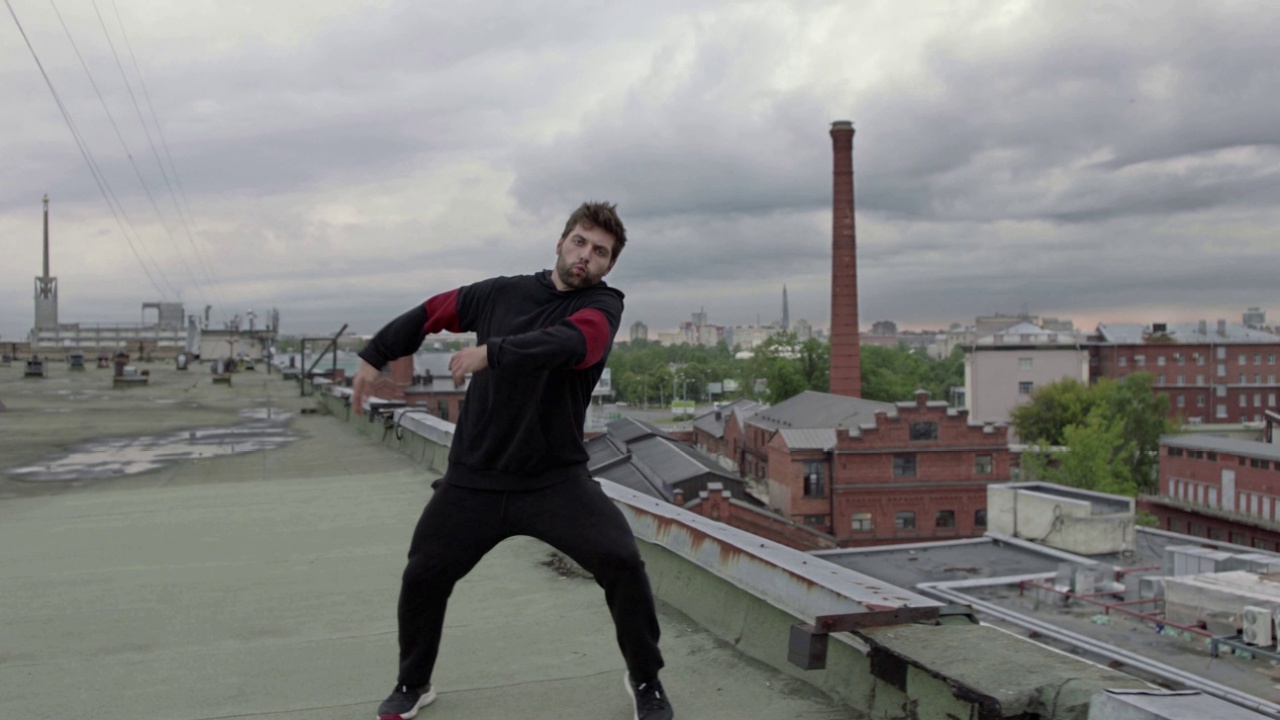}%
		\label{fig:teaser_top_in}%
	\end{subfigure}\hfill 
	\begin{subfigure}{0.195\textwidth}%
		\includegraphics[trim=4cm 0 4cm 3cm, clip, width=\textwidth]{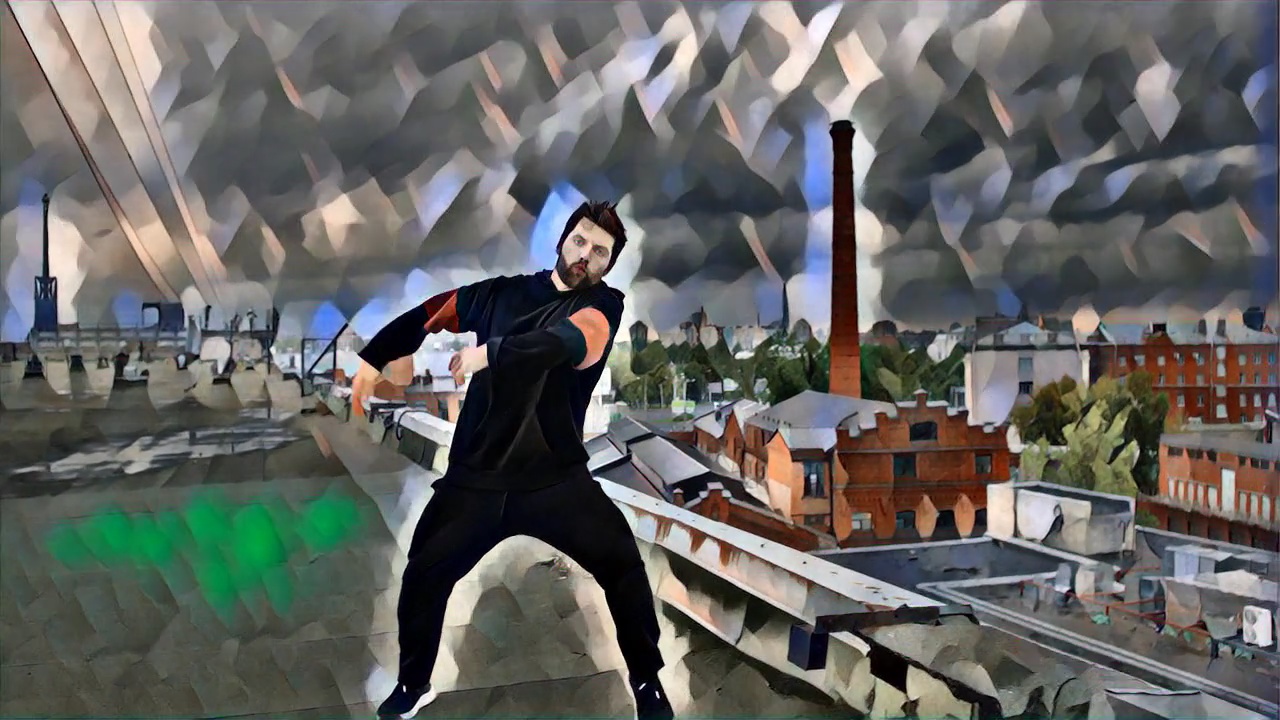}%
		\label{fig:teaser_top_pr}%
	\end{subfigure}\hfill 
	\begin{subfigure}{0.195\textwidth}%
		\includegraphics[trim=4cm 0 4cm 3cm, clip, width=\textwidth]{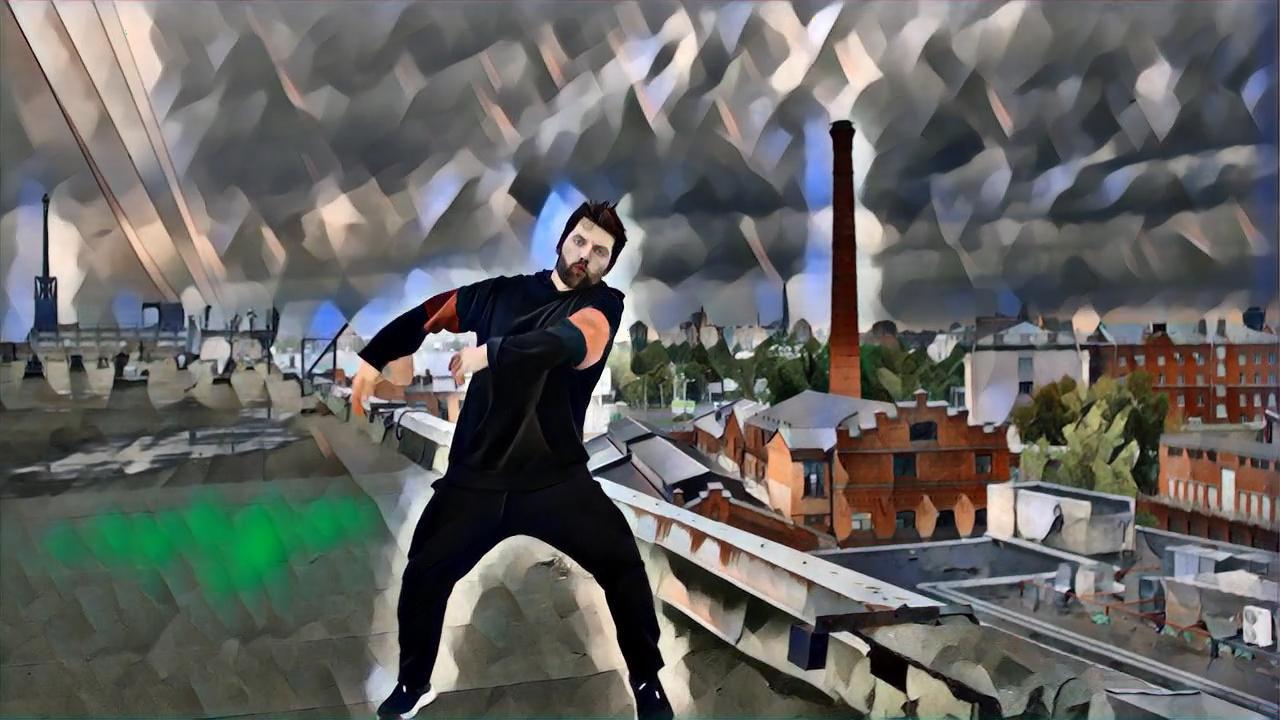}%
		\label{fig:teaser_top_our}%
	\end{subfigure}\hfill 
	\begin{subfigure}{0.195\textwidth}%
		\includegraphics[trim=4cm 0 4cm 3cm, clip, width=\textwidth]{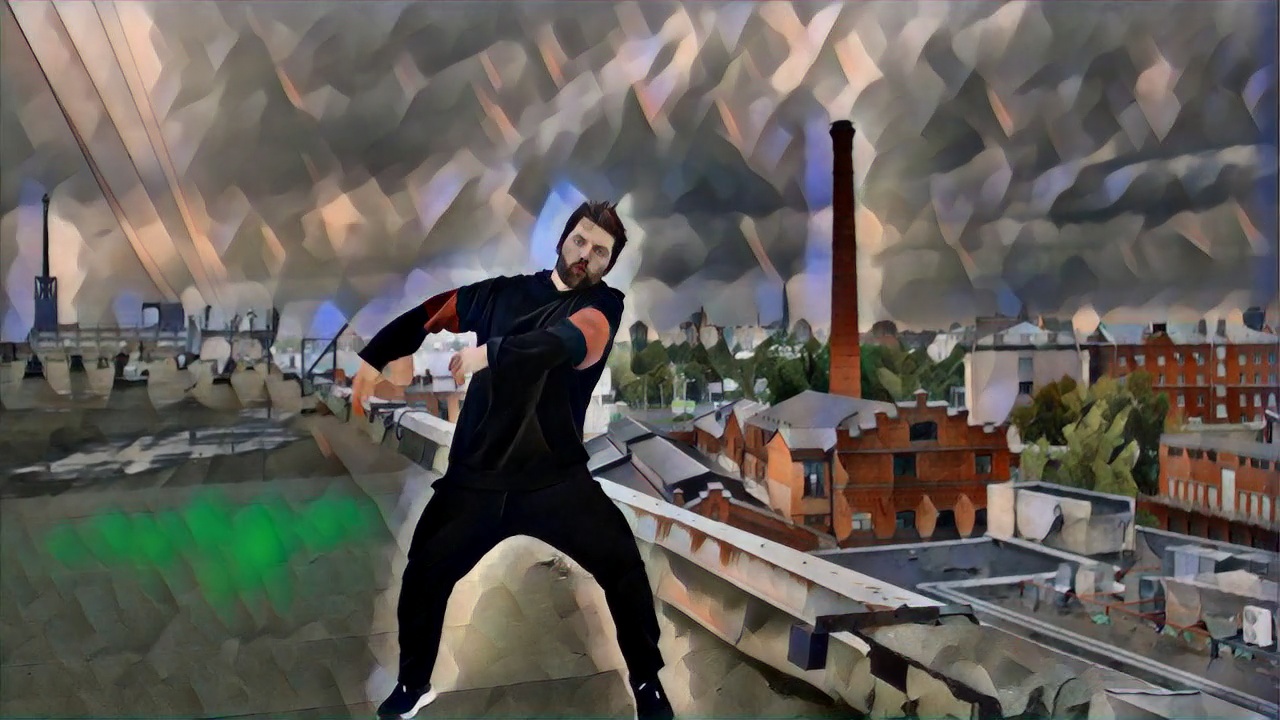}%
		\label{fig:teaser_top_lai}%
	\end{subfigure}\hfill 
	\begin{subfigure}{0.195\textwidth}%
		\includegraphics[trim=4cm 0 4cm 3cm, clip, width=\textwidth]{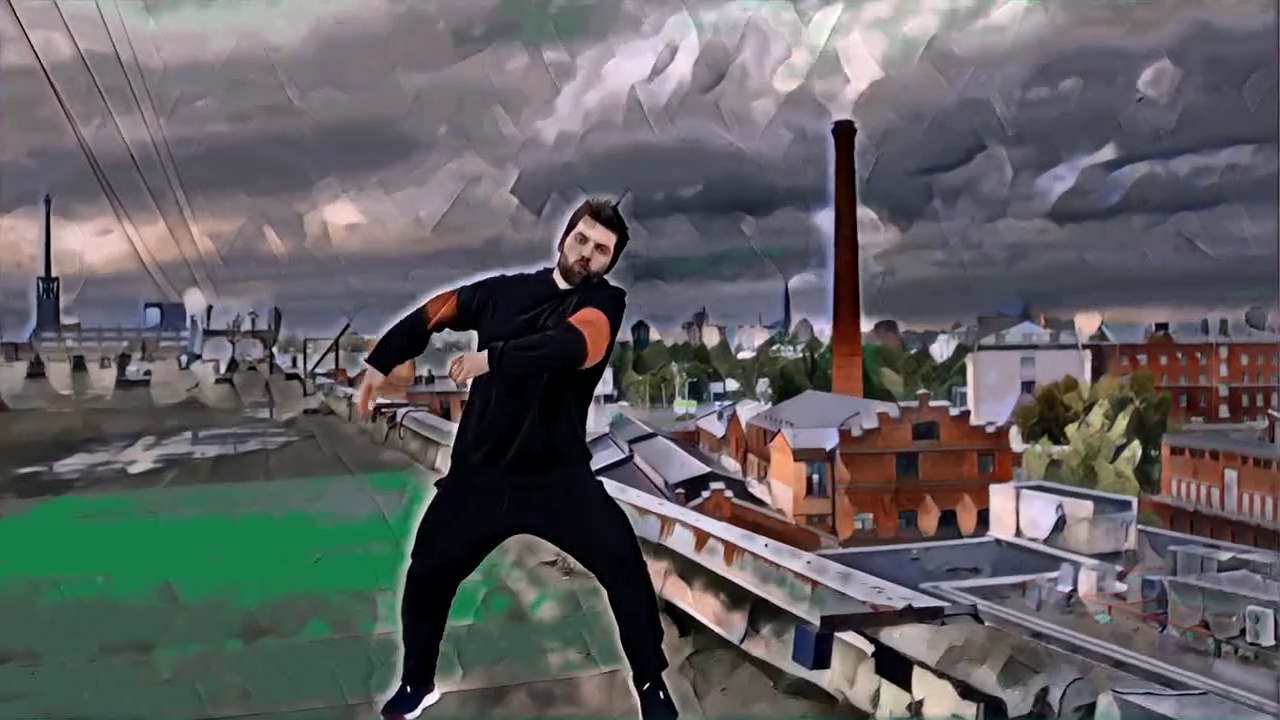}%
		\label{fig:teaser_top_bon}%
	\end{subfigure} \\
	\begin{subfigure}{0.195\textwidth}%
		\includegraphics[trim=0cm 0 8cm 3cm, clip, width=\textwidth]{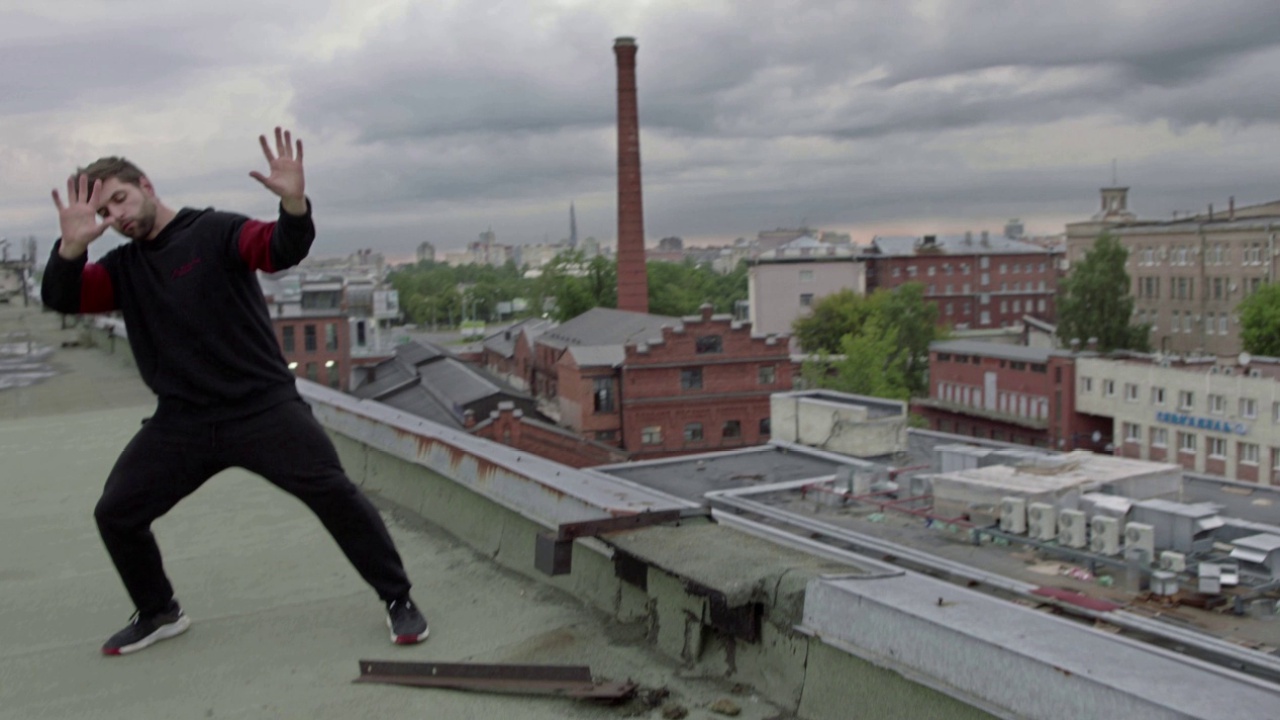}%
		\label{fig:teaser_mid_in}%
	\end{subfigure}\hfill 
	\begin{subfigure}{0.195\textwidth}%
		\includegraphics[trim=0cm 0 8cm 3cm, clip, width=\textwidth]{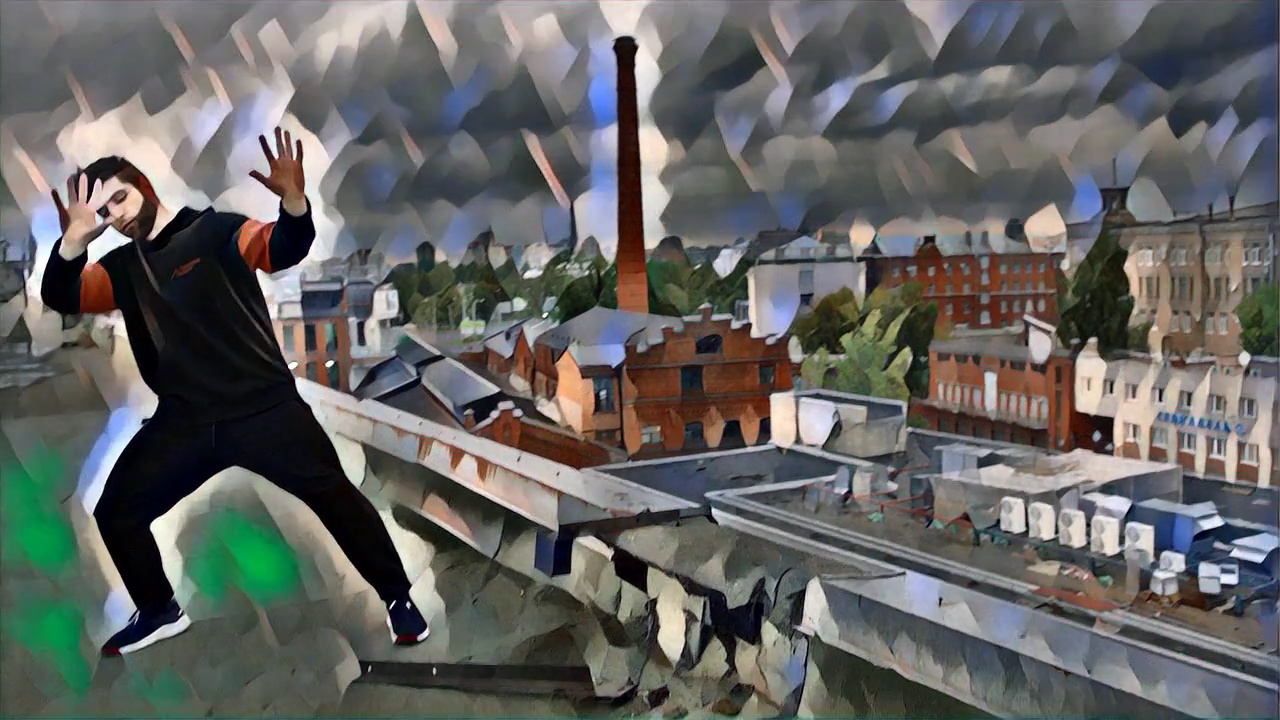}%
		\label{fig:teaser_mid_pr}%
	\end{subfigure}\hfill 
	\begin{subfigure}{0.195\textwidth}%
		\includegraphics[trim=0cm 0 8cm 3cm, clip, width=\textwidth]{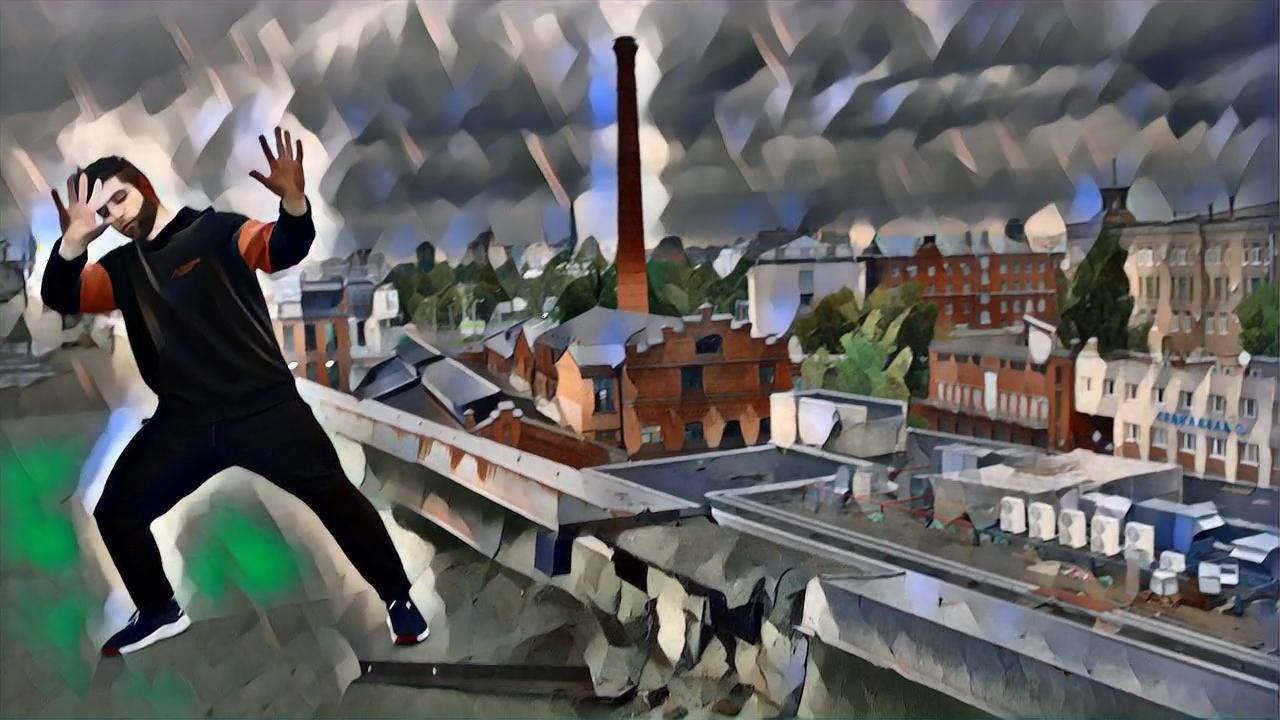}%
		\label{fig:teaser_mid_our}%
	\end{subfigure}\hfill 
	\begin{subfigure}{0.195\textwidth}%
		\includegraphics[trim=0cm 0 8cm 3cm, clip, width=\textwidth]{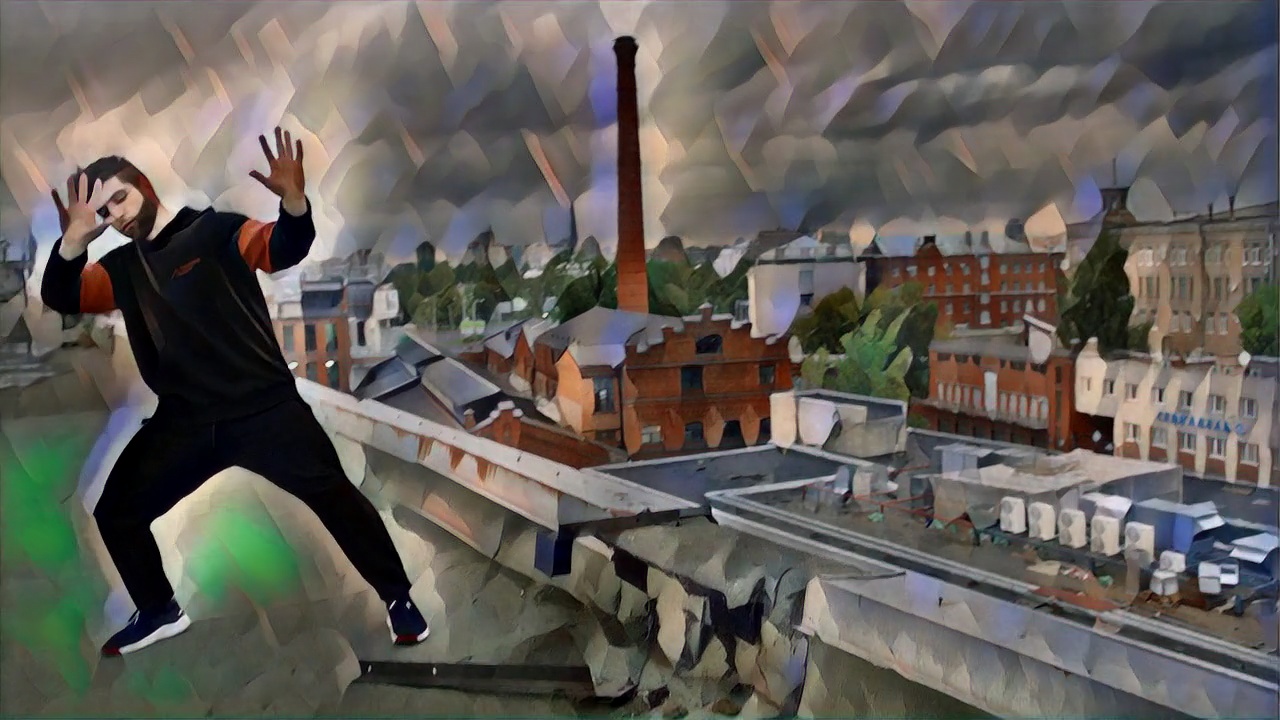}%
		\label{fig:teaser_mid_lai}%
	\end{subfigure}\hfill 
	\begin{subfigure}{0.195\textwidth}%
		\includegraphics[trim=0cm 0 8cm 3cm, clip, width=\textwidth]{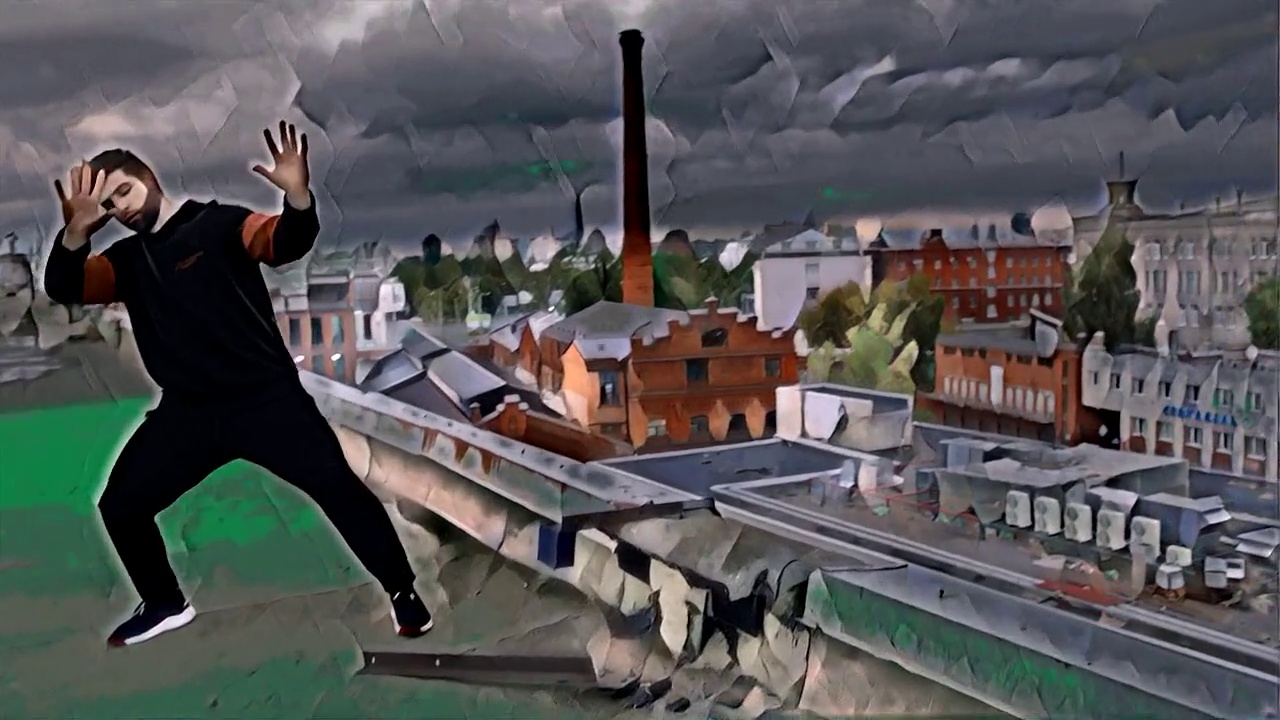}%
		\label{fig:teaser_mid_bon}%
	\end{subfigure} \\
	\begin{subfigure}{0.195\textwidth}%
		\includegraphics[trim=0cm 0 8cm 0, clip, width=\textwidth]{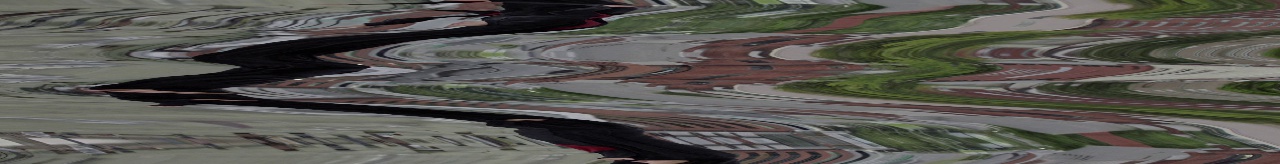}%
		\subcaption{Input}
		\label{fig:teaser_btm_in}%
	\end{subfigure}\hfill 
	\begin{subfigure}{0.195\textwidth}%
		\includegraphics[trim=0cm 0 8cm 0, clip, width=\textwidth]{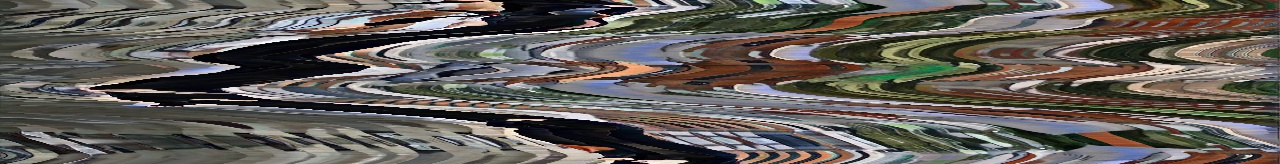}%
		\subcaption{Processed}
		\label{fig:teaser_btm_pr}%
	\end{subfigure}\hfill 
	\begin{subfigure}{0.195\textwidth}%
		\includegraphics[trim=0cm 0 8cm 0, clip, width=\textwidth]{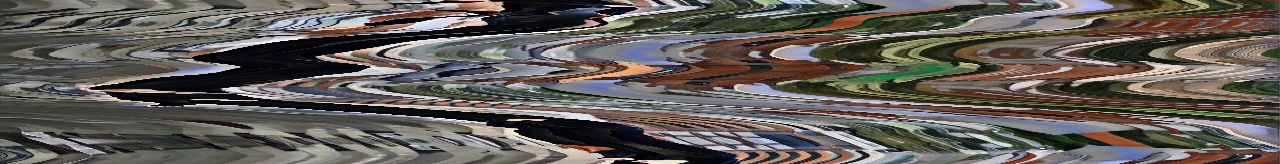}%
		\subcaption{Ours}
		\label{fig:teaser_btm_our}%
	\end{subfigure}\hfill 
	\begin{subfigure}{0.195\textwidth}%
		\includegraphics[trim=0cm 0 8cm 0, clip, width=\textwidth]{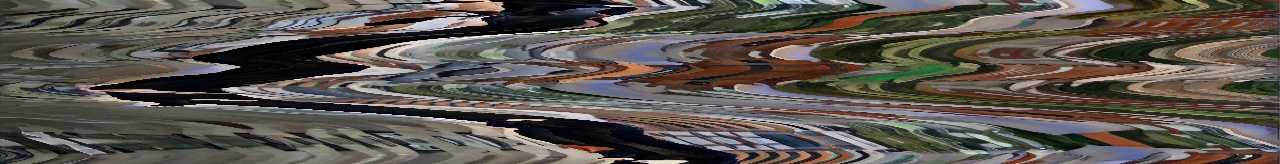}%
		\subcaption{Lai~\etal~\cite{Lai_Learning2018}}
		\label{fig:teaser_btm_lai}%
	\end{subfigure}\hfill 
	\begin{subfigure}{0.195\textwidth}%
		\includegraphics[trim=0cm 0 8cm 0, clip, width=\textwidth]{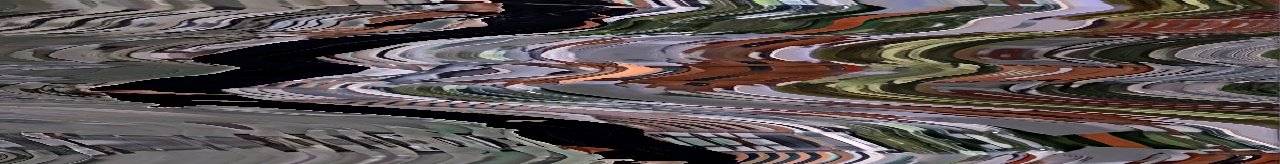}%
		\subcaption{Bonneel~\etal~\cite{Bonneel_Blind2015}}
		\label{fig:teaser_btm_bon}%
	\end{subfigure}
\caption{For the top-row: first two columns depict (a) input and (b) processed result for frame-24,  columns three to five depict the corresponding consistent output using (c) Ours (d) Lai's, and (e) Bonneel's method.
For the mid-row: depict the corresponding results for frame-80.
For the bottom row: we show the \ac{TSI} for the entire video sequence depicting long-term temporal similarity with the per-frame processed output. 
Note, that our method is able to preserve the look and feel of the per-frame processed result in comparison to the method of Lai~\etal which suffers from color bleeding artifacts while the stylized textures are lost for the output of Bonneel~\etal. Please see the supplementary material for video results.}
\label{fig:teaser}
\end{figure*}

For thousands of years, paintings have served as a tool for visual communication and expression.
However, it was not until the late \nth{20} century that computers were used to simulate paintings~\cite{Haeberli_Paint1990}.
In the course of following decades, the field of artistic stylization~\cite{Kyprianidis_SOTA2013} has significantly developed and  extended by learning-based methods, such as \acfp{NST}~\cite{Semmo_NST2017, Jing_Review2020}.
Even though a large number of image stylization techniques exist, extending these to video remains challenging. 
A major obstacle in this regard is the enforcement of temporal coherence between stylized video frames. 
With the proliferation of video streaming applications, stylizing video streams has also become popular, however, the requirements of low-latency processing add additional challenges.
Most of the existing methods, to address the above, can be classified into one of the following four categories:

\begin{description}

\item[Style Specific.] A common approach is to develop a specific method for a particular artistic style and exploit its characteristics for temporal coherency \cite{Bousseau_Video2007, Noris_Temporal_2011}.
Such methods work effectively for the specific target style, however, do not generalize well. 
Many of these specialized approaches have been discussed by B{\'e}nard \etal~\cite{Benard_Coherent_2013}.

\item[Coherent Noise.] Another class of techniques adopts and transforms a generic, temporally-coherent noise function to yield a visually plausible stylized output~\cite{Benard_Noise_2010, Kass_Noise_2011}.
Compared to target-based coherence enforcement \cite{Bousseau_Video2007}, these apply to a wider range of techniques but are limited to scenarios with rapid temporal changes. 

\item[Stylization by Example.] More recently, authors have adopted a stylization-by-example approach to support a wide range of stylization techniques~\cite{Benard_Example_2013, Jamrivska_Stylizing_2019, Texler_Interactive_2020, Futschik_STALP_2021}. 
However, this approach requires the paring of the complete video and keyframe marking.
Thus, by design, it does not apply to video streams.

\item[Consistent Video Filtering.] One can also enable the stylization of video streams using consistent video filtering techniques. 
Existing approaches are either not well-suited for \ac{IB-AR}~\cite{Bonneel_Blind2015, Yao_Occlusion2017} (\Cref{fig:teaser}) or do not provide interactive consistency control~\cite{Lai_Learning2018, Thimonier_learning2021}, which is an essential requirement for artistic rendering~\cite{Fivser_Color2014}.
Currently, the only method that provides interactive consistency control is limited to offline processing and requires pre-processing~\cite{Shekhar_Consistent2019}. 

\end{description}

We aim to develop a temporal consistency enforcement approach for artistic stylization techniques that provides (1) interactive consistency control and (2) online processing to facilitate the application to video streams.

A determining factor towards the slow performance of existing online and interactive consistent video filtering technique~\cite{Bonneel_Blind2015} is the costly step of the optical-flow computation.
Previous works using learning-based methods are able to achieve a considerable accuracy for optical-flow estimation~\cite{Teed_RAFT2020, jiang2021learning}.
However, we argue that such high accuracy is not particularly necessary to enforce temporal consistency for artistic stylization tasks. 
To validate our conjecture, we conduct a user study, wherein the participants prefer the final consistent video output generated using our flow network as compared to that being obtained using \ac{SOTA} approaches.

In contrast to accuracy, less attention has been paid to improving the run-time performance of optical-flow estimation, which is essential for online-interactive editing. 
To this end, we develop a lite optical-flow neural network that runs at a high-speed (approx. 80 \ac{FPS} on mid-tier desktop GPUs) while maintaining sufficient accuracy. 
The compact network is also deployable on mobile devices (iPhones and iPads) where it runs at interactive frame rates (24 \ac{FPS} on iPad Pro 2020).
We use the optical-flow output from the above network to warp neighboring processed frames (for local consistency) and previous consistent output (for global consistency), which allows for interactive global and local temporal consistency control. 
Our approach is able to stabilize incoming video streams in real-time with one frame latency on a consumer desktop GPU at HD resolutions, and, using a fast preset, also in full HD. 

To summarize we present the following contributions: 
\begin{enumerate}
	\item A novel approach for making per-frame stylized videos temporally consistent via an adaptive combination of local and global consistency features which allows for interactive consistency control. 
	\item A lite optical-flow network, to achieve interactive performance, that runs at 80 \ac{FPS} on a mid-tier desktop PC and at 24 \ac{FPS} on a mobile device while achieving reasonable accuracy. 
\end{enumerate}  

Note, that we define artistic stylization as the adaptation of colors, textures, and strokes.
While our approach is effective for most image-based stylization techniques (\eg \acp{NST}, algorithmic filtering), it cannot handle significant shape or content inconsistencies between frames introduced by semantically-driven image synthesis (\eg image-to-image diffusion-based models~\cite{rombach2022high}).
Flow-based warping is insufficient to enforce consistency in such cases.


\begin{figure*}[tb]
\centering
\includegraphics[trim={0.1cm 10.2cm 0.1cm 0.1cm},clip,width=0.935\textwidth]{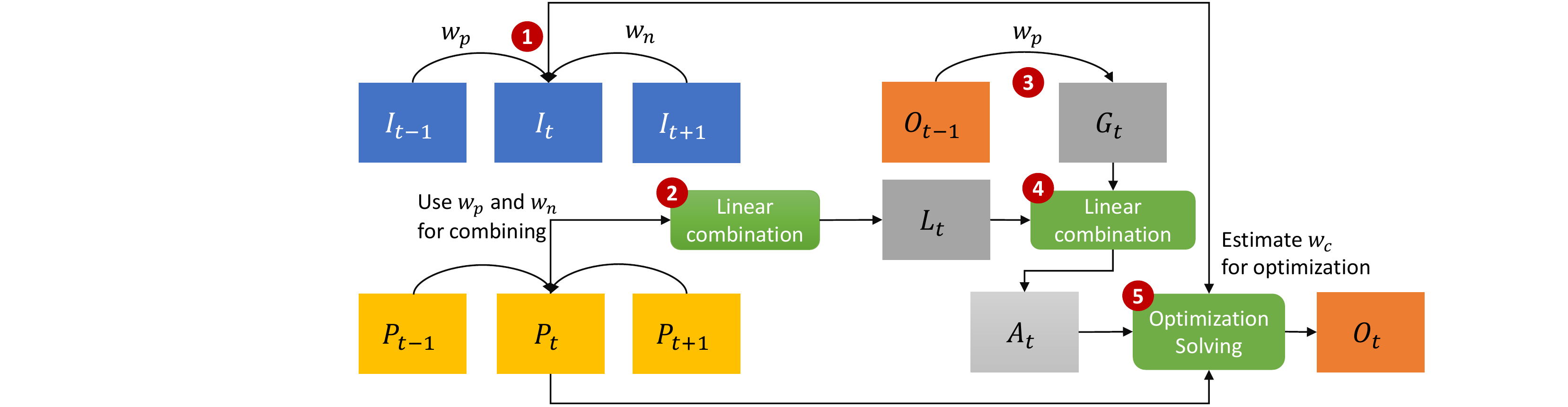}
\caption{Schematic overview of our approach: (1) We start by calculating the warping weights $w_p$ and $w_n$ by applying \Cref{eqn:local_consist_wt} on the input image sequence $I_{t-1}$, $I_t$, $I_{t+1}$. (2) The computed weights are used to linearly combine the per-frame processed sequence $P_{t-1}$, $P_t$, and $P_{t+1}$ to obtain the locally consistent image $L_t$, see \Cref{eqn:local_consist}. (3) To obtain the globally consistent version $G_t$ we warp the output at previous time instance $O_{t-1}$ as depicted in \Cref{eqn:global_consist}. (4) The local and global consistent images, $L_t$ and $G_t$, are linearly combined to obtain a temporally smooth version $A_t$, see \Cref{eqn:local_global_consist}. (5) To include high-frequency details from the per-frame processed result, $A_t$ and $P_t$ are adaptively combined via the optimization in \Cref{eqn:optimize} using the weights $w_c$ (\Cref{eqn:final_con_wt}) to obtain the final result $O_t$.}
\label{fig:flow_diagram}
\end{figure*}

\section{Background \& Related Work}
\label{Sec:RelatedWork}

\paragraph*{Consistent Video Filtering.} 
Lang~\etal~\cite{Lang_Practical2012} propose a solution to enforce temporal consistency for a large class of optimization-based problems via iterative filtering along the motion path. 
Dong~\etal~\cite{Dong_Region2015} address the problem of temporal inconsistency for enhancement algorithms by dividing individual video frames into multiple regions and performing a region-based spatio-temporal optimization. 
Bonneel~\etal~\cite{Bonneel_Blind2015} was the first to present a generalized approach for consistent video filtering which is agnostic to the type of filtering applied on individual video frames.
The method combines gradient-based characteristics of the per-frame processed result with the warped version of the previous-frame output using a gradient-domain-based optimization scheme.
Yao~\etal~\cite{Yao_Occlusion2017} propose a similar approach however considers multiple key-frames for warping-based consistency to avoid problems due to occlusion.  
Both of the approaches assume that the gradient of the processed video is similar to that of the input video and thus cannot handle artistic rendering tasks where new gradients resembling brush strokes are generated as part of the stylization process.
Moreover, due to slow optical-flow computation, they are non-interactive in nature. 
Shekhar~\etal~\cite{Shekhar_Consistent2019} employs a similar formulation as Bonneel~\etal, with the difference of using a temporally denoised version of the current frame for consistency guidance.
However, the temporal denoising requires the complete video as input making the method offline in nature.
Lai~\etal~\cite{Lai_Learning2018} propose the first learning-based technique in this context.
The authors employ perceptual loss to enforce similarity with the processed frames and for consistency make use of short-term and long-term
temporal losses.
Thimonier~\etal~\cite{Thimonier_learning2021} employs a ping-pong loss and a corresponding training procedure for temporal consistency. 
Both learning-based techniques are faster than their optimization-based counterpart since they do not perform optical-flow computation at inference time.
However, these learning-based techniques do not allow to control of the degree of consistency in the final output which is vital for the task of stylization. 
Thus, the above-discussed methods are either non-interactive/offline or do not provide any consistency control at inference time.
Our approach addresses these limitations~(\Cref{tab:consist_cntrl_compare}). 


\paragraph*{Optical Flow for Consistent Filtering.} Both Bonneel~\etal and Yao~\etal use the PatchMatch algorithm~\cite{Barnes_Patch2009} for flow-based warping, however, the slow performance of PatchMatch makes them non-interactive.
Lai~\etal use FlowNet 2.0~\cite{FlowNet2} for flow-based warping to design their short-term and long-term temporal consistency losses. 
FlowNet 2.0 is on par with the quality of state-of-the-art classical methods, however, due to a large number of parameters and operations, achieves only interactive frame rates even on high-end desktop \acp{GPU}.
An improved compact optical-flow \ac{CNN} is proposed by Sun~\etal\cite{PWCNet} -- PWC-Net.
It combines coarse-to-fine estimation with pyramidal image features, correlation, warping, and \ac{CNN}-based estimation.
Furthermore, a refinement \ac{CNN} is stacked at the end to improve the final flow estimate.
PWC-Net is orders of magnitude smaller than FlowNet 2.0 and runs at real-time frame rates using desktop \acp{GPU}.
Liu~\etal\cite{ARFlow} employ their approach to train a similar architecture in an unsupervised setting and achieve reasonable accuracy -- ARFlow.
LiteFlowNet and its successor LiteFlowNet2, both proposed by Hui~\etal\cite{LiteFlowNet,LiteFlowNet2}, have similar compact architectures. Further improvement in accuracy is achieved by models using iterative refinements, such as RAFT~\cite{Teed_RAFT2020} and transformer modules such as GMA~\cite{jiang2021learning}, however, they heavily trade runtime for accuracy. Based on a runtime-accuracy comparison (see \Cref{subsec:liteopticalflow}), we select PWC-Net as a base network to develop a "Lite" flow network with improved performance for interactive consistent filtering.


\paragraph*{Temporal Consistency for Video Stylization.}
Litwinowicz~\cite{Litwinowicz_Processing1997} describes a technique to apply an impressionist effect on images and videos. 
For enforcing temporal coherence, optical flow was used to transform the brush strokes from one frame to the next.  
Winnem\"{o}ller~\etal~\cite{Winnemoeller_RT2006} develop a real-time video and image abstraction framework. 
The authors employ soft quantization that spreads over a larger area, thus significantly reducing temporal incoherence.
Bousseau~\etal~\cite{Bousseau_Video2007} advects texture in forward and backward directions using optical flow for coherent water-colorization of videos. 
Noris~\etal~\cite{Noris_Temporal_2011} preserve the geometric richness of the sketched style in each frame while allowing to successful decrease the temporal noise to a desirable rate.
Fi\v{s}er~\etal~\cite{Fivser_Color2014} propose similar temporal noise control for hand-colored animations.
Noris~\etal and Fi\v{s}er~\etal come close to our vision of providing control over the extent of temporal noise.
However, the method of Noris~\etal can only handle sketchy animations while Fi\v{s}er~\etal requires a clean animation as input onto which temporal noise, extracted from hand-colored examples, is added.  
In comparison, our approach works on already stylized videos and can handle a broad range of stylization techniques.
Numerous such specialized video-based approaches have been discussed by B{\'e}nard~\etal~\cite{Benard_Coherent_2013}.
The above classical \ac{IB-AR} techniques approximate rendering primitives by modifying traditional image filters.
Most often, they use low-level image features for modeling and fail to model structures resembling a particular style. 
Recently, deep \acp{CNN} were successfully used to transfer high-level style attributes from a painting onto a given image~\cite{Gatys_NST2016}. 
Various methods have been proposed to extend the above for videos~\cite{Huang_Real2017, Chen_Coherent2017, Gupta_Characterizing2017, Ruder_Artistic2018, Li_Learning2019, Puy_Flexible2019, Deng_Tang_Dong_Huang_Ma_Xu_2021}. 
Ruder~\etal~\cite{Ruder_Artistic2018} proposes a novel initialization technique and loss functions for consistent stylized output even in cases with large motion and strong occlusion.
The methods of Gupta~\etal~\cite{Gupta_Characterizing2017}, Chen~\etal~\cite{Chen_Coherent2017}, and Huang~\etal~\cite{Huang_Real2017} enforce consistency via certain formulations of temporal loss and use optical-flow based warping only during the training phase thus achieving fast performance. 
Li~\etal~\cite{Li_Learning2019} proposes a method for arbitrary style transfer and shows its applicability to real-time video style transfer by applying style features to consecutive frames using a shallow autoencoder. 
However, we show that our approach, applied to their per-frame processed videos is able to significantly reduce flickering and is more consistent than their stabilized version (see supplementary).
Puy and Pérez~\cite{Puy_Flexible2019} develop a flexible deep \ac{CNN} for controllable artistic style transfer that allows for the addition of a temporal regularizer at testing time to remove the flickering artifacts. 
The above method comes closest in terms of providing some consistency control at test time for \ac{NST}-based methods.
However, they cannot handle classical stylization techniques. 
Keyframe-based Stylization (KBS)~\cite{Benard_Example_2013, Jamrivska_Stylizing_2019, Texler_Interactive_2020, Futschik_STALP_2021} caters to both classical and neural paradigms via priors involving keyframe-based warping.
Nonetheless, it is usually applied as an offline process involving pre-training on the input video.
Moreover, we show that our approach is able to interactively stabilize online KBS approaches such as ~\cite{Texler_Interactive_2020}.
We aim to propose a generic solution that is agnostic to the type of stylization and provides online performance and interactive consistency control. 
\section{Method}
\label{Sec:Method}

\subsection{Temporal Consistency Enforcement}

Given an input video stream $\dots I_{t-1},\;I_t,\;I_{t+1}, \dots$ and its per-frame processed version $\dots P_{t-1},\;P_t,\;P_{t+1}, \dots$,
we seek to find a temporally consistent output $\dots O_{t-1},\;O_t,\;O_{t+1} \dots$.
Our method is agnostic to the stylization technique $f$ applied to each frame, where $P_t\;=\;f(I_t)$.
However, it is necessary for $f$ to not introduce significant shape or content inconsistencies between consecutive frames, as the changes in the stylized frames should correspond to the optical flow (calculated based on the content). 
We initialize the consistent output for the first frame as its per-frame processed result \ie $O_1 = P_1$.
To obtain the output for subsequent frames ($O_t$ at any given instance $t$) we require only a snippet of input ($I_{t-1},I_t, I_{t+1}$) and processed streams ($P_{t-1},P_t, P_{t+1}$), and the consistent output at the previous instance $O_{t-1}$.
For enforcing consistency, we solve the following gradient-domain optimization scheme:
\begin{equation}
\begin{aligned}
\label{eqn:optimize}
E(O_t) = \int_{\Omega}  \bigg( \underbrace{ {|| \nabla O_t - \nabla P_t ||}^2 }_\textrm{data} \; + \;
\underbrace{ w_c{||O_t - A_t||}^2 }_\textrm{smoothness} \;  \bigg)\;d\Omega.
\end{aligned}
\end{equation}
\noindent where $\Omega$ represents the image domain. The \emph{data} term in this optimization enforces similarity with the per-frame processed result~$P_t$ in the gradient domain.
The gradient-based data term ensures that we borrow only the necessary details from the per-frame processed results (in the form of edges) while avoiding inconsistencies.
Thus, high-frequency details are taken from $P_t$ and the \emph{smoothness} term enforces temporal consistency where low-frequency content is taken from the image $A_t$.
The optimization formulation in \Cref{eqn:optimize} is commonly known as \textit{screened Poisson equation} and has been successfully employed for various image editing applications~\cite{Bhat_Fourier_2008, Bhat_Gradient_2010}.
In the context of consistent video filtering, it was first used by Bonneel~\etal~\cite{Bonneel_Blind2015} followed by Shekhar~\etal~\cite{Shekhar_Consistent2019} (\Cref{tab:consist_constraints}). 
However, our novelty is the way in which we construct our \textit{smoothness} term that, unlike previous approaches, considers both \textit{global} and \textit{local} consistency aspects.   
Our novel smoothness term is able to better preserve the color and textures in the stylized output while providing both short-term and long-term temporal consistency.

\begin{figure*}[tb]
	\begin{subfigure}{0.195\textwidth}%
		\includegraphics[trim=14.5cm 13.5cm 18cm 2cm, clip, width=\textwidth]{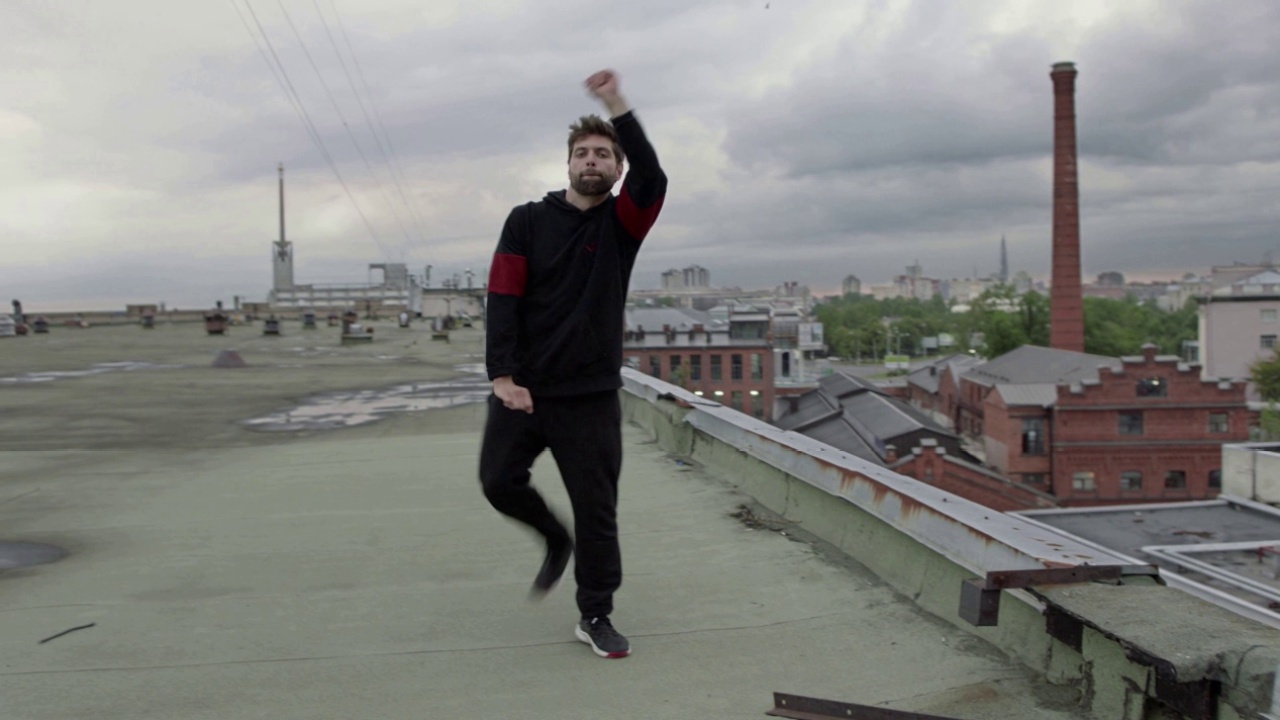}%
		\subcaption{Input}
		\label{fig:modes_top_in}%
	\end{subfigure}\hfill 
	\begin{subfigure}{0.195\textwidth}%
		\includegraphics[trim=14.5cm 13.5cm 18cm 2cm, clip, width=\textwidth]{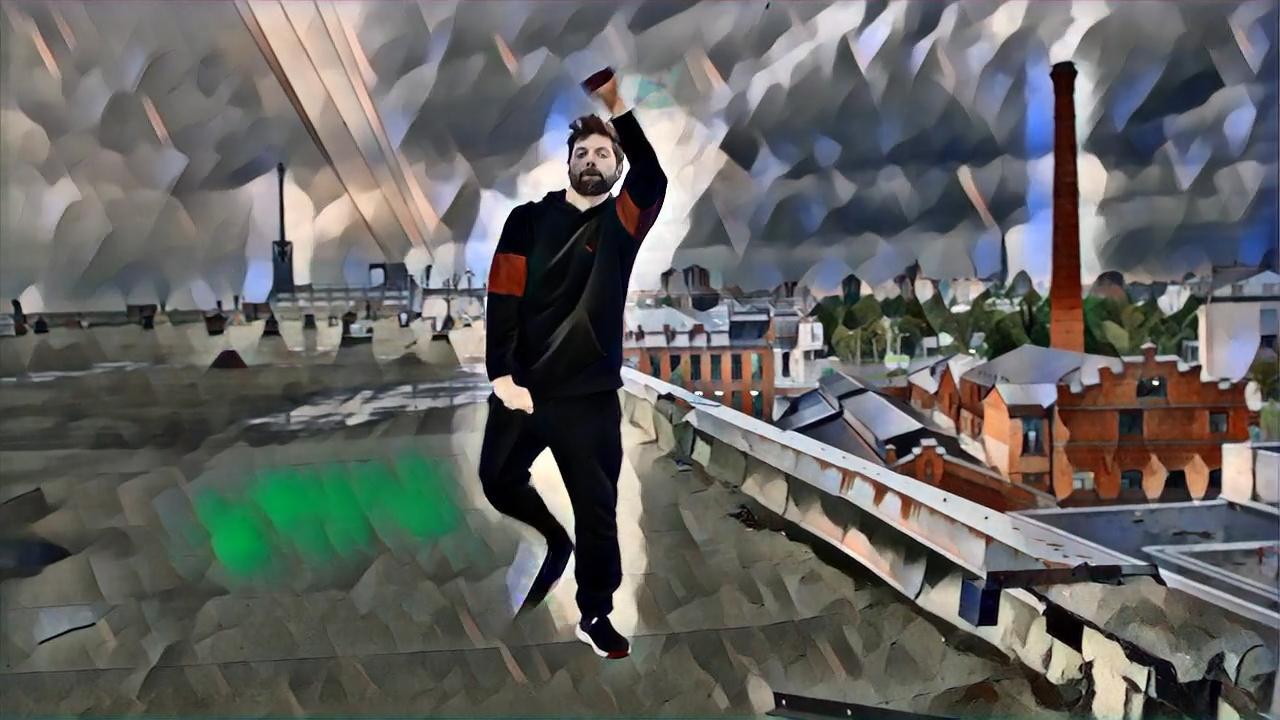}%
		\subcaption{$k_1 = 0.3$}
		\label{fig:modes_top_wp_03}%
	\end{subfigure}\hfill 
	\begin{subfigure}{0.195\textwidth}%
		\includegraphics[trim=14.5cm 13.5cm 18cm 2cm, clip, width=\textwidth]{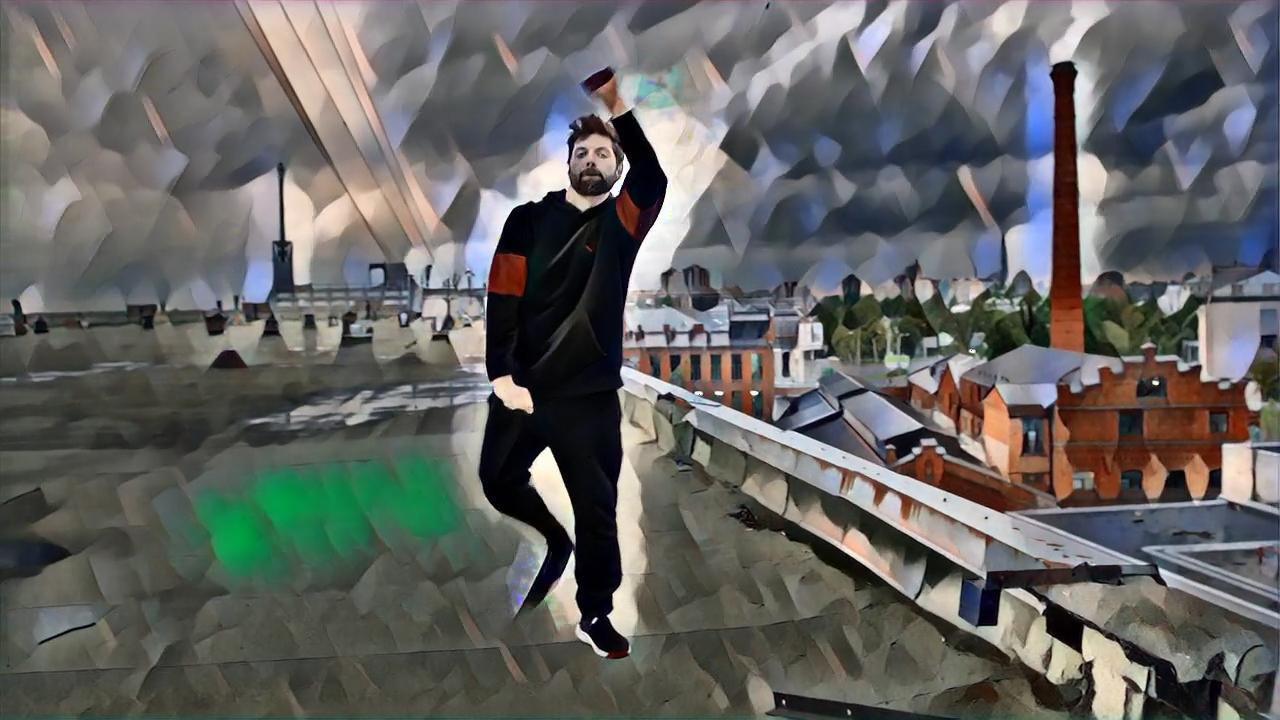}%
		\subcaption{$k_1 = 0.5$}
		\label{fig:modes_top_wp_05}%
	\end{subfigure}\hfill 
	\begin{subfigure}{0.195\textwidth}%
		\includegraphics[trim=14.5cm 13.5cm 18cm 2cm, clip, width=\textwidth]{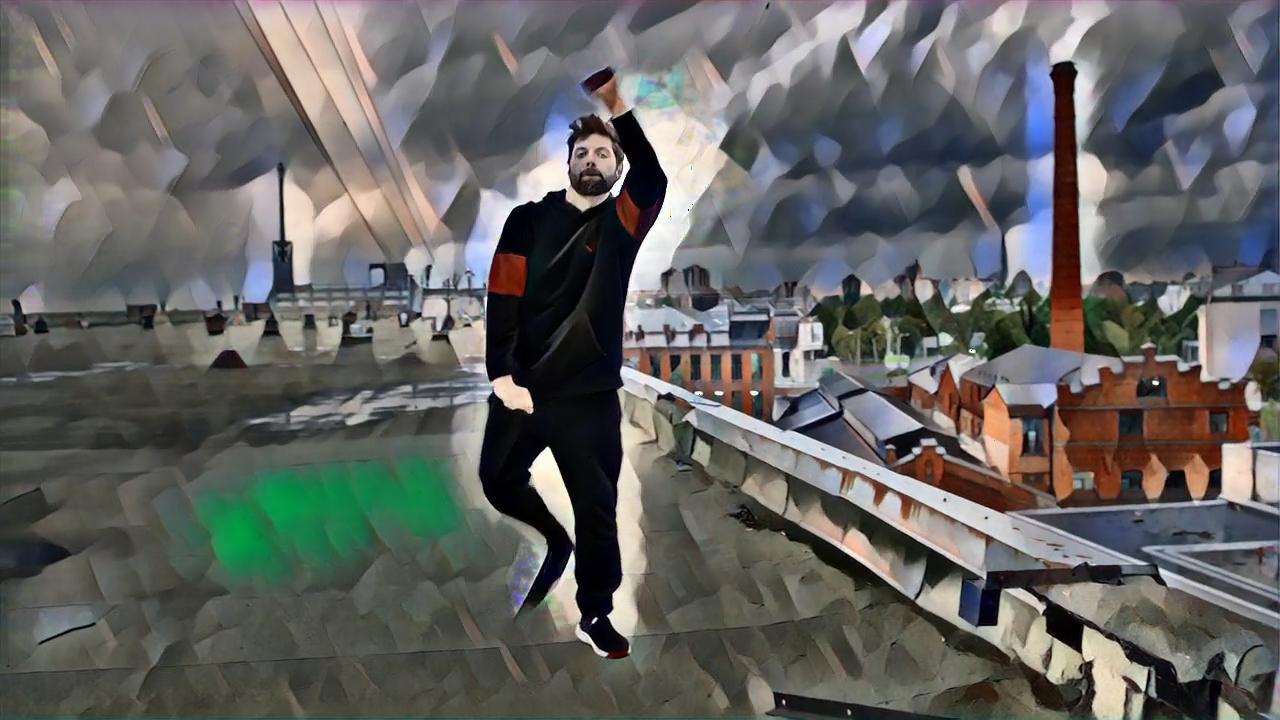}%
		\subcaption{$k_1 = 0.7$}
		\label{fig:modes_top_wp_07}%
	\end{subfigure}\hfill 
	\begin{subfigure}{0.195\textwidth}%
		\includegraphics[trim=14.5cm 13.5cm 18cm 2cm, clip, width=\textwidth]{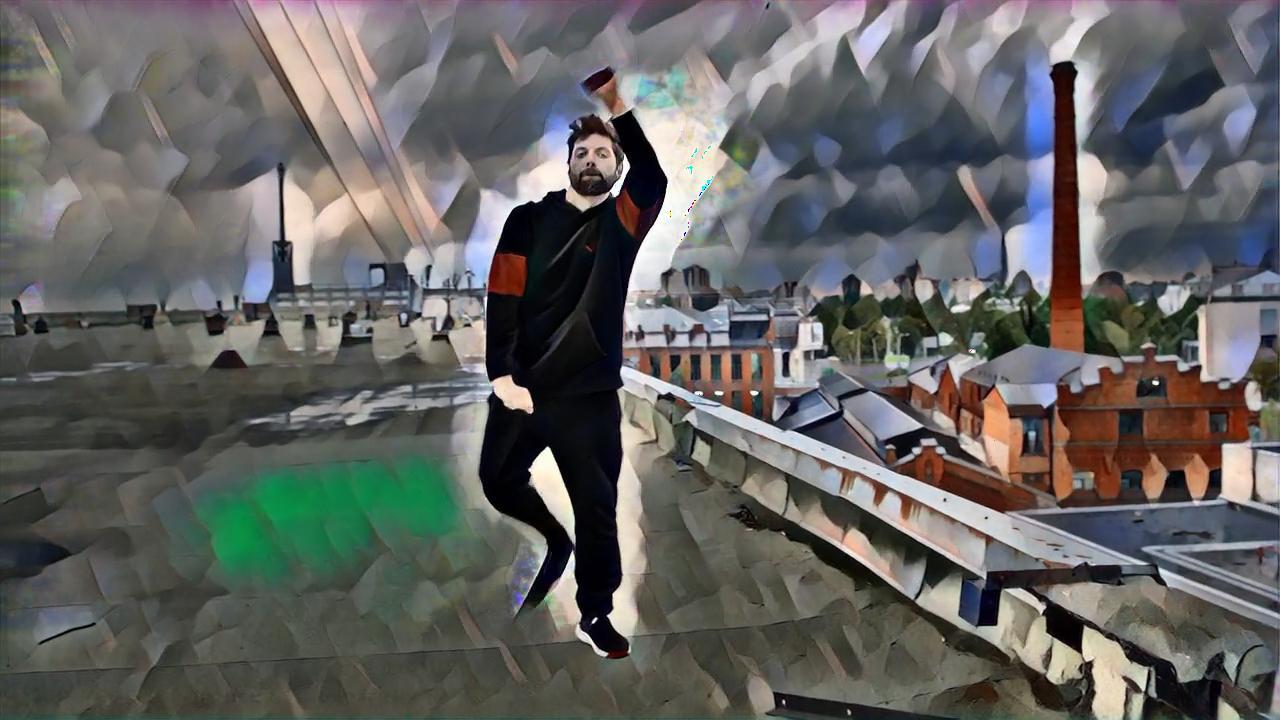}%
		\subcaption{$k_1 = 0.9$}
		\label{fig:modes_top_wp_09}%
	\end{subfigure} \\
	\begin{subfigure}{0.195\textwidth}%
		\includegraphics[trim=14.5cm 13.5cm 18cm 2cm, clip, width=\textwidth]{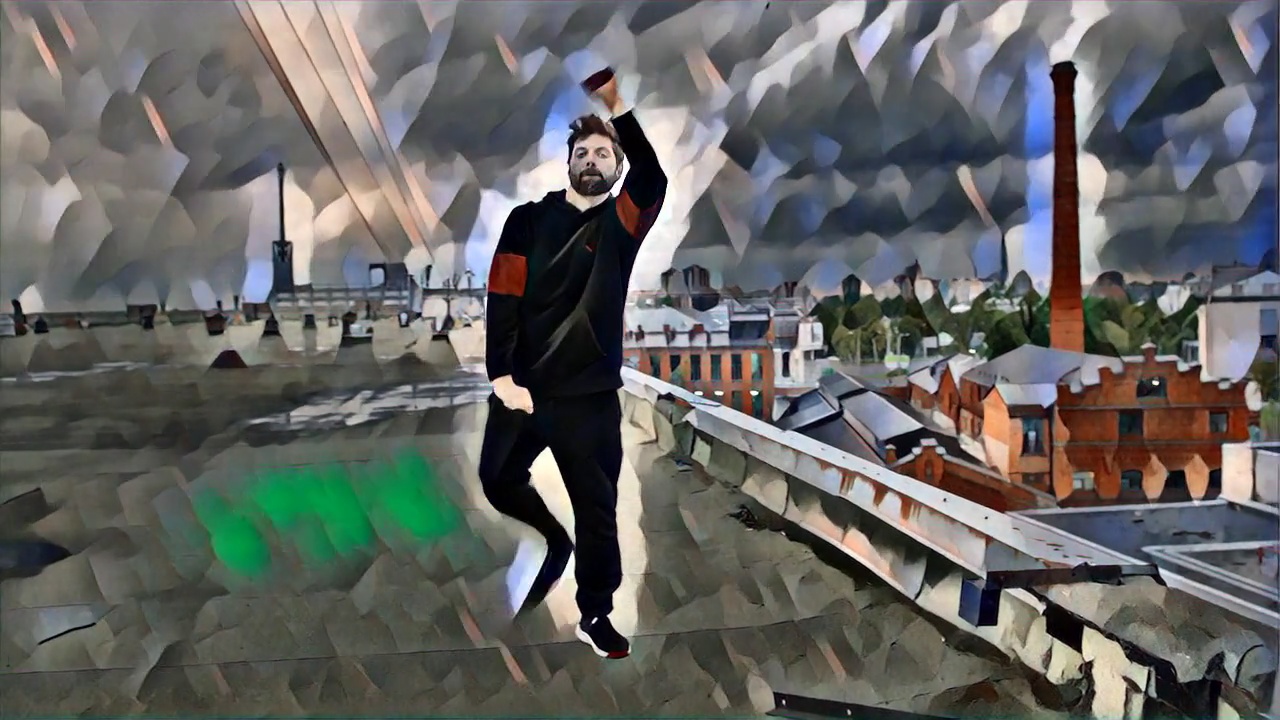}%
		\subcaption{Processed}
		\label{fig:modes_btm_pr}%
	\end{subfigure}\hfill 
	\begin{subfigure}{0.195\textwidth}%
		\includegraphics[trim=14.5cm 13.5cm 18cm 2cm, clip, width=\textwidth]{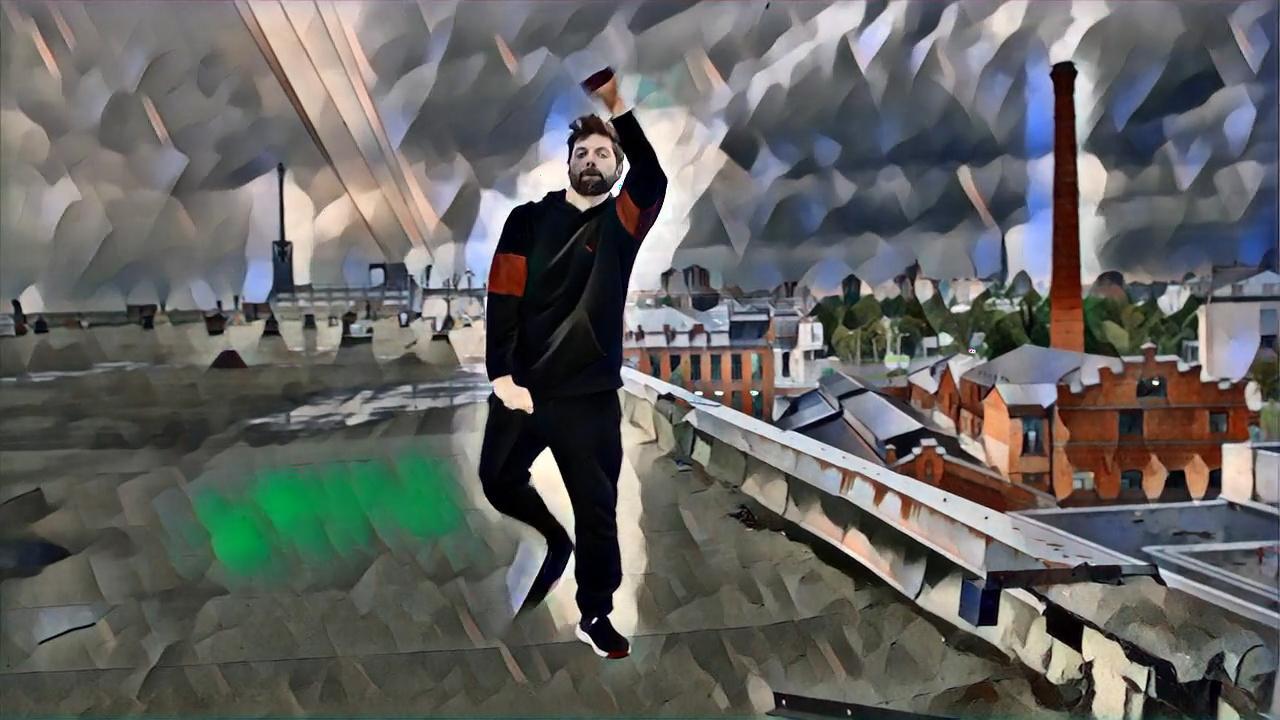}%
		\subcaption{$\lambda = 0.1$}
		\label{fig:modes_lam_01}%
	\end{subfigure}\hfill 
	\begin{subfigure}{0.195\textwidth}%
		\includegraphics[trim=14.5cm 13.5cm 18cm 2cm, clip, width=\textwidth]{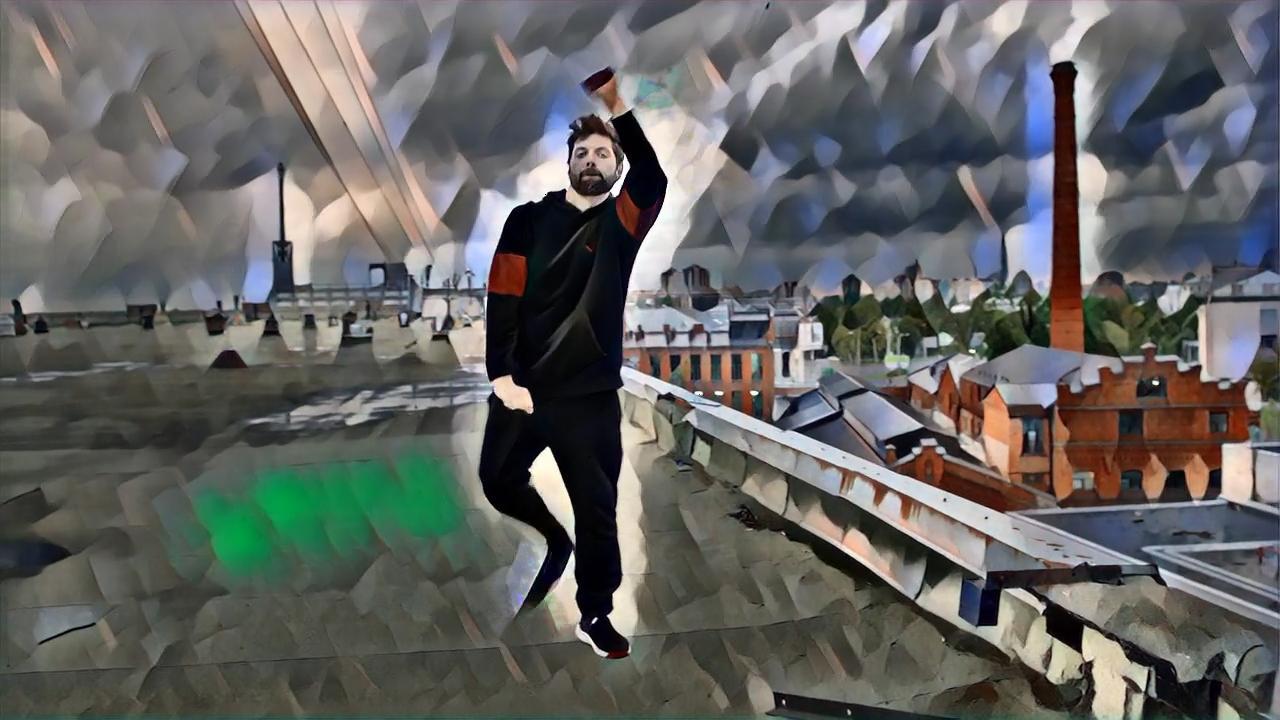}%
		\subcaption{$\lambda = 1.0$}
		\label{fig:modes_lam_1}%
	\end{subfigure}\hfill 
	\begin{subfigure}{0.195\textwidth}%
		\includegraphics[trim=14.5cm 13.5cm 18cm 2cm, clip, width=\textwidth]{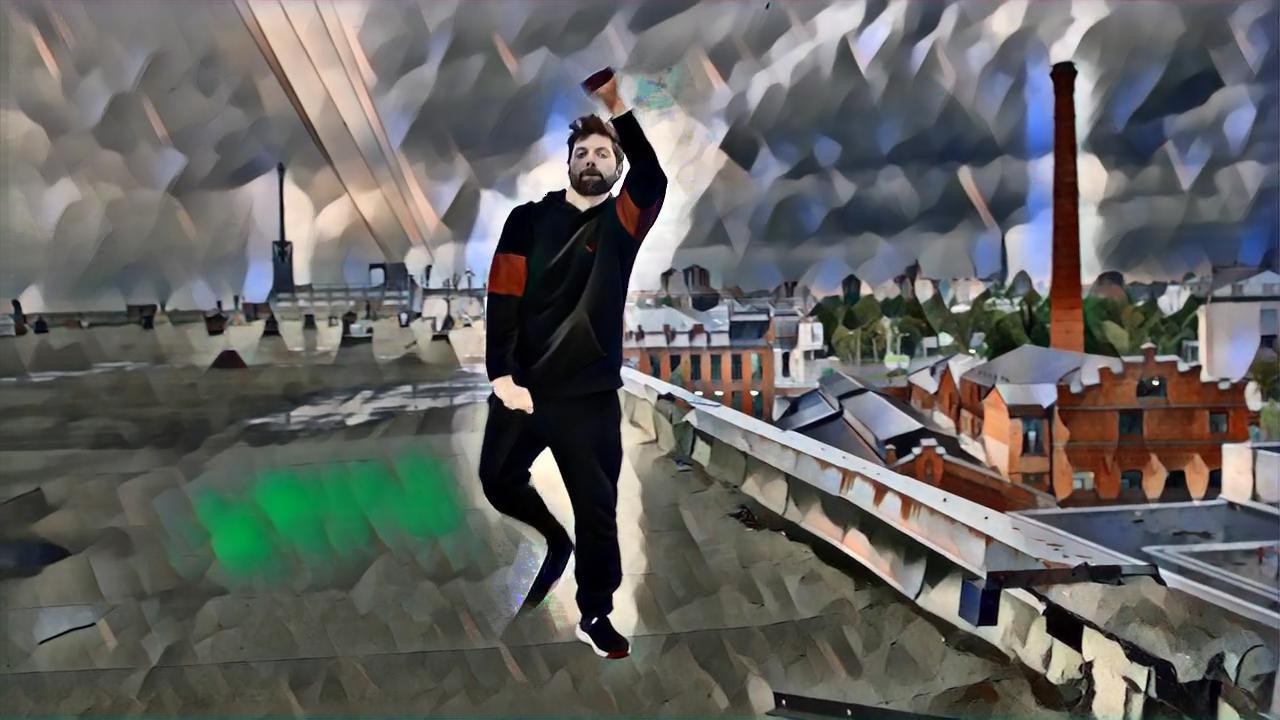}%
		\subcaption{$\lambda = 5.0$}
		\label{fig:modes_lam_5}%
	\end{subfigure}\hfill 
	\begin{subfigure}{0.195\textwidth}%
		\includegraphics[trim=14.5cm 13.5cm 18cm 2cm, clip, width=\textwidth]{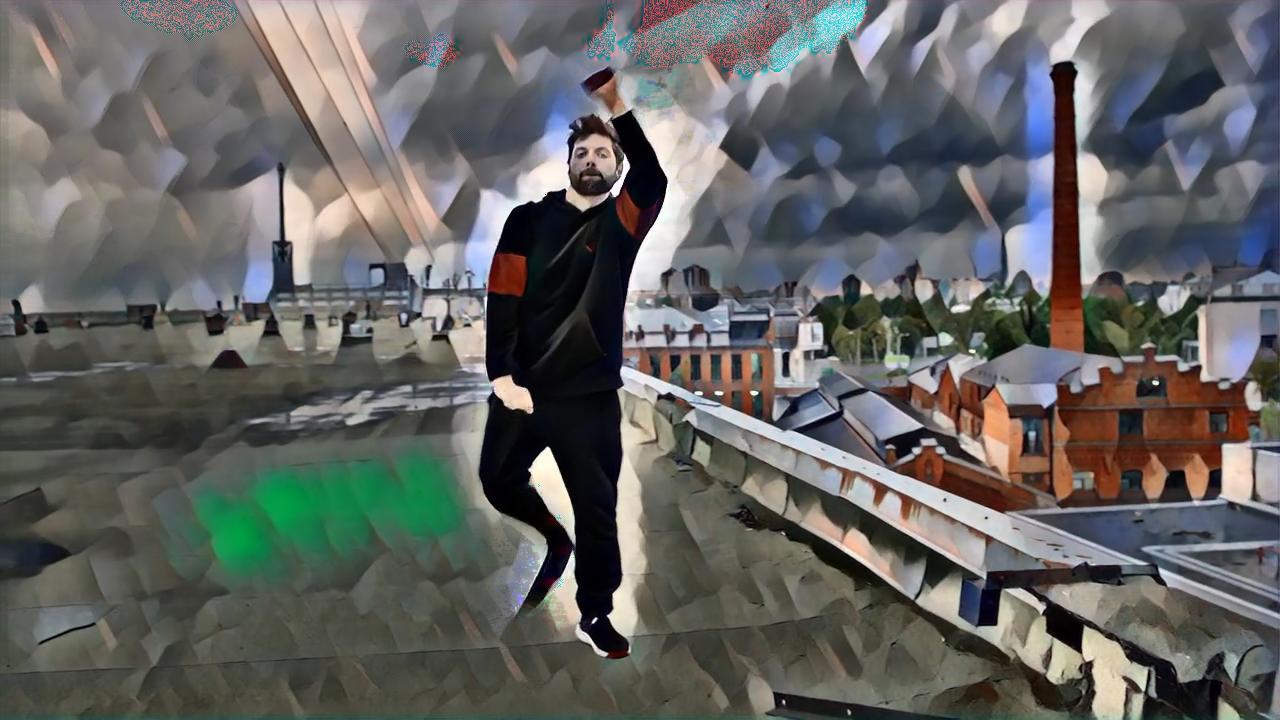}%
		\subcaption{$\lambda = 7.06$}
		\label{fig:modes_lam_7}%
	\end{subfigure} 
	\caption{The level of consistency in the final output can be controlled via parameters $k_1$ and $\lambda$. Here we show how the final result varies by increasing these, for lower values the consistency is negligible and the results (\Cref{fig:modes_top_wp_03} and \Cref{fig:modes_lam_01}) visually look similar to the per-frame processed output (\Cref{fig:modes_top_wp_03}). For higher values, we start observing artifacts due to ghosting and/or optimization (\Cref{fig:modes_top_wp_09} and \Cref{fig:modes_lam_7}).}
	\label{fig:modes_of_cc}
\end{figure*}


\begin{table}[t]
	\caption{Constituent elements of smoothness term in \Cref{eqn:optimize} for different methods. Here, $w_s$ and $T_d$ refers to saliency-based weights and temporally-denoised image respectively, introduced by Shekhar~\etal~\cite{Shekhar_Consistent2019}}
	\label{tab:consist_constraints}
	\centering
	\begin{tabular}{|l|l|l|}
		\hline
		Method                            & Weight     & Consistent Image          \\ \hline
		Ours                              & $w_c$                  & $A_t$                     \\ 
		Bonneel~\etal~\cite{Bonneel_Blind2015} & $w_p$                  & $\Gamma(O_{t-1})$                  \\
		Shekhar~\etal~\cite{Shekhar_Consistent2019} & $w_s$  & $T_d$ \\ \hline
	\end{tabular}
\end{table}


\paragraph*{Local Consistency.} For enforcing temporal consistency at a local level, we use optical flow to warp neighboring per-frame processed results to the current time instance $t$. 
This is performed by computing an adaptive combination of (1) warped previous per-frame processed image $\Gamma(P_{t-1})$, (2) warped next per-frame processed image $\Gamma(P_{t+1})$, and (3) the current per-frame processed image $P_t$, where $\Gamma$ is the warping function.
By including both backward and forward warping in our formulation, we are able to significantly reduce artifacts due to occlusion and flow inaccuracies. 
The linear combination of (1), (2), and (3) gives us a locally consistent version $L_t$ where,
\begin{equation}
\label{eqn:local_consist}
L_t = (1 - ( w_p +  w_n)) \cdot P_{t} \; + \; w_p \cdot \Gamma(P_{t-1}) \; + \; w_n \cdot \Gamma(P_{t+1}). 
\end{equation}
The weights $w_p$ and $w_n$ capture the inaccuracies in the warping of previous and next frames respectively and are defined as follows:
\begin{equation}
\label{eqn:local_consist_wt}
\begin{aligned}
w_p &= \exp \left(-\alpha{||I_{t} - \Gamma(I_{t-1})||}^2\right) \;\text{and} \\
w_n &= \exp \left(-\alpha{||I_{t} - \Gamma(I_{t+1})||}^2\right). \\
\end{aligned}
\end{equation}
\noindent
In order to also incorporate contribution from $P_t$, we clamp the weights $w_p$ and $w_n$ as follows: $w_p \in [0, k_1]$ and $w_n \in [0, k_2]$, where $k_1$ and $k_2$ are two constants and their sum is less than one, \ie $0 < (k_1 + k_2) < 1$. 
The locally consistent image sequence given by $L_t$ has improved temporal consistency over the per-frame processed output, however, it still has visible flickering artifacts. 
Thus, the reduction in flickering due to the warping of only one temporal neighbor is not sufficient.
To further improve consistency, one can warp more neighboring frames around the current time instance $t$.
As we increase the temporal window size for such an adaptive combination it has a denoising effect leading to further reduction in flickering.  
The temporal denoising performed by Shekhar~\etal~\cite{Shekhar_Consistent2019}, for enforcing consistency, can be considered as a specific example of the above scenario. 
However, for interactive stylization, warping more frames to the current instance is not feasible due to time constraints.
Moreover, in the case of video streams, we do not have frames to warp from the forward temporal direction.

\paragraph*{Global Consistency.} 
In order to overcome this limitation, existing techniques \cite{Bonneel_Blind2015, Lai_Learning2018} adopt a global approach.    
For global consistency, one can consider the previous stabilized output $O_{t-1}$ and enforce similarity with its warped version $G_t$ where,   
\begin{equation}
\label{eqn:global_consist}
G_t =  \Gamma(O_{t-1}).
\end{equation}
To enforce only global temporal smoothness, we replace $A_t$ with $G_t$ in \Cref{eqn:optimize}.
Further, in order to compensate for optical-flow inaccuracies, the smoothness term is weighted using $w_p$ (\ie $w_c = w_p$) in \Cref{eqn:optimize}.

However, considering only global consistency for flicker reduction leads to a loss of stylization (in terms of colors and textures) and local temporal variations in the final output.
Moreover, in this case, any warping error (due to flow inaccuracies) or noise (as part of the stylization process) keeps getting propagated to future frames.
Due to the above factors, such an approach only gives plausible results where the gradients of the original video are similar to the gradients of the processed video. 
The above does not hold for the task of stylization where stylistic elements such as brush strokes, textures, or stroke textons \cite{zhu2005textons}, in general, can vary largely between frames even for small changes in input gradient.

\paragraph*{Combining Local and Global Consistency.}
For preserving local temporal variations (in terms of look and feel) while significantly reducing flickering artifacts, we linearly combine globally and locally consistent images $G_t$ and $L_t$ respectively,
\begin{equation}
\label{eqn:local_global_consist}
A_t = w_p \cdot G_t \; + \; (1 - w_p) \cdot L_t.  
\end{equation}
We use the adaptively combined image $A_t$ as our reference for consistency while enforcing temporal smoothness in \Cref{eqn:optimize}.
The upper limit of weight $w_p$ (\ie $k_1$) can be increased to increase the influence of global-temporal smoothness and vice versa. 
Further, the influence of the smoothness term is controlled by per-pixel consistency weights~$w_c$.
We would like to invoke the smoothness term only when the warping accuracy is sufficiently high.
To this end, we construct a warped version of the input image similar to $L_t$ as, 
\begin{equation}
	\label{eqn:warp_input}
	{A^I}_t = (1 - ( w_p +  w_n)) \cdot I_{t} \; + \; wp \cdot \Gamma(I_{t-1}) \; + \; w_n \cdot \Gamma(I_{t+1}). 
\end{equation} 
Only when the input image $I_t$ is similar to ${A^I}_t$, the smoothness term is invoked.
To measure this similarity, we use the weight $w_c$, 
\begin{equation}
	\label{eqn:final_con_wt}
	w_c = \lambda \cdot \exp \left(-\alpha{||I_{t} - {A^I}_t||}^2\right).
\end{equation}
The parameter $\lambda$ is used to scale up or down the weight $w_c$. 

\paragraph*{Consistency Control Modes.}

The above adaptive combination of local and global consistency provides two different ways of consistency control in the final output. 
By increasing the upper limit of $w_{p}$, \ie $k_1$ we can increase the proportion of global consistency in the adaptively combined image $A_t$ and vice versa.
On the other hand, the optimization parameter $\lambda$ dictates how close the output $O_t$ will be to the adaptively combined image $A_t$.
Thus, the level of consistency in the final output can be controlled in two different ways: 
(1) by setting the upper limit of parameter $w_p$, \ie $k_1$ or (2) by scaling the weight parameter $\lambda$.  
For low values of $k_1$ (\Cref{fig:modes_top_wp_03}), the consistency enforced is negligible and the final result resembles the per-frame processed output (\Cref{fig:modes_btm_pr}).
However, for higher values, we start observing noisy ghosting artifacts (\Cref{fig:modes_top_wp_09}). 
The higher values for $k_1$ translate to using only global consistency which results in the accumulation of flow inaccuracies visualized as ghosting artifacts.
Similarly, for lower values of $\lambda$ (\Cref{fig:modes_lam_01}), the final result is visually similar to the per-frame processed output (\Cref{fig:modes_btm_pr}).
However, for higher values, the optimization becomes unstable  resulting in noisy optimization-based artifacts. (\Cref{fig:modes_lam_7}). 

\paragraph*{Optimization Solver.}
The energy terms in \Cref{eqn:optimize} are smooth and convex in nature, which allows a straightforward energy minimization with respect to $O_t$. 
To this end, we employ an iterative approach thus avoiding: ($i$) storage of a large matrix in memory and ($ii$) further estimating its inverse.
Moreover, an iterative approach allows us to stop the solver once we have achieved visually plausible results. 
An iterative update ${O_t}^{j+1}$ is obtained by employing \ac{SGD} with momentum \cite{Qian_Momentum_1999},
\begin{equation}
	{O_t}^{j+1} = {O_t}^{j} - \eta \nabla E({O_t}^{j}) + \kappa ({O_t}^{j} - {O_t}^{j-1}).
\end{equation}
where $\eta$ and $\kappa$ are the step size parameters, $\nabla E$ is the energy gradient with respect to ${O_t}$, and $j$ is the iteration count.
For most of our experiments, $\eta = 0.15$ and $\kappa = 0.2$ yield plausible results.
We consider the trade-off between performance \vs accuracy as stopping criteria and do not compute energy residue for this purpose.
To obtain a consistent output while having interactive performance, we empirically determine $150$ iterations to be sufficient.
The optimization is stable for the given parameter settings and early stopping is only employed for computational gain.

An integral aspect common to both our \textit{local} and \textit{global} consistency is the warping function $\Gamma$. 
Apart from the number of solver iterations, for interactive performance the above warping should also happen at a fast rate -- which in turn necessitates fast optical-flow estimation.

\begin{figure}[tb]
\centering
\includegraphics[trim=0cm 3cm 0.5cm 3cm, clip, width=0.9\textwidth]{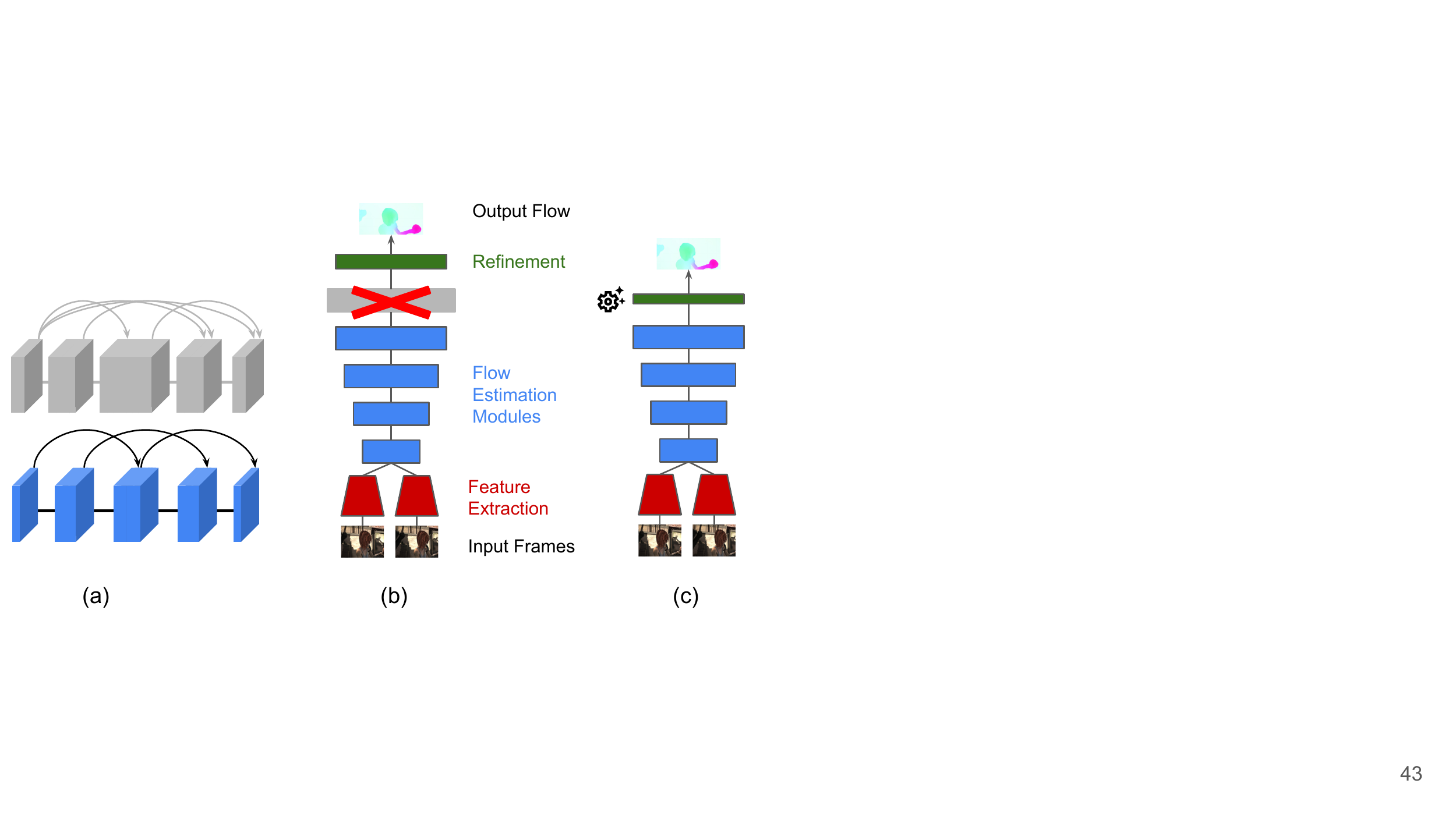}
\caption{Modification of the PWC-Net~\cite{PWCNet} architecture for real-time performance. We apply the following network compression steps: (a) Replace DenseNet connections with light ones, (b) Reduce the number of flow estimators, and (c) Replace dense connections in the refinement module with separable convolutions.}
\label{RTOF:Fig:Full_pipeline}
\end{figure}

\subsection{Lite Optical-Flow Network}
\label{subsec:liteopticalflow}
We aim to obtain a flow network capable of running at high-speed on consumer hardware with reasonable accuracy. 
To this end, we start by selecting an existing \ac{CNN}-based optical flow estimation technique, based on accuracy \vs run-time analysis. 
After the selection of a base network, we perform further optimization steps to increase the performance as outlined in~\Cref{RTOF:Fig:Full_pipeline}.

\begin{figure}[tb]
	\centering
\begin{tikzpicture}

\begin{axis}[
height=0.315\textwidth,
tick align=outside,
tick pos=left,
width=0.5\textwidth,
x grid style={white!69.0196078431373!black},
xlabel={Sintelfinal-test EPE (lower=better)},
xmin=0, xmax=9,
xtick style={color=black},
xticklabel={\pgfmathprintnumber{\tick}px},
y grid style={white!69.0196078431373!black},
ylabel={FPS (higher=better)},
ymin=0, ymax=87,
ytick style={color=black}
]
\addplot [
  mark=*,
  only marks,
  scatter,
  scatter/@post marker code/.code={%
  \endscope
},
  scatter/@pre marker code/.code={%
  \expanded{%
  \noexpand\definecolor{thispointdrawcolor}{RGB}{\drawcolor}%
  \noexpand\definecolor{thispointfillcolor}{RGB}{\fillcolor}%
  }%
  \scope[draw=thispointdrawcolor, fill=thispointfillcolor]%
},
  visualization depends on={value \thisrow{draw} \as \drawcolor},
  visualization depends on={value \thisrow{fill} \as \fillcolor}
]
table{%
x  y  draw  fill
5.74 10.4 31,119,180 31,119,180
8.36 13.77 31,119,180 31,119,180
4.9 29.97 31,119,180 31,119,180
5.67 36.16 31,119,180 31,119,180
4.69 35.21 31,119,180 31,119,180
4.4 6.16 31,119,180 31,119,180
2.86 7.03 31,119,180 31,119,180
2.47 5.85 31,119,180 31,119,180
7.43 84.0 255,0,0 255,0,0
};
\draw (axis cs:5.1,10.45) ++(5pt,2pt) node[
  scale=0.8,
  anchor=base west,
  text=black,
  rotate=0.0
]{flownet2};
\draw (axis cs:7.4,13.99) ++(5pt,2pt) node[
  scale=0.8,
  anchor=base west,
  text=black,
  rotate=0.0
]{spynet};
\draw (axis cs:4.7,28.0) ++(5pt,2pt) node[
  scale=0.8,
  anchor=base west,
  text=black,
  rotate=0.0
]{pwcnet};
\draw (axis cs:5.5,36.0) ++(5pt,2pt) node[
  scale=0.8,
  anchor=base west,
  text=black,
  rotate=0.0
]{arflow};
\draw (axis cs:4.69,35.0) ++(-25pt,5pt) node[
  scale=0.8,
  anchor=base west,
  text=black,
  rotate=0.0
]{liteflownet2};
\draw (axis cs:4.1,6.22) ++(5pt,2pt) node[
  scale=0.8,
  anchor=base west,
  text=black,
  rotate=0.0
]{vcn};
\draw (axis cs:2.4,7.2) ++(5pt,2pt) node[
  scale=0.8,
  anchor=base west,
  text=black,
  rotate=0.0
]{raft};
\draw (axis cs:1.39,4.2) ++(5pt,2pt) node[
  scale=0.8,
  anchor=base west,
  text=black,
  rotate=0.0
]{gma};
\draw (axis cs:7.3,79) ++(5pt,2pt) node[
scale=0.8,
anchor=base west,
text=black,
rotate=0.0
]{Ours};
\end{axis}

\end{tikzpicture}

	\caption{Accuracy \vs run-time performance of existing methods measured on \mbox{Sintel Final (Test set)~\cite{SintelDataset}}. The \ac{EPE} metric measures Euclidean distance (in pixels) between ground truth and predicted optical flow vectors. Note how our method achieves a high \ac{FPS} while being accurate enough for temporal consistency enforcement.}
	\label{RTOF:Fig:EvalExist}
\end{figure}
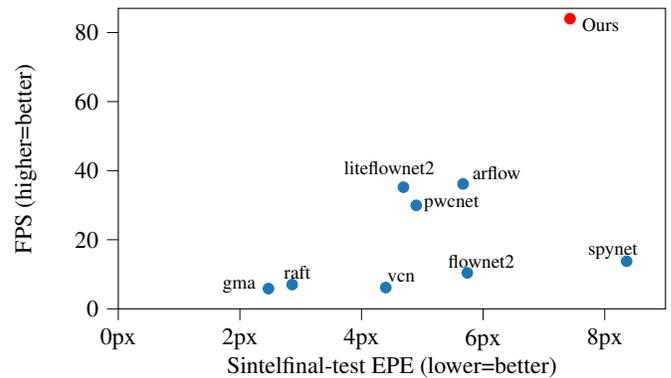

\begin{figure*}[tb]
	\centering
	\begin{subfigure}[t]{0.5\linewidth}
	\vskip 0pt
\begin{tikzpicture}

\begin{axis}[
height=0.55\textwidth,
tick align=outside,
tick pos=left,
width=1.0\columnwidth,
x grid style={white!69.0196078431373!black},
xlabel={(a) Sintelfinal-train \ac{EPE} (lower=better)},
xmin=1.5, xmax=5.5,
xtick style={color=black},
xticklabel={\pgfmathprintnumber{\tick}px},
y grid style={white!69.0196078431373!black},
ylabel={FPS (higher=better)},
ymin=15, ymax=100,
ytick style={color=black}
]
\addplot [
  mark=*,
  only marks,
  scatter,
  scatter/@post marker code/.code={%
  \endscope
},
  scatter/@pre marker code/.code={%
  \expanded{%
  \noexpand\definecolor{thispointdrawcolor}{RGB}{\drawcolor}%
  \noexpand\definecolor{thispointfillcolor}{RGB}{\fillcolor}%
  }%
  \scope[draw=thispointdrawcolor, fill=thispointfillcolor]%
},
  visualization depends on={value \thisrow{draw} \as \drawcolor},
  visualization depends on={value \thisrow{fill} \as \fillcolor}
]
table{%
x  y  draw  fill
2.507458 29.97 31,119,180 31,119,180
2.825573 40.26 255,0,0 255,0,0
3.903354 43.49 255,0,0 255,0,0
3.104323 53.49 255,0,0 255,0,0
3.46062 81.28 255,0,0 255,0,0
4.236251 56.07 255,0,0 255,0,0
3.659757 46.86 255,0,0 255,0,0
3.229143 79.12 255,0,0 255,0,0
3.169556 83.53 0,128,0 0,128,0
3.443847 86.61 255,0,0 255,0,0
};
\addplot [semithick, white!50.1960784313725!black]
table {%
2.507458 29.97
2.825573 40.26
};
\addplot [semithick, white!50.1960784313725!black]
table {%
2.825573 40.26
3.229143 79.12
};
\addplot [semithick, white!50.1960784313725!black]
table {%
3.229143 79.12
3.169556 83.53
};
\draw (axis cs:2.10,22) ++(5pt,2pt) node[
  scale=0.8,
  anchor=base west,
  text=black,
  rotate=0.0
]{pwcnet};
\draw (axis cs:2.7,40.26) ++(5pt,2pt) node[
  scale=0.8,
  anchor=base west,
  text=black,
  rotate=0.0
]{our-5light};
\draw (axis cs:3.903354,43.49) ++(5pt,0pt) node[
  scale=0.8,
  anchor=base west,
  text=black,
  rotate=0.0
]{our-5light-5sep};
\draw (axis cs:2.95,54.5) ++(5pt,2pt) node[
  scale=0.8,
  anchor=base west,
  text=black,
  rotate=0.0
]{our-5light-2sep};
\draw (axis cs:3.46062,81.28) ++(5pt,2pt) node[
  scale=0.8,
  anchor=base west,
  text=black,
  rotate=0.0
]{our-4light-1sep};
\draw (axis cs:4.236251,56.07) ++(5pt,2pt) node[
  scale=0.8,
  anchor=base west,
  text=black,
  rotate=0.0
]{our-5light-c50};
\draw (axis cs:3.659757,46.86) ++(5pt,2pt) node[
  scale=0.8,
  anchor=base west,
  text=black,
  rotate=0.0
]{our-5light-c75};
\draw (axis cs:3.229143,79.12) ++(-2pt,-8pt) node[
  scale=0.8,
  anchor=base west,
  text=black,
  rotate=0.0
]{our-4light};
\draw (axis cs:3.169556,83.53) ++(-60pt,5pt) node[
  scale=0.8,
  anchor=base west,
  text=black,
  rotate=0.0
]{our-4light-sepref};
\end{axis}

\end{tikzpicture}
	\label{RTOF:Fig:DesktopRunTimeVsAccuracyFine}
	\end{subfigure}
    \hfill
    \begin{subfigure}[t]{0.45\linewidth}
    \vskip 6pt
	\begin{footnotesize}
	    \centering
		\begin{tabular}{|l |p{32mm}| l|}
			\hline
			Modifier	            & Description                                                                      & Default \\
			\hline
			-$N$light & $N$ light~\cite{ARFlow} flow estimators.                    & 5 dense  \cite{PWCNet} \\
			-$M$sep					& last $M$ flow estimators use depthwise separable convolutions~\cite{MobileNets}. & standard convs. \\
			-sepref					& refinement uses depthwise separable convolutions~\cite{MobileNets}.	           & standard convs. \\
			-c$P$					& use $P$\% of channels.                                         & 100\% \\ 
			\hline
		\end{tabular}
	\end{footnotesize}
	\caption*{(b) Legend of our \ac{CNN} variants.}
	\end{subfigure}
\caption{Accuracy \vs run-time performance of our \ac{CNN} variants on desktop, measured on Sintel Final (Train)~\cite{SintelDataset}. Optimization steps that lead to significant improvement in run-time are connected by a line. Our architectural modifications to PWC-Net~\cite{PWCNet} are detailed on the right, e.g., our-4light-sepref denotes a 4 light flow estimators and refinement using depthwise separable convolutions. We achieve a high accuracy on Sintel training data, however, for testing data the accuracy is low, see \Cref{RTOF:Fig:EvalExist}.
}
\label{RTOF:Fig:DesktopRunTimeVsAccuracyGraph}
\end{figure*}

\paragraph*{Base Network Selection for Compression.} 

In \Cref{RTOF:Fig:EvalExist}, we compare several well-known optical methods to find a base network candidate that best matches our runtime/accuracy requirements. We employ the following models for this: FlowNet 2.0~\cite{FlowNet2}, 
SpyNet~\cite{Ranjan_SpyNet2017}, LiteFlowNet2~\cite{LiteFlowNet2}, PWCNet~\cite{PWCNet}, ARFlow~\cite{ARFlow}, 
VCN~\cite{Yang_VCN2019}, RAFT~\cite{Teed_RAFT2020}
and finally GMA~\cite{jiang2021learning} (state-of-the-art in terms of \ac{EPE}-based accuracy). Our experiments are carried out on an Nvidia RTX 2070 \ac{GPU}, which we deem to be a good representative of a current mid-to higher-end consumer GPU. Under a constraint of interactive performance on consumer hardware,  LiteFlowNet2~\cite{LiteFlowNet2} and PWC-Net~\cite{PWCNet} offer the best trade-off between run-time performance and accuracy~(\Cref{RTOF:Fig:EvalExist}).
LiteFlowNet2~\cite{LiteFlowNet2} is already an optimized version of FlowNet 2.0~\cite{FlowNet2}, in comparison PWC-Net~\cite{PWCNet} has more potential for optimization/compression.
Moreover, recently it has been shown that PWC-Net can achieve similar accuracy to RAFT when trained on a large-scale synthetic dataset~\cite{Sun_AutoFlow2021} and that PWC-Net achieves favorable trade-offs \vs other state-of-the-art methods when selecting for runtime performance or higher image resolutions~\cite{sun2022disentangling}.
Hence, we select PWC-Net for further compression.

\paragraph*{Optimized Network Architecture.} We start with the base architecture of PWC-Net. 
As the first compression step, we reduce the computationally expensive DenseNet~\cite{DenseNet} connections in the flow estimators to retain connections only in the last two layers ("-light" in \Cref{RTOF:Fig:DesktopRunTimeVsAccuracyGraph}b). %
Similar to LiteFlowNet2~\cite{LiteFlowNet2}, we remove the fifth flow estimator -- operating on the highest resolution -- as it heavily trades off run-time for only a marginal increase in accuracy (compare "4light" vs "5light" in \Cref{RTOF:Fig:DesktopRunTimeVsAccuracyGraph}b). %
We replace the standard convolutions in the refinement by depthwise separable convolutions~\cite{MobileNets} ("-sepref" in \Cref{RTOF:Fig:DesktopRunTimeVsAccuracyGraph}b). %
Moreover, we also explore reducing the number of channels~\cite{MobileNets}, but find that reducing channels results in a worse trade-off as compared to other optimizations.

\paragraph*{Training.} 
For training, we follow the original PWC-Net~\cite{PWCNet} schedule.
However, we find that weighting the multi-scale losses equally, instead of exponentially~\cite{PWCNet,LiteFlowNet,LiteFlowNet2,Yang_VCN2019}, improves accuracy. 
For our experiments on the desktop system, we use PyTorch~\cite{PyTorch} and take inspiration from the implementation by Niklaus~\cite{ImplPWCNet}.
Similar to PWC-Net~\cite{PWCNet}, we train our mobile architecture on the training dataset schedule FlyingChairs~\cite{FlowNet} $\rightarrow$ FlyingThings3D~\cite{FlyingThings3D}$\rightarrow$ Sintel~\cite{SintelDataset}.
In the supplementary material, we provide training settings for each stage in detail.
We employ a multi-scale loss~\cite{PWCNet} applied to each flow estimator and optimize using the AdamW optimizer~\cite{AdamW} with $\beta_1 = 0.09$, $\beta_2 = 0.99$, and $l_2$ weight regularization with trade-off $\gamma = 0.0004$. 
Furthermore, extensive dataset augmentation is applied to prevent model overfitting.
We refer to the supplementary material for more details. 


\paragraph*{Our Final Model.}
We analyze various optimization options and chose \enquote{\textit{our-4light-sepref}} as our final model for desktop systems as it provides the best trade-off between accuracy \vs run-time. 
As depicted in \Cref{RTOF:Fig:DesktopRunTimeVsAccuracyGraph}a, our method improves the run-time performance of PWC-Net from 30~\ac{FPS} to 85~\ac{FPS} -- a speed-up of factor 2.8.
For Sintel training data, the accuracy drops by $\approx$ 0.5px in \ac{EPE} terms, however for test data the drop in accuracy is significant where the final \ac{EPE} is 7.43, see \Cref{RTOF:Fig:EvalExist}.
Nevertheless, the accuracy is sufficient enough for enforcing warping-based consistency.
To validate our design decisions, we conduct an extensive ablation study in which we vary the architectural and training choices -- please see the supplementary for details. 
Furthermore, we tune our architecture for optical flow calculation on mobile devices using channel pruning and quantization, which we also detail in the supplementary material. 
Here, we improve run-time performance from 2.8~\ac{FPS} to 24~\ac{FPS} (iPad Pro 2020), and 1.5~\ac{FPS} to 13~\ac{FPS} (iPad Air) – an improvement of factor 8. 
Next to showing the general applicability of optical flow \acp{CNN} on mobile devices, this demonstrates that real-time on-device stabilization of videos using our presented approach will become feasible with a further moderate increase in mobile GPU computing power.
A fast optical-flow-based warping enables our framework to interactively control the degree of consistency and generate visually plausible results.

\begin{table}[tb]
	\caption{Runtime performance in milliseconds per frame. We measure the total processing time (without disk IO) and the individual stages on two GPU models (Nvidia GTX 1080Ti and RTX 3090). \textdagger Fast preset. Downscales flow computation by $2\times$ and only uses 50 iterations of stabilization instead of 150. }
	\label{tab:runtimeperformancesystem}
	\begin{center}
		\begin{tabular}{|r@{\hskip3pt}|l@{\hskip5pt}l@{\hskip3pt}|l@{\hskip5pt}l@{\hskip3pt}|l@{\hskip5pt}l@{\hskip3pt}|} 
			\hline 
			Task  & \multicolumn{2}{c|}{\text {Optical flow}} & \multicolumn{2}{c|}{\text {Stabilization}} &  \multicolumn{2}{c|}{\text {Total}}  \\
			$\downarrow$ Res. / GPU              &  1080Ti &   3090      &  1080Ti &   3090  & 1080Ti  &   3090 \\
			\hline 
			$1920 \times 1080$ px & $66.8$ &  $40.0$  &  $184.1$ & $42.7$ & $250.8$ & $82.7$ \\
			$1280 \times 720$ px & $31.3$  &  $19.7$   & $86.5$    & $21.1$& $117.8$ & $40.8$ \\
			$640 \times 480$ px & $12.6$   &  $6.2$   & $20.6$   & $6.3$ & $33.2$ & $12.5$ \\
			\hline
                $1920 \times 1080$ px\textdagger & $26.2$ &  $13.0$  &  $71.1$ & $16.5$ & $97.3$ & $29.5$ \\
                \hline 
		\end{tabular}
	\end{center}
\end{table}

\section{Experimental Results}

\begin{figure*}[tb]
\centering
\input{graphics/evaluation/compare_flow_sintelfinal_coarse.tex}
\caption{Optical flow estimated using the synthetic Sintel dataset\cite{SintelDataset}.}
\label{RTOF:Fig:QualityFlowSintelCoarse}
\end{figure*} 

\begin{figure*}[tb]
\centering
\input{graphics/evaluation/compare_flow_davis_coarse.tex}
\caption{Optical flow estimated for the real-world dataset DAVIS~\cite{Davis2017Dataset}.}
\label{RTOF:Fig:QualityFlowDavisCoarse}
\end{figure*}

\newcommand{\imagewithspy}[3]{
    \begin{tikzpicture}[every node/.style={inner sep=0,outer sep=0}]
        \begin{scope}[ 
			inner sep = 0pt,outer sep=0,spy using outlines={rectangle, red, magnification=2}
                ]
        \node (n0)  { \includegraphics[trim=4cm 0 4cm 3cm, clip, width=\textwidth]{#1}};
        \spy [red,size=0.8cm] on (#2,#3) in node[anchor=south east,inner sep=0pt] at (n0.south east);
        \end{scope}
    \end{tikzpicture}
}

\begin{figure*}[tb]  
	\begin{subfigure}{0.195\textwidth}%
    \imagewithspy{}{0.9cm}{0.45cm}
	\label{fig:rslts_cow_in_2}%
	\end{subfigure}\hfill%
	\begin{subfigure}{0.195\textwidth}%
    \imagewithspy{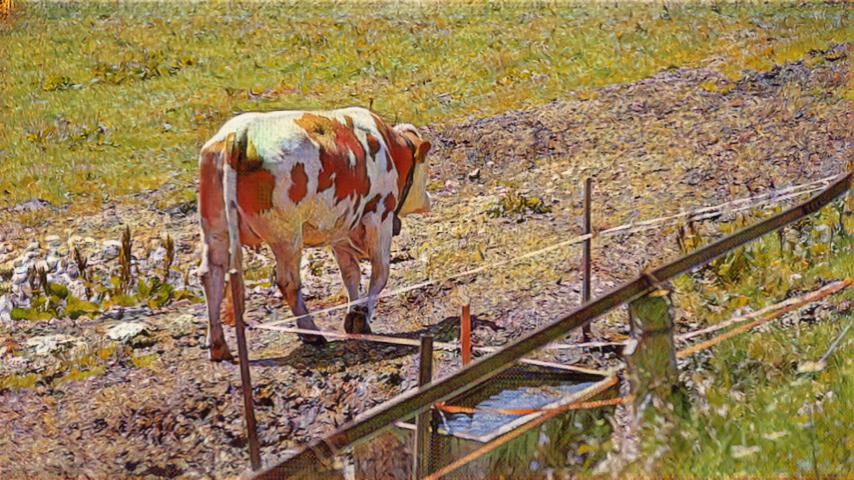}{0.9cm}{0.45cm}
	\label{fig:rslts_cow_pr_2}%
	\end{subfigure}\hfill 
	\begin{subfigure}{0.195\textwidth}%
	\imagewithspy{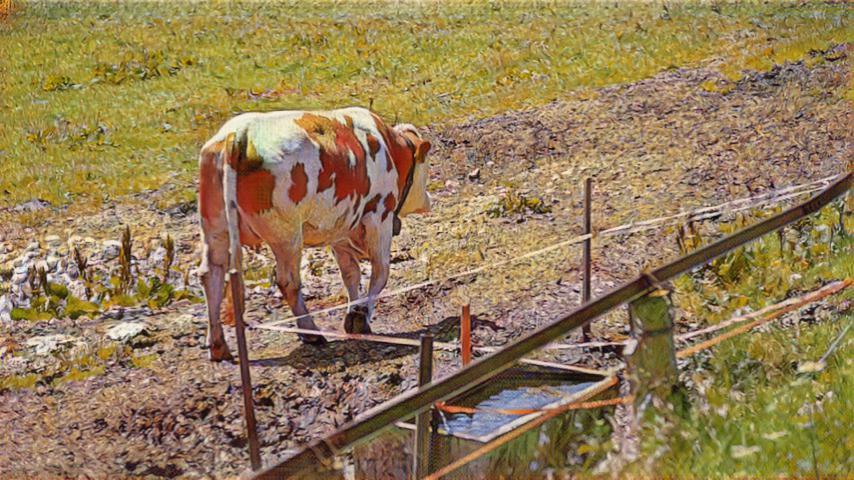}{0.9cm}{0.45cm}
    \label{fig:rslts_cow_ours_2}%
	\end{subfigure}\hfill 
	\begin{subfigure}{0.195\textwidth}%
	\imagewithspy{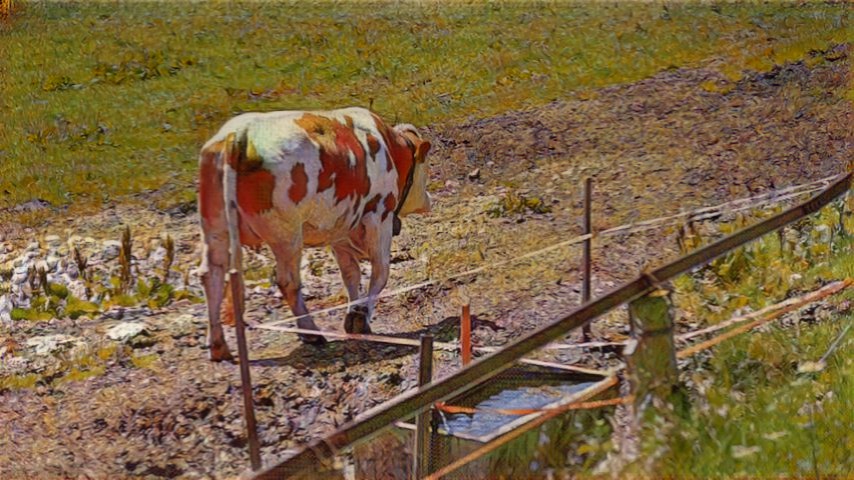}{0.9cm}{0.45cm}
    \label{fig:rslts_cow_lai_2}%
	\end{subfigure}\hfill%
	\begin{subfigure}{0.195\textwidth}%
	\imagewithspy{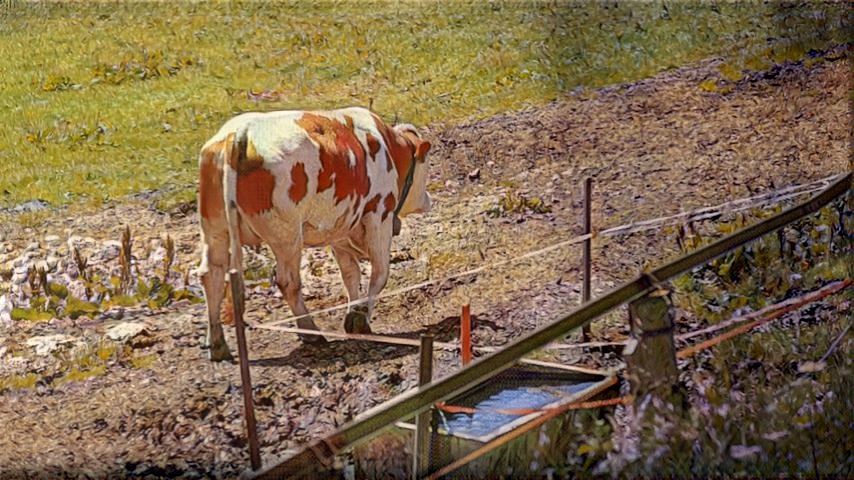}{0.9cm}{0.45cm}
    \label{fig:rslts_cow_bonn_2}%
	\end{subfigure}\\[-2ex]
    \begin{subfigure}{0.195\textwidth}%
   	\imagewithspy{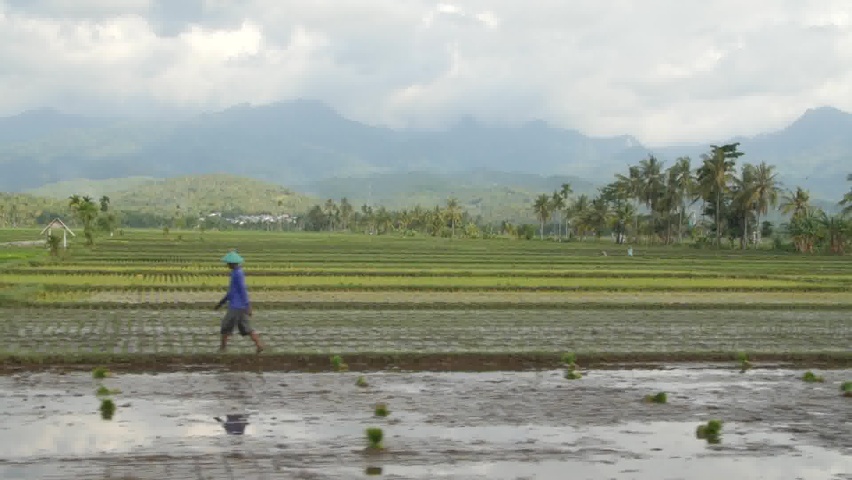}{-1.2cm}{-0.05cm}
    	\label{fig:rslts_farm_in_2}%
    \end{subfigure}\hfill 
    \begin{subfigure}{0.195\textwidth}%
    \imagewithspy{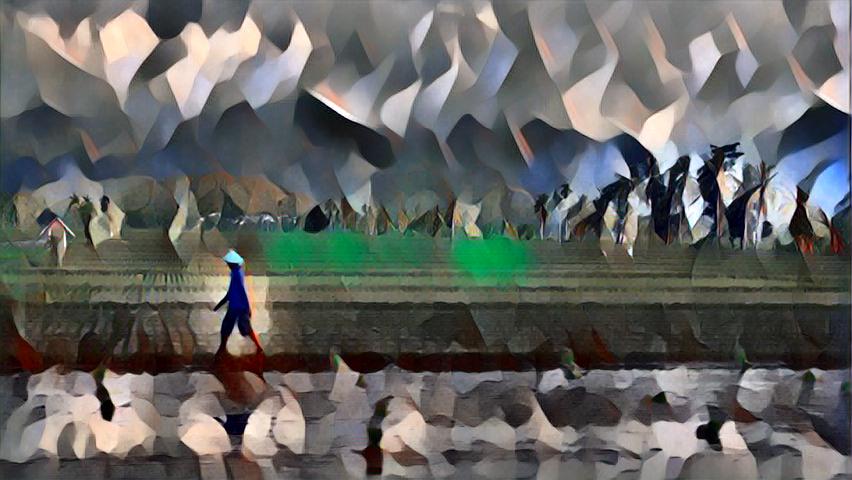}{-1.2cm}{-0.05cm}
    	\label{fig:rslts_farm_pr_2}%
    \end{subfigure}\hfill 
    \begin{subfigure}{0.195\textwidth}%
    \imagewithspy{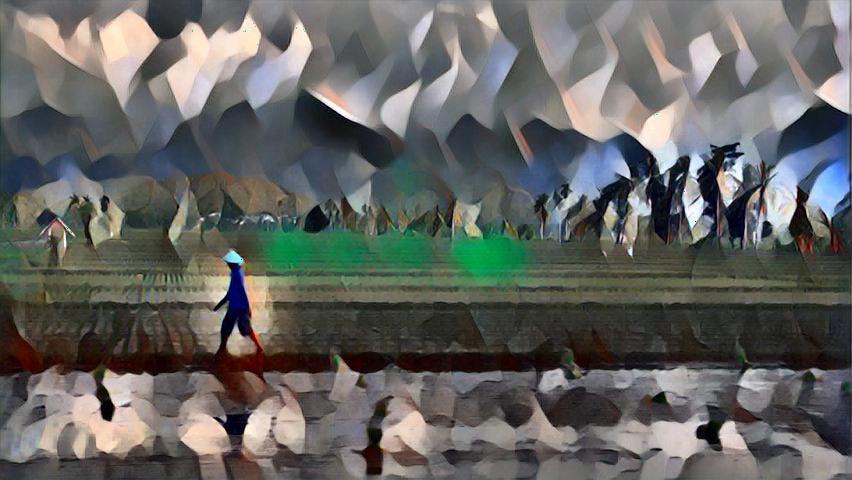}{-1.2cm}{-0.05cm}
   	\label{fig:rslts_farm_ours_2}%
    \end{subfigure}\hfill 
    \begin{subfigure}{0.195\textwidth}%
    \imagewithspy{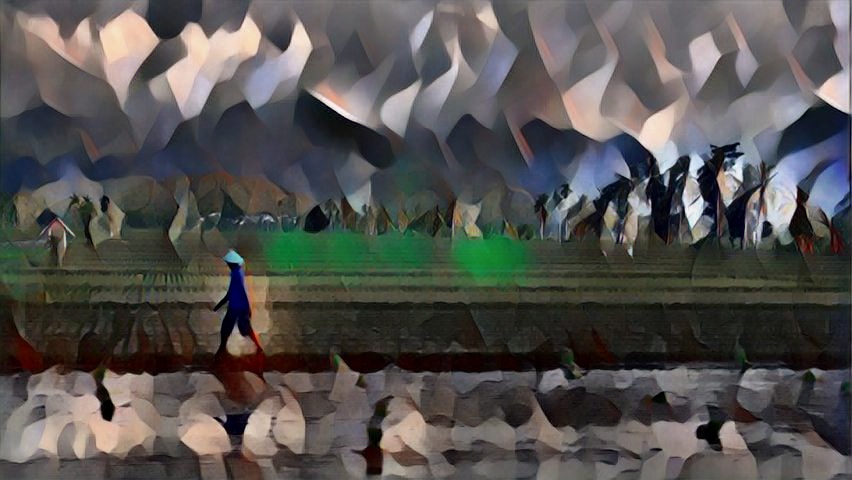}{-1.2cm}{-0.05cm}
   	\label{fig:rslts_farm_lai_2}%
    \end{subfigure}\hfill 
    \begin{subfigure}{0.195\textwidth}%
    \imagewithspy{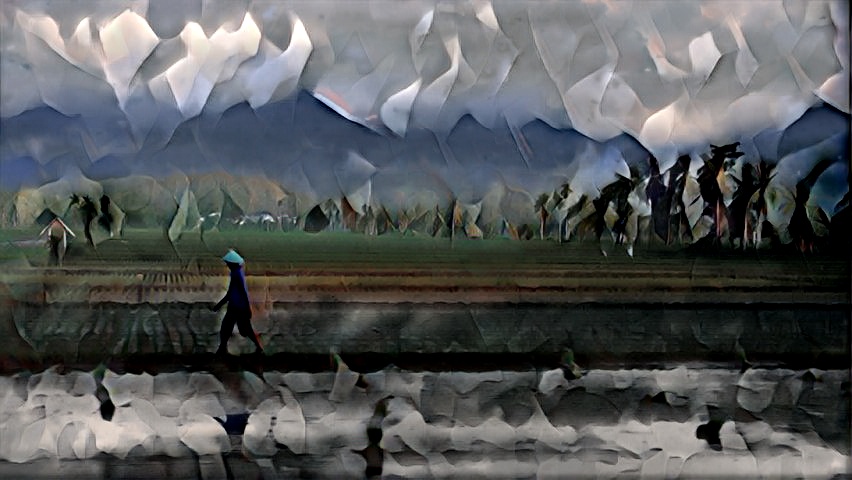}{-1.2cm}{-0.05cm}
   	\label{fig:rslts_farm_bonn_2}%
    \end{subfigure} \\[-2ex]
    \begin{subfigure}{0.195\textwidth}%
    \imagewithspy{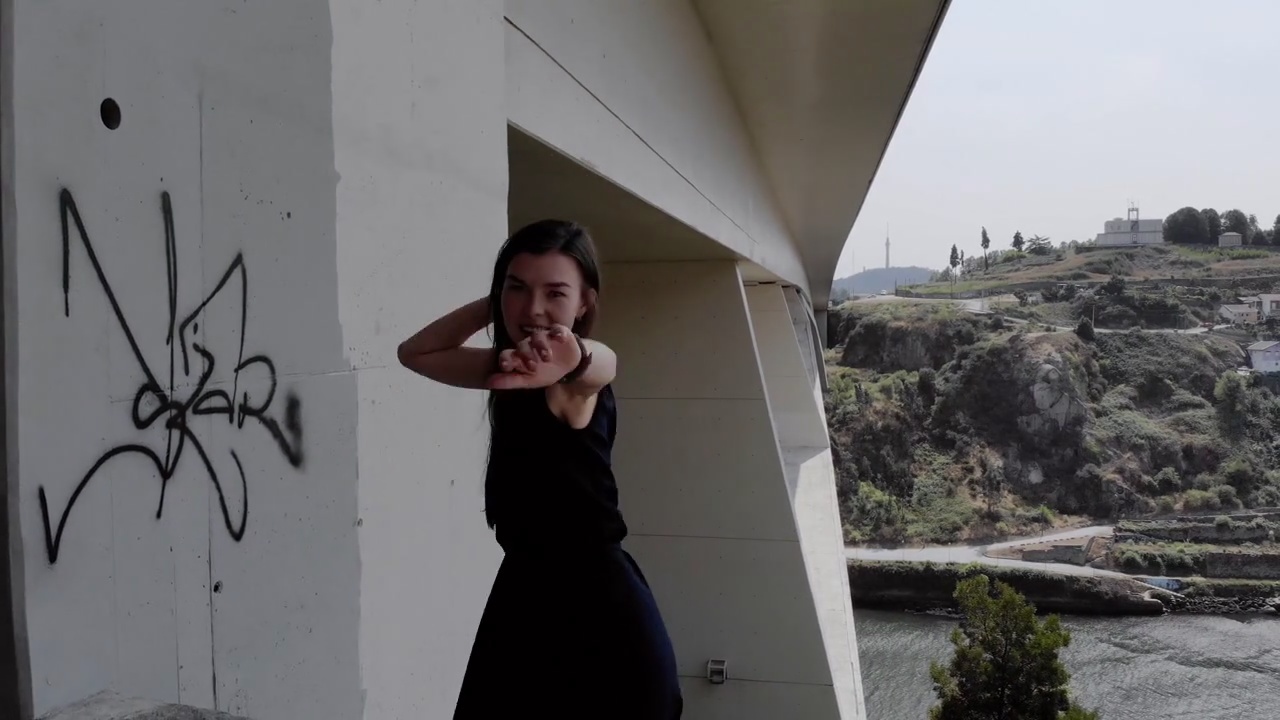}{-0.22cm}{0.3cm}%
    	\subcaption{Input}
    	\label{fig:rslts_woman_in_1}%
    \end{subfigure}\hfill 
    \begin{subfigure}{0.195\textwidth}%
    \imagewithspy{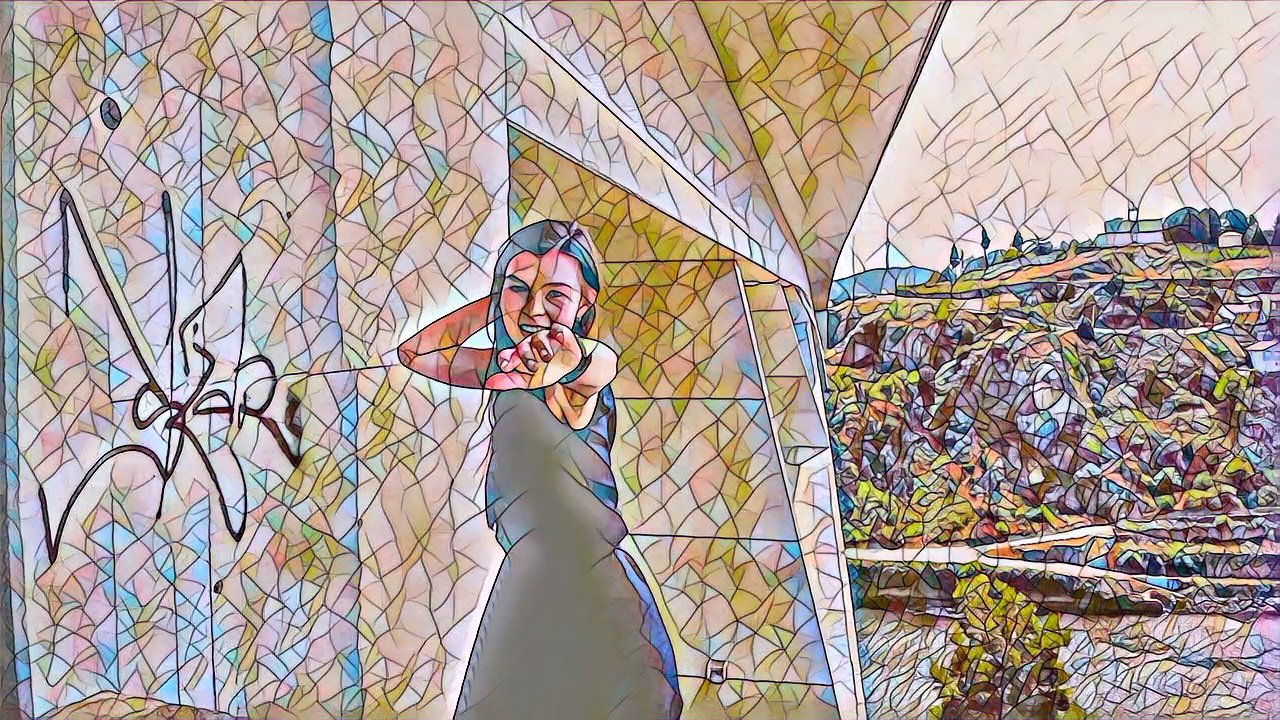}{-0.22cm}{0.3cm}%
        \subcaption{Processed}
    	\label{fig:rslts_woman_pr_1}%
    \end{subfigure}\hfill 
    \begin{subfigure}{0.195\textwidth}%
    \imagewithspy{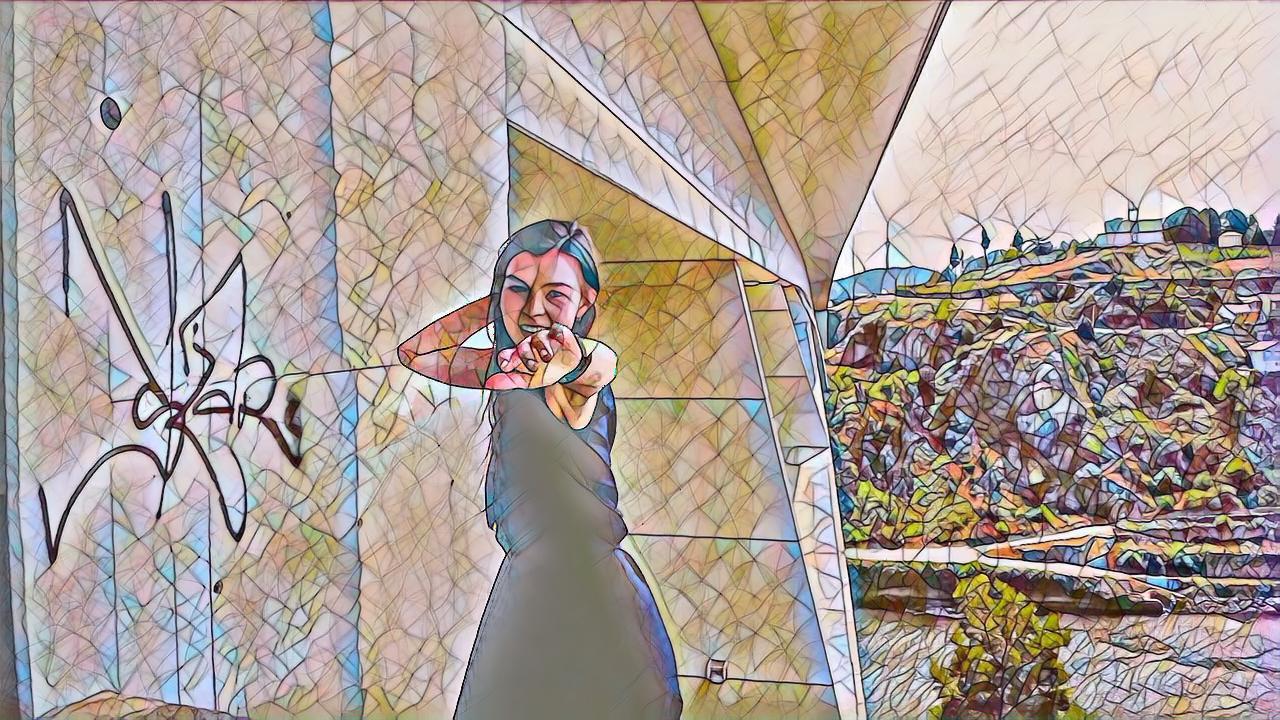}{-0.22cm}{0.3cm}%
        \subcaption{Ours}
    	\label{fig:rslts_woman_ours_1}%
    \end{subfigure}\hfill 
    \begin{subfigure}{0.195\textwidth}%
    \imagewithspy{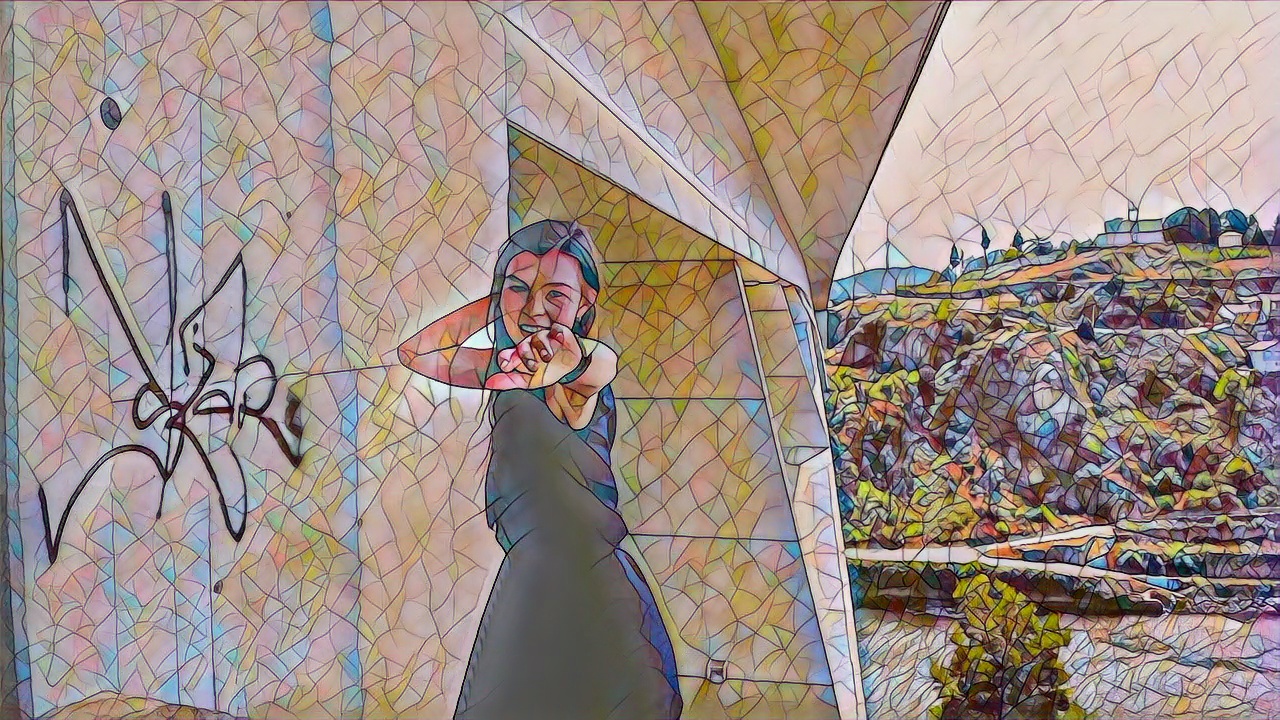}{-0.22cm}{0.3cm}%
        \subcaption{Lai~\etal~\cite{Lai_Learning2018}}
    	\label{fig:rslts_woman_lai_1}%
    \end{subfigure}\hfill 
    \begin{subfigure}{0.195\textwidth}%
    \imagewithspy{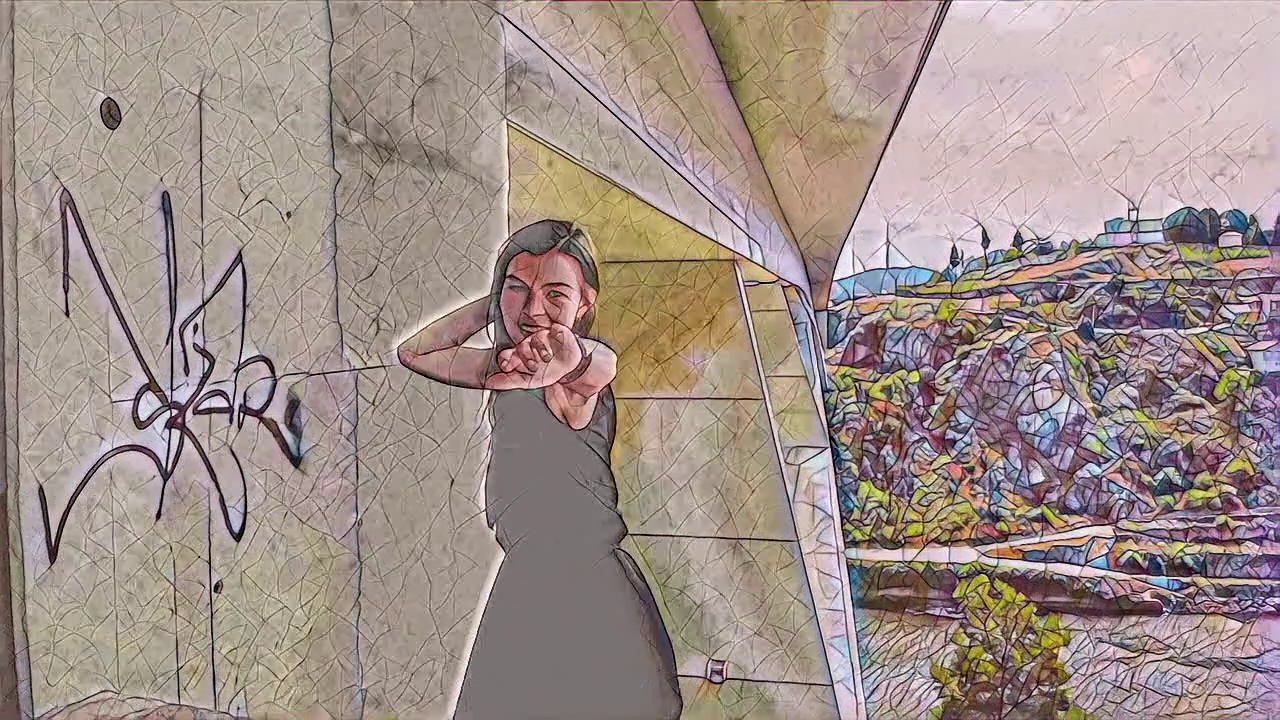}{-0.22cm}{0.3cm}%
        \subcaption{Bonneel~\etal~\cite{Bonneel_Blind2015}}
    	\label{fig:rslts_woman_bonn_1}%
    \end{subfigure}
	\caption{Comparing our results with Lai~\etal~\cite{Lai_Learning2018} and Bonneel~\etal~\cite{Bonneel_Blind2015} for three different video sequences. Note how the consistent output for Lai~\etal and Bonneel~\etal look different from the corresponding per-frame processed results.}
	\label{fig:results}
\end{figure*}
\subsection{Implementation Details}
All our experiments were performed on a consumer PC 
with an AMD Ryzen 1920X 12-Core CPU, 48 GB of RAM, and a Nvidia GTX 1080Ti and RTX 3090 graphics cards with VRAMs of 11 GB and 24 GB respectively.
We implement a real-time video-consistency framework in C++, using ONNXRuntime for cross-platform acceleration of our lite optical-flow network, and implement the stabilization code using Nvidia CUDA (v11). 
In \Cref{tab:runtimeperformancesystem}, we measure the runtime performance of our system. 
We find that an incoming stream of frames can be stabilized at real-time performance for VGA resolution even on low- and mid-tier GPUs and higher-tier GPUs (such as a RTX 3090) can stabilize HD at common video frame rates (approx. 24~\ac{FPS}) and full-HD resolutions at interactive frame rates (> 10~\ac{FPS}) (\Cref{tab:runtimeperformancesystem}).
We also test a fast preset that uses less iterations and computes optical flow on half-sized inputs, and find that a full-HD video stream can be processed in real-time at the cost of minor additional flickering - see the supplementary video for a comparison.
We implement a graphical user interface that allows for real-time decoding and stabilization of stylized video streams, where the stabilization parameters can be interactively controlled, see the supplementary video for a demonstration.

\subsection{Parameter Settings}
Initially, we tune the parameters of our consistency framework towards achieving a low warping error (\Cref{tab:warp_error}).
We refer to this setting as \textit{Ours-objective} with the following parameter values $k_1 = k_2 = 0.3$, $\alpha = 10 \times 10^3$, and $\lambda = 0.7$. %
However, we observed that even though the warping error indicated good temporal stability, subjective flickering, and artifacts were noticeable.
Unlike existing approaches, our framework allows for interactive parameter adjustment. Thus, a parameter set that subjectively produces well-stabilized results on a broad range of tasks and videos was obtained experimentally. %
As our final version, we use the values of  $k_1 = 0.3$, $k_2 = 0.5$, $\alpha = 6.5 \times 10^3$, and $\lambda =2.0$ to generate all the images in the paper and the videos provided in the supplementary.
We further compare \textit{Ours-objective} settings with our final version as part of our user study to validate our parameter choices. %
The consistent outputs obtained using the above parameter settings are compared against state-of-the-art approaches thereby showcasing its efficacy.

\subsection{Optical Flow Results}
We visualize optical flow on frames from the Sintel \cite{SintelDataset} dataset in \Cref{RTOF:Fig:QualityFlowSintelCoarse} and compare it to state-of-the-art methods. 
All depicted methods have been fine-tuned on Sintel. We find that our optimized method has more blurry motion boundaries  and misses estimating certain details accurately (\eg the right hand, however, PWCNet also fails at this), but still captures the overall motion direction of objects correctly with a smooth flow field.
\Cref{RTOF:Fig:QualityFlowDavisCoarse} shows results for real-world videos on the DAVIS dataset \cite{Davis2017Dataset} (no ground-truth flow available).
We find that some real-world image phenomena, such as complex/ambiguous occlusions (\eg bus behind the tree) are not well-handled by state-of-the-art methods like RAFT~\cite{Teed_RAFT2020} or PWC-Net~\cite{PWCNet}, similarly, such results are also degraded for our optimized method. 
Besides the stronger blurred motion boundaries, we find that our network generally performs well and is also robust for real-world videos.

\subsection{Consistent Outputs}
We use videos from DAVIS~\cite{Perazzi_2016} dataset and other open-source videos (taken from \cite{Videvo} and \cite{Pexels}) for comparison. 
For per-frame stylization, we employ the following stylization techniques: Fast \ac{NST}~\cite{Johnson_Perceptual_2016}, WCT~\cite{Li_Universal_2017}, and CycleGAN~\cite{Zhu_Unpaired_2017}. 
The results for the method of Lai~\etal and Bonneel~\etal on videos taken from DAVIS~\cite{Perazzi_2016} and Videvo (\cite{Videvo}) are borrowed from the \textit{results-dataset} provided by Lai~\etal. 
For other videos, we employ the source code provided by the authors to generate the results. 
We compare our consistent outputs with that of Bonneel~\etal~\cite{Bonneel_Blind2015} and Lai~\etal~\cite{Lai_Learning2018} in \Cref{fig:results}. 
Among the three competing methods Bonneel~\etal is the least effective in preserving the underlying style for the final output (compare the second column with the fifth one in \Cref{fig:results}).
Hyper-parameter tuning in the above method (with only global consistency) can provide a certain degree of consistency control. 
However, by employing both global and local consistency we achieve finer consistency control while being similar to the per-frame-processed result.
For the method of Lai~\etal, we observe some color bleeding or darkening in the output frames (compare the second column with the fourth one in \Cref{fig:results}).
In comparison, we are able to preserve the style, color, and textures, while being consistent.

\section{Evaluation}

\subsection{Quantitative Evaluation}

\begin{table*}[tb]
	\caption{Quantitative evaluation on perceptual distance using SSIM (higher = more similar to per-frame processed result).}
	\label{table:ssim}
	\begin{center}
		\begin{tabular}{|l|ccc|ccc|}
        \hline
			& \multicolumn{3}{c|}{\text { DAVIS }} & \multicolumn{3}{c|}{\text { VIDEVO }} \\
			Task &  {\cite{Bonneel_Blind2015}} & \text {\cite{Lai_Learning2018}} & Ours & {\cite{Bonneel_Blind2015}} & \text {\cite{Lai_Learning2018}} & Ours\\
			\hline
			CycleGAN/photo2ukiyoe~\cite{Zhu_Unpaired_2017} & 0.693 & 0.781 & \textbf{0.978} & 0.626 & 0.743 & \textbf{0.980} \\
			CycleGAN/photo2vangogh~\cite{Zhu_Unpaired_2017} & 0.707 & 0.792 & \textbf{0.961} & 0.679 & 0.789 & \textbf{0.965} \\
			fast-neural-style/rain-princess~\cite{Johnson_Perceptual_2016} & 0.553 & 0.799 & \textbf{0.921} & 0.491 & 0.796 & \textbf{0.920} \\
			fast-neural-style/udnie~\cite{Johnson_Perceptual_2016} & 0.597 & 0.785 & \textbf{0.956} & 0.579 & 0.747 & \textbf{0.959} \\
			WCT/antimonocromatismo~\cite{Li_Universal_2017} & 0.389 & 0.811 & \textbf{0.915} & 0.388 & 0.761 & \textbf{0.914} \\
			WCT/asheville~\cite{Li_Universal_2017} & 0.329 & 0.801 & \textbf{0.904} & 0.348 & 0.771 & \textbf{0.901} \\
			WCT/candy~\cite{Li_Universal_2017} & 0.289 & 0.763 & \textbf{0.882} & 0.310 & 0.738 & \textbf{0.885} \\
			WCT/feathers~\cite{Li_Universal_2017} & 0.418 & 0.863 & \textbf{0.891} & 0.415 & 0.848 & \textbf{0.888} \\
			WCT/sketch~\cite{Li_Universal_2017} & 0.370 & 0.845 & \textbf{0.923} & 0.370 & 0.833 & \textbf{0.922} \\
			WCT/wave~\cite{Li_Universal_2017} & 0.358 & 0.700 & \textbf{0.902} & 0.352 & 0.637 & \textbf{0.899} \\
			\hline
			Average & 0.470 & 0.794 & \textbf{0.923} & 0.456 & 0.766 & \textbf{0.923} \\
        \hline   
		\end{tabular}
	\end{center}
\end{table*}

 \begin{table*}[!h]
     \caption{Flow warping error average over tasks shown in \Cref{table:ssim}. A per-task breakdown is shown in the supplementary. Note that the slightly higher warping error (lower is better) of our method  is subjectively not noticeable as we show in a user study. }
    \label{tab:warp_error}
    \begin{center}
    \begin{tabular}{|l|cccccc|}
        \hline
        Dataset & $V_p$ & \cite{Bonneel_Blind2015} & \text{\cite{Lai_Learning2018}} & Ours & Ours-RAFT & Ours-DIS \\
        \hline
        DAVIS & 0.056 & 0.034 & 0.040 & 0.046 & 0.045 & 0.050 \\ 
        VIDEVO & 0.051 & 0.036 & 0.036 & 0.042 & 0.042 & 0.046 \\
        \hline
    \end{tabular}
    \end{center}
\end{table*}
            
Following Lai \etal \cite{Lai_Learning2018}, we measure the similarity between per-frame processed output and stabilized results, and the temporal warping error between consecutive stabilized frames.

For the former, we report the similarity in the form of the SSIM metric in \Cref{table:ssim}. We achieve significantly higher similarity scores than the methods of Bonneel \etal \cite{Bonneel_Blind2015} and Lai \etal \cite{Lai_Learning2018}.
Following \cite{Bonneel_Blind2015} and \cite{Lai_Learning2018}, we also measure the temporal warping error between a frame $V_{t}$ and the warped consecutive frame $\hat{V}_{t+1}$, defined as:
\begin{equation}
E_{\mathrm{warp}}\left(V_t, V_{t+1}\right)=\frac{1}{\sum_{i=1}^N M_t^{(i)}} \sum_{i=1}^N M_t^{(i)}\left\|V_t^{(i)}-\hat{V}_{t+1}^{(i)}\right\|_1,
\end{equation}
where $M_t \in \{0,1\}$ is a non-occlusion mask \cite{Lai_Learning2018, Ruder_Artistic2018}, indicating non-occluded regions. The warped frame $\hat{V}_{t+1}$ is obtained by calculating the optical flow (using GMA \cite{jiang2021learning}) between frames $V_t,V_{t+1}$, and applying a backwards warping to frame $V_{t+1}$. We compute $E_{\mathrm{warp}}$ for every frame of a video and then averaged to obtain the warping error of a video $E_{\mathrm{warp}}(V)$. In \Cref{tab:warp_error} we report the average warping error per dataset (see the supplementary for a per-task breakdown). We find that the warping error is slightly higher than that of Bonneel \etal\cite{Bonneel_Blind2015} and Lai \etal\cite{Lai_Learning2018}. However, as Lai \etal \cite{Lai_Learning2018} notes, results with high temporal stability (expressed by a low warping error) can also be achieved via temporally smoothing the video, which can be seen in various results of Bonneel \etal\cite{Bonneel_Blind2015}. Our qualitative results in the form of a user study \Cref{subsec:qualitative} further substantiate the divide between warping error (as a stability metric) and perceived stability.

\paragraph*{Using other Optical Flow Computations.}
\label{par:other_flows}
We also tested other optical flow methods within our pipeline which were either faster~\cite{Kroeger_DIS_2016} or more accurate~\cite{Teed_RAFT2020}.
For the fast optical method by Kroeger~\etal~\cite{Kroeger_DIS_2016}(DIS) the final output is less consistent than ours in both objective and subjective metrics. 
Using DIS for our stabilization, the average warp-error is higher (\Cref{tab:warp_error}), and perceptual-similarity with the per-frame processed result is lower than ours (0.9 in SSIM over DAVIS and VIDEVO). 
Visually, DIS-stabilized results show significantly more flickering, validating our design choice for the optical flow. A much more accurate optical flow is given by the method of Teed~\etal~\cite{Teed_RAFT2020} (RAFT) at the cost of slow computation.  The stabilized results obtained using RAFT look visually indistinguishable from the one obtained using our flow; the average warp-error is the same or marginally lower (\Cref{tab:warp_error}), while the perceptual-similarity is the same in terms of SSIM as in \Cref{table:ssim}. 
Due to visually unnoticeable and metric-wise only minor differences for RAFT, we conjecture that there will not be any significant improvement in output quality for even more accurate flow methods.

\begin{figure}[tb]
	\centering
	\includegraphics[trim={3.9cm 5.2cm 8.8cm 1.5cm},clip,width=1.0\columnwidth]{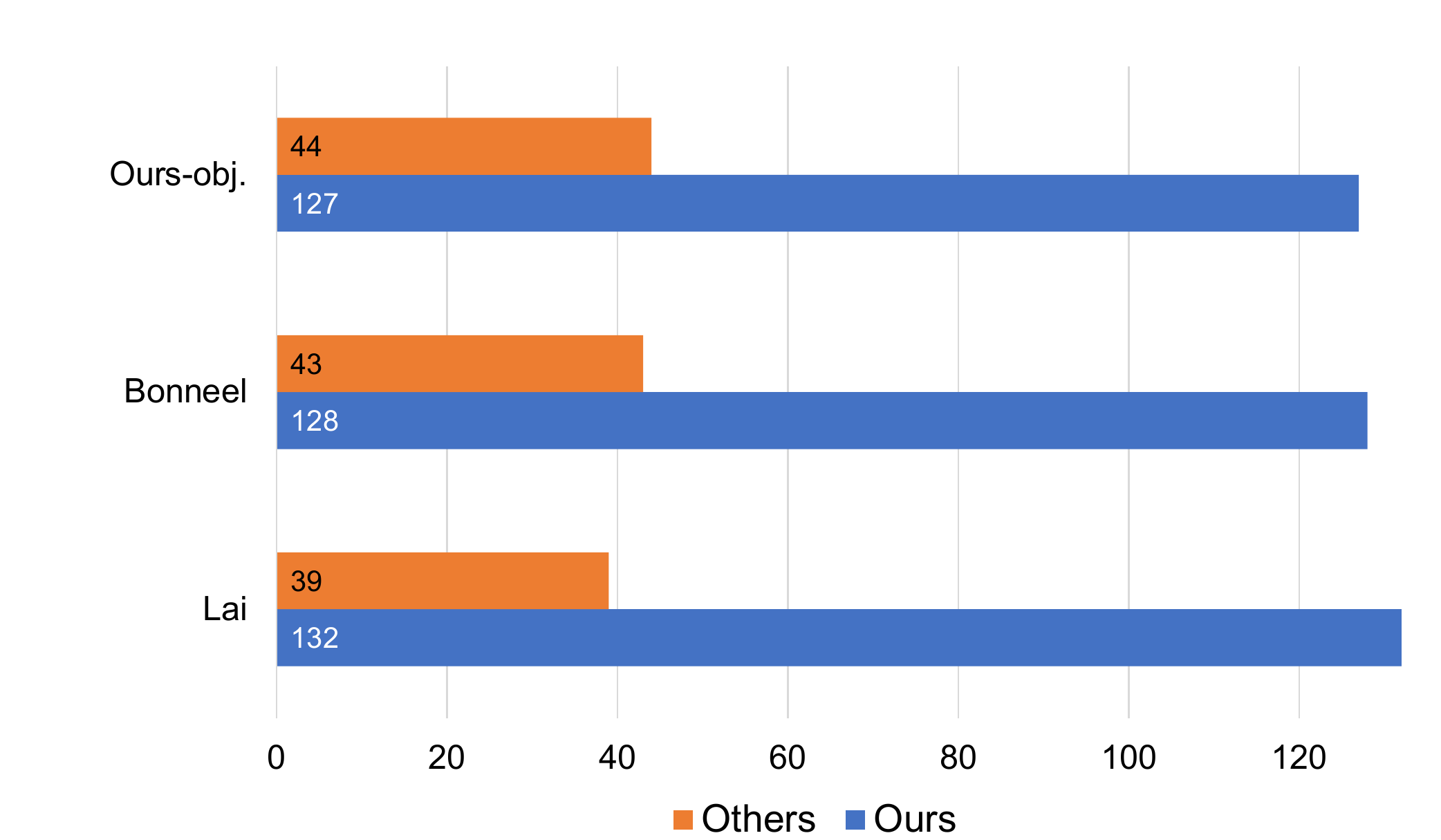}
	\caption{Statistics of the user study results on removal of temporal flickering from per-frame stylized videos. For $19$ participants and $9$ different videos we compare our method against Bonneel~\etal, Lai~\etal, and \textit{Ours-objective} through a total of $171$ randomized A/B tests.}
	\label{fig:user_study}
\end{figure}

\subsection{Qualitative Evaluation}
\label{subsec:qualitative}

For qualitative evaluation, we perform a subjective user study where we ask participants to compare the temporally-consistent result obtained using our method with that of Lai~\etal~\cite{Lai_Learning2018}, Bonneel~\etal~\cite{Bonneel_Blind2015}, and \textit{Ours-objective} -- a different parameter setting of ours.  
We use $9$ different videos for this purpose: $3$ from DAVIS~\cite{Perazzi_2016}, $3$ from Videvo~\cite{Videvo}, and $3$ from Pexels~\cite{Pexels} datasets respectively.
For each of the above videos we stylize them using either the Fast \ac{NST}~\cite{Johnson_Perceptual_2016} (in the styles of \textit{udnie}, \textit{rain-princess}, and \textit{mosaic}) or WCT~\cite{Li_Universal_2017} (in the styles of \textit{wave} and \textit{antimono}) or Cycle-GAN (in the styles of \textit{photo2vangogh} and \textit{photo2ukiyoe}). 
For each sample, we show the input video and its per-frame stylized version on the top row of the user-study interface for inference.
In the bottom row, we show two different versions of the temporally stabilized output where one of them is ours.  
We ask the participants to select the output which best preserves: ($i$) temporal consistency and ($ii$) similarity with the per-frame processed video. 
For $9$ videos and $3$ other competing methods, each user sees a total of $27$ blind A/B tests which are shown in a randomized order to each participant.  
In total, $19$ persons ($3$ female and $16$ male) between the ages of $22$ to $43$ years participated in the study. 
\Cref{fig:user_study} shows that our method surpasses all others  by a large margin. 
It was interesting to observe that for certain cases the method of Bonneel~\etal which degrades the processed style significantly was still preferred by users over others due to its high consistency quality. 
We also show subjective user preference for our method over the methods of Shekhar~\etal~\cite{Shekhar_Consistent2019} and Thimonier~\etal \cite{Thimonier_learning2021} via another user study in the supplementary material.
Shekhar~\etal which is completely based on temporal denoising is prone to introduce motion blur artifacts and is less efficient in terms of enforcing global consistency.
Similar to Lai~\etal, Thimonier~\etal introduces artifacts in the form of color bleeding and darkening, see the supplementary video. 

\subsection{Comparison to Other Methods}
\label{subsec:compareother}

\newcommand{\imagewithspyeval}[3]{
    \begin{tikzpicture}[every node/.style={inner sep=0,outer sep=0}]
        \begin{scope}[
			inner sep = 0pt,outer sep=0,spy using outlines={rectangle, red, magnification=3}
                ]
        \node (n0)  { \includegraphics[trim=0.2cm 0cm 3cm 0cm, clip, width=\textwidth]{#1}};
        \spy [red,size=0.8cm] on (#2,#3) in node[anchor=south east,inner sep=0pt] at (n0.south east);
        \end{scope}
    \end{tikzpicture}
}

\begin{figure}
     \centering
     \begin{subfigure}{0.49\linewidth}
     \imagewithspyeval{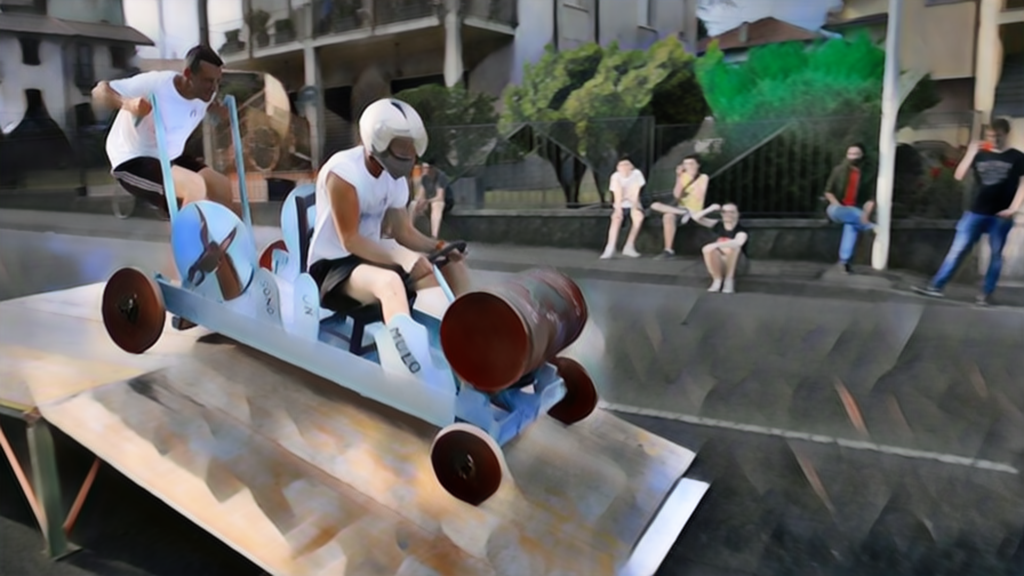}{0.3cm}{0.0cm}
          \vspace{-0.4cm}\caption{Bilat + gauss ~\cite{Texler_Interactive_2020} }
     \end{subfigure}
    \begin{subfigure}{0.49\linewidth}
     \imagewithspyeval{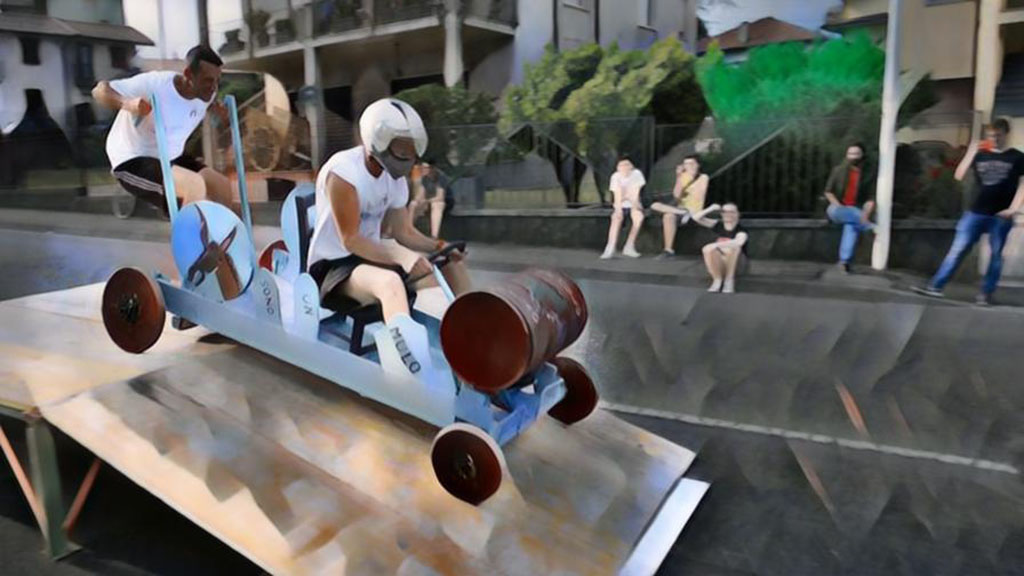}{0.3cm}{0.0cm}
          \vspace{-0.4cm}\caption{ Ours }
     \end{subfigure}
     \caption{Comparing temporal stabilization for keyframe-based stylization (KBS) \cite{Texler_Interactive_2020}. We compare the stabilization approach proposed by Texler~\etal~\cite{Texler_Interactive_2020} to our method. To obtain the unstabilized KBS video, we select three keyframes (1st, middle, last) from a video from DAVIS, apply for style transfer, and train a KBS model \cite{Texler_Interactive_2020} on them, which is then used to stylize the remaining the frames. }
     \label{fig:oursstabs-vs-texler}
 \end{figure}

\paragraph*{Keyframe-based Stylization (KBS).} The goal of KBS is to propagate the style from a few selected keyframes to the rest of the sequence, usually with per-sequence pretraining, running in an offline scenario. Our focus on the other hand is blind video consistency, i.e., we assume to have no control over the per-frame stylization process and consider a video stream as our application scenario. However, some KBS methods such as Texler~\etal~\cite{Texler_Interactive_2020} or Futschik~\etal~\cite{Futschik_STALP_2021}  are capable of running in online scenarios, under the assumption that the video stream is similar enough to the pre-trained frames (\eg a webcam stream). 
The results however often exhibit temporal flickering. 
To combat this, Texler~\etal propose a method for temporal stabilization that involves pre-filtering frames using a motion-compensated bilateral filter and encoding location data using a mixture of Gaussians. 
In \Cref{fig:oursstabs-vs-texler} and the supplementary video, we compare their stabilization approach to ours on KBS-stylized videos. 
For videos stylized using \ac{NST}, visual similarities are apparent between both stabilized versions, though our method displays superior detail preservation. However, for complex-structured styles that introduce texture into homogeneous regions, such as pencil drawings, their Gaussian-mixture-based stabilization improves texture adherence. In contrast, our method may lead to over-smoothing due to inaccurate flow computation in such featureless regions (see supplementary video). 
This effect can be mitigated to a certain degree by employing a lower temporal consistency factor ( $\lambda$).
Compared to Gaussian-mixture-based stabilization our approach runs at least an order of magnitude faster, making it better suited for interactive scenarios such as KBS-stylizing and stabilizing an incoming video stream.  
Futschik~\etal improves on the temporal consistency of Texler~\etal by considering additional frames from the video during training. However, this makes it less applicable to out-of-domain videos (i.e., content not seen during training) which is common in video streams.
Our method can also effectively stabilize such videos as shown in the supplementary video.

\newcommand{\imagewithspyevaltwo}[3]{
    \begin{tikzpicture}[every node/.style={inner sep=0,outer sep=0}]
        \begin{scope}[ 
			inner sep = 0pt,outer sep=0,spy using outlines={rectangle, red, magnification=2}
                ]
        \node (n0)  { \includegraphics[trim=0cm 0cm 0cm 0cm, clip, width=\textwidth]{#1}};
        \spy [red,size=1.2cm] on (#2,#3) in node[anchor=south east,inner sep=0pt] at (n0.south east);
        \end{scope}
    \end{tikzpicture}
}

 \begin{figure}
     \centering
     \begin{subfigure}{0.49\linewidth}
     \imagewithspyevaltwo{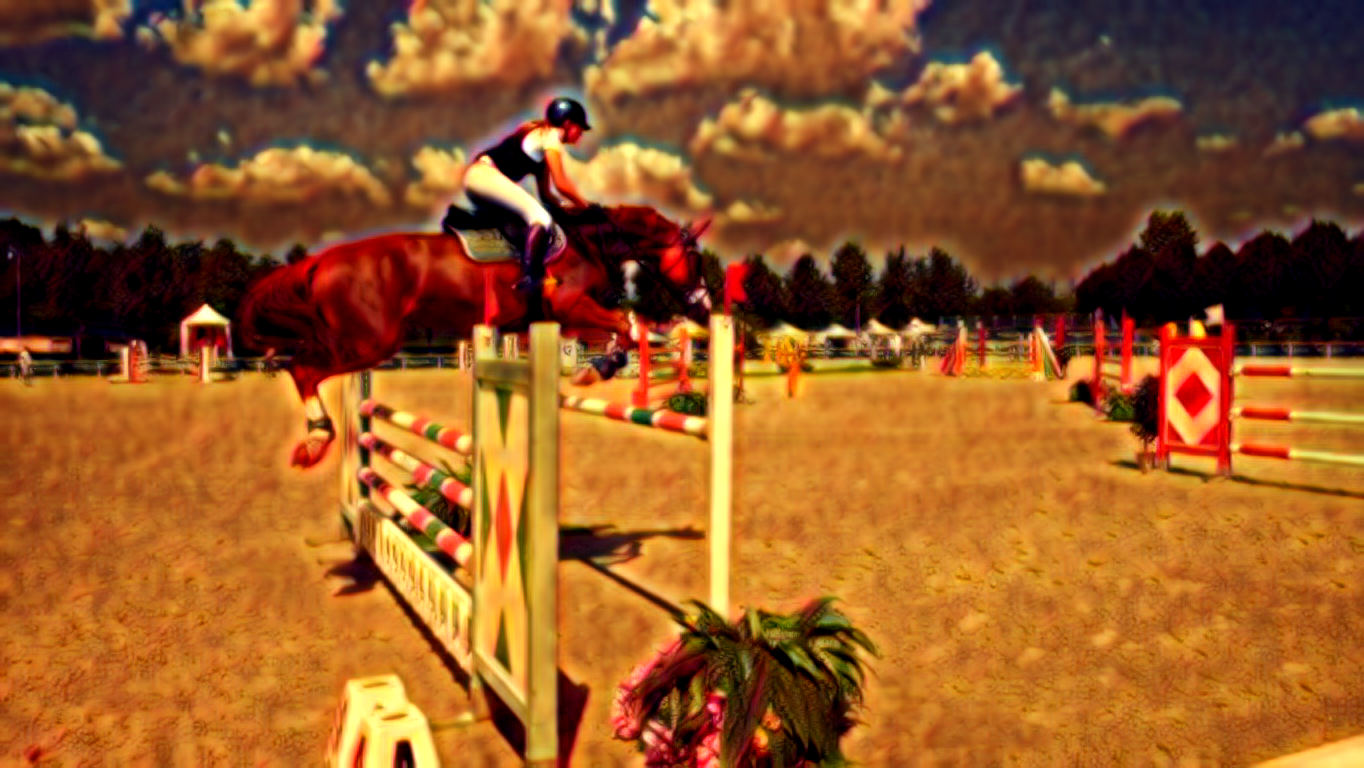}{0.4cm}{0.7cm}
     \vspace{-0.4cm}\caption{Video Style Transfer~\cite{Li_Learning2019} }
     \end{subfigure}
    \begin{subfigure}{0.49\linewidth}
     \imagewithspyevaltwo{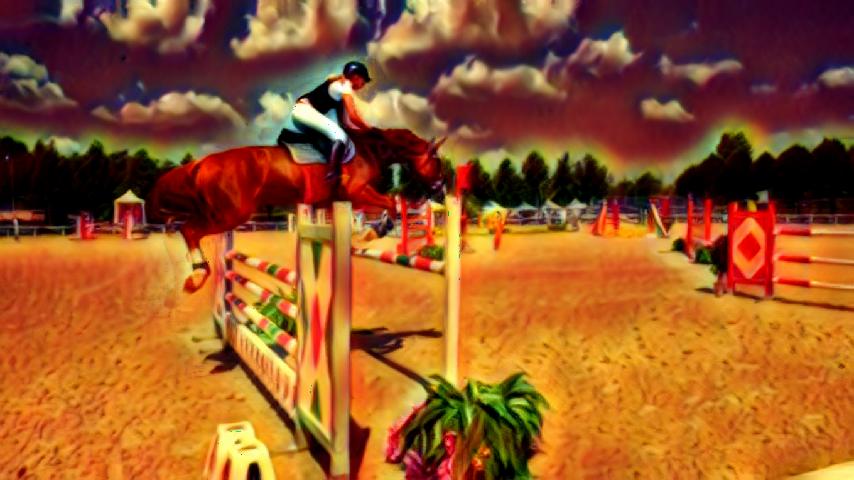}{0.4cm}{0.7cm}
     \vspace{-0.4cm}\caption{ Per-Frame NST + Ours }
     \end{subfigure}
     \caption{Comparing to video style transfer \cite{Li_Learning2019}. We compare the implicit stabilization of their video style transfer technique to their per-frame NST stabilized with our approach.}
     \label{fig:oursstab-vs-li-nst}
 \end{figure}

\paragraph*{Video Style Transfer.} In \Cref{fig:oursstab-vs-li-nst}, we compare our method to the arbitrary style transfer for videos of Li~\etal~\cite{Li_Learning2019}. Despite their method having full control over the stylization process, their results exhibit more temporal flickering and blurring, particularly in smooth regions such as the sky. Their video style transfer method also tends to under-stylize image features compared to their per-frame style transfer. Please see the supplementary video for a video-based comparison.

\section{Discussion}
\label{Sec:Discussion}

\newcommand{\imagewithspyevaldisc}[3]{
    \begin{tikzpicture}[every node/.style={inner sep=0,outer sep=0}]
        \begin{scope}[ 
			inner sep = 0pt,outer sep=0,spy using outlines={rectangle, red, magnification=2}
                ]
        \node (n0)  { \includegraphics[trim=5cm 0.5cm 0cm 0.5cm, clip, width=\textwidth]{#1}};
        \spy [red,size=0.7cm] on (#2,#3) in node[anchor=south east,inner sep=0pt] at (n0.south east);
        \end{scope}
    \end{tikzpicture}
} 

\newcommand{\imagewithspyevaldisctwo}[3]{
    \begin{tikzpicture}[every node/.style={inner sep=0,outer sep=0}]
        \begin{scope}[ 
			inner sep = 0pt,outer sep=0,spy using outlines={rectangle, red, magnification=2}
                ]
        \node (n0)  { \includegraphics[trim=3.5cm 0.3cm 0cm 0.3cm, clip, width=\textwidth]{#1}};
        \spy [red,size=0.7cm] on (#2,#3) in node[anchor=south east,inner sep=0pt] at (n0.south east);
        \end{scope}
    \end{tikzpicture}
}

\begin{figure}
    \centering
    \begin{subfigure}{0.325\linewidth}
        \imagewithspyevaldisc{}{0.35cm}{0.1cm}
        \label{fig:stablediffusion:sub1}
    \end{subfigure}
    \hfill
    \begin{subfigure}{0.325\linewidth}
        \imagewithspyevaldisc{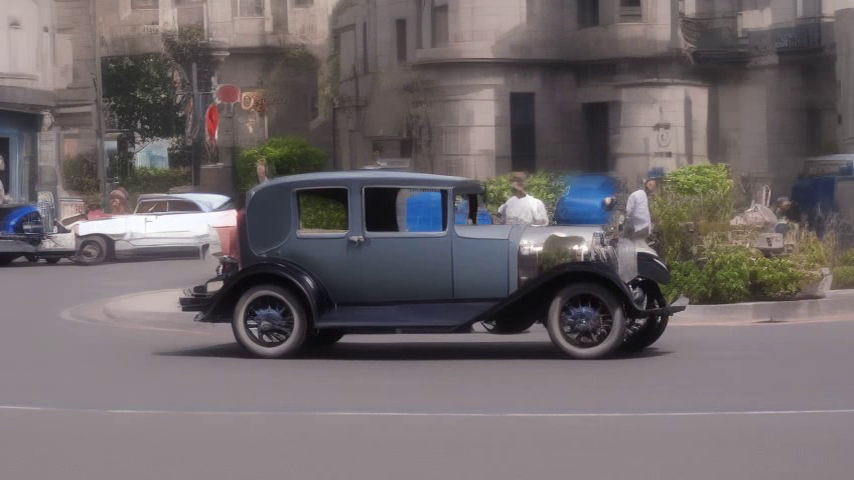}{0.35cm}{0.1cm}
        \label{fig:stablediffusion:sub2}
    \end{subfigure}
    \hfill
    \begin{subfigure}{0.325\linewidth}
        \imagewithspyevaldisctwo{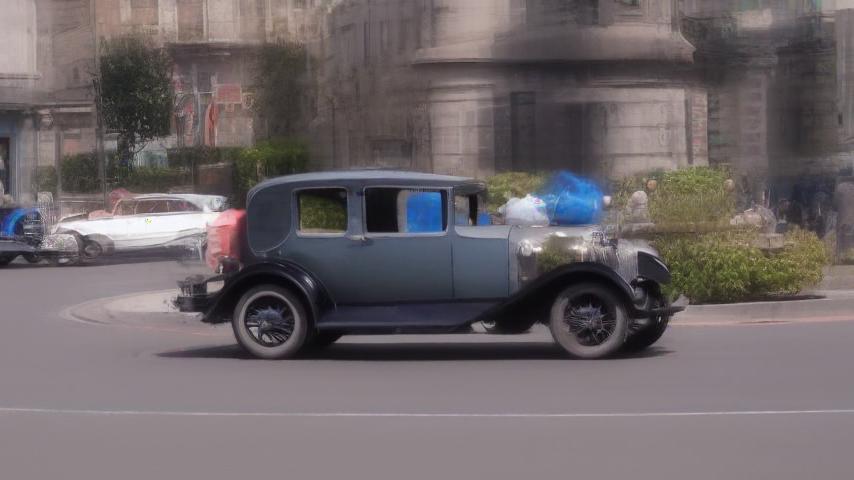}{0.35cm}{0.1cm}
        \label{fig:stablediffusion:sub3}
    \end{subfigure}\\[-2.2ex]
    \begin{subfigure}{0.33\linewidth}
        \imagewithspyevaldisc{}{0.1cm}{0.1cm}
        \vspace{-0.5cm}\caption{Input frame}
        \label{fig:stablediffusion:sub4}
    \end{subfigure}
    \hfill
    \begin{subfigure}{0.325\linewidth}
        \imagewithspyevaldisc{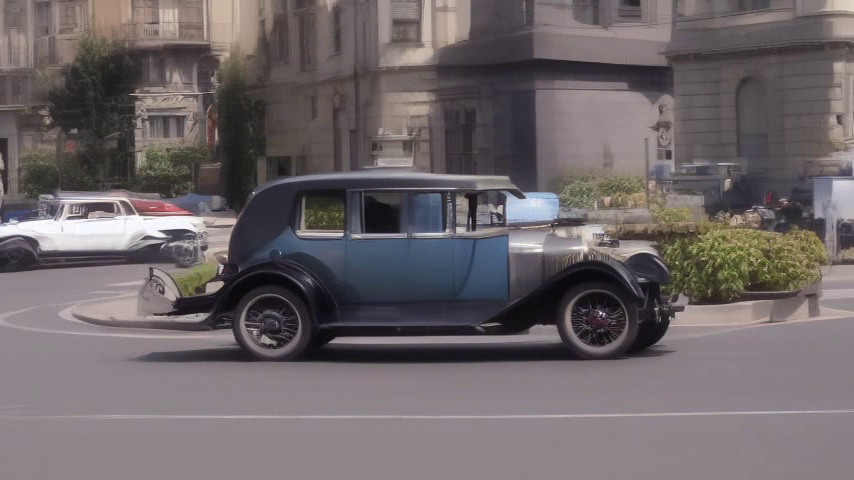}{0.1cm}{0.1cm}
        \vspace{-0.5cm}\caption{Per-frame I2I-SD}
        \label{fig:stablediffusion:sub5}
    \end{subfigure}
    \hfill
    \begin{subfigure}{0.325\linewidth}
        \imagewithspyevaldisctwo{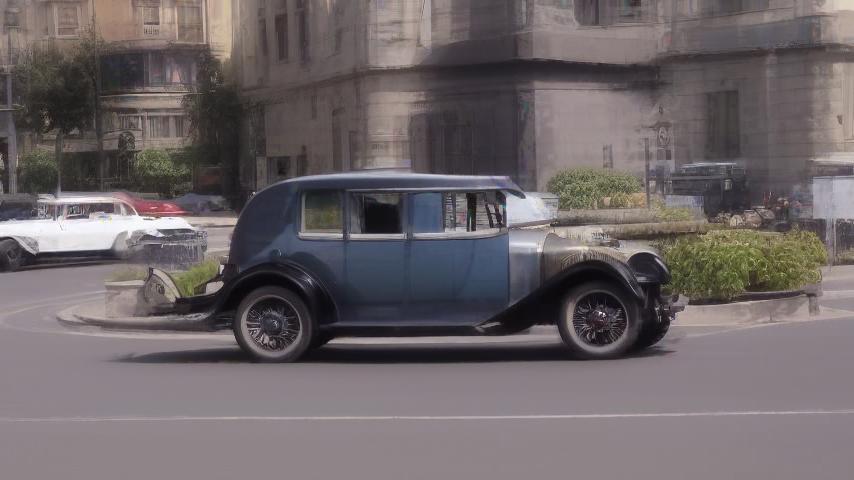}{0.1cm}{0.1cm}
        \vspace{-0.5cm}\caption{Stabilized (ours)}
        \label{fig:stablediffusion:sub6}
    \end{subfigure}
    \vspace{-0.2cm}
    \caption{Results on per-frame img2img stable diffusion (I2I-SD) \cite{rombach2022high} applied with the prompt "a 1920s car in a roundabout". Latent codes are interpolated with the previous frames to improve consistency. We use CfG scale $=7.5$ and a denoising factor of 0.4.  }
    \label{fig:stablediffusion}
\end{figure}

Our approach takes a video pair as an input: ($i$) the original and ($ii$) its per-frame stylized version. 
We assume that the stylization is based on the input image gradients and appears as variations in the form of colors and/or textures.
Thereby, we employ the original video as a guide for enforcing consistency.
However, for text-guided generative arts such as recent diffusion model-based approaches \cite{ramesh2022hierarchical,rombach2022high} the stylized frames are often only weakly correlated with the original input, and we cannot handle such cases. In \Cref{fig:stablediffusion} we provide an example of a per-frame stylization using stable diffusion \cite{rombach2022high}, in which despite using a latent pre-initialization from previous frames, new details are hallucinated in every frame, which cannot be effectively removed by our method, resulting in a blurry output video.

For the evaluation, we mainly use CNN-based stylization techniques. 
However, our approach can also handle classical stylization algorithms~\cite{Kyprianidis_SOTA2013}, we show a few such examples in the supplementary.
Our local consistency component comprising a convex combination of temporal neighbors can be seen as a crude form of local temporal denoising.
Previously it has been shown that temporal denoising is effective in enforcing consistency~\cite{Shekhar_Consistent2019}.
We conjecture that efficient temporal denoising combined with flow-based warping can further improve temporal stabilization not only for stylization but also for other tasks. We show examples for such non-stylization tasks, particularly for image enhancement (DBL \cite{gharbi2017deep}) and intrinsic decomposition, in the supplementary. 

We start with the assumption that temporal flickering is not completely undesirable for the task of stylization and thus we provide interactive consistency control. 
However, during the subjective user study, we observed that participants had different tolerance levels for flickering in the foreground as compared to that in the background. 
As part of future work, one can use depth-based or saliency-based masks to vary the consistency control parameters spatially for a more visually pleasing result. 

\begin{figure}[htb]
	\begin{subfigure}{0.49\columnwidth}%
		\includegraphics[trim=5.5cm 1.0cm 2cm 2cm, clip, width=\textwidth]{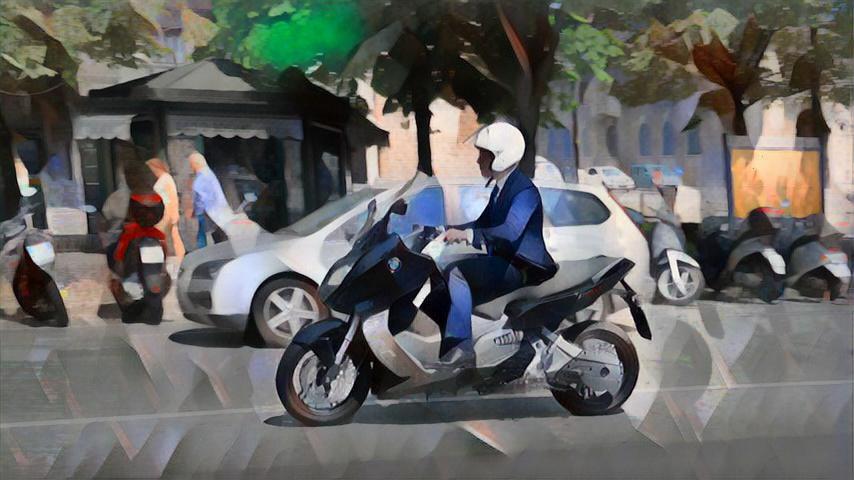}%
		\subcaption{$k_1 = 0.5$}
		\label{fig:wp_05}%
	\end{subfigure}\hfill 
	\begin{subfigure}{0.49\columnwidth}%
		\includegraphics[trim=5.5cm 1.0cm 2cm 2cm, clip, width=\textwidth]{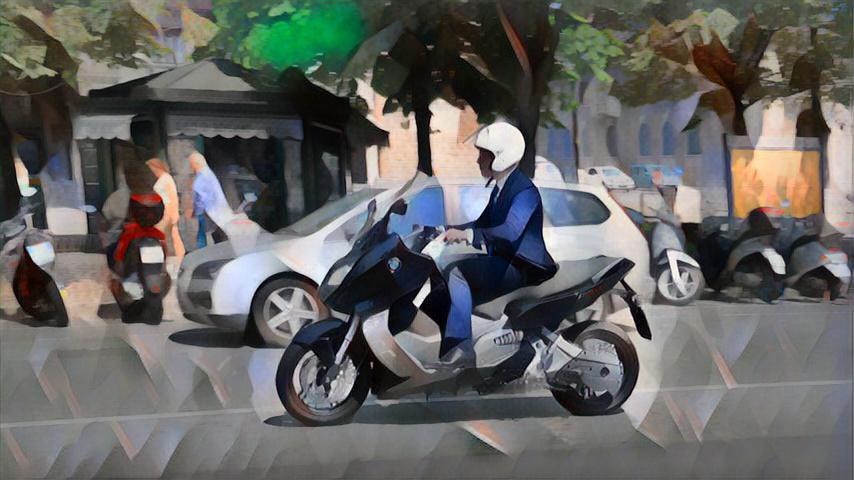}%
		\subcaption{$k_1 = 0.1$}
		\label{fig:wp_03}%
	\end{subfigure}
	\caption{The ghosting artifacts on the rear wheel of the scooter are significant in the final output for $k_1 = 0.5$, however, they are significantly reduced for $k_1 = 0.1$.}
	\label{fig:limitation}
	\vspace{-0.3cm}
\end{figure}

\paragraph*{Limitation.}
Our approach tends to have ghosting artifacts for fast-moving objects where the object motion between consecutive frames is large (\Cref{fig:limitation}).
The above can be reduced by reducing the upper limit of $w_{p}$ (\ie $k_1$), however, such a reduction also reduces consistency in the final output. 
We argue that since we provide interactive control of parameters the above trade-off between artifacts vs. consistency will not significantly hinder its usability.

\section{Conclusions}
\label{Sec:Conclusions}

We propose an approach that makes per-frame stylized videos temporally coherent irrespective of the underlying stylization applied to individual frames. 
At this, we introduce a novel temporal consistency term that combines local and global consistency aspects.
We maintain similarity with the per-frame processed result by minimizing the difference in the gradient domain.
Unlike previous approaches, we provide interactive consistency control by computing optical flow on the incoming video stream at high speed and with sufficient accuracy for stabilization. 
The fast optical-flow inference is achieved by developing a lightweight flow network architecture based on PWC-Net. 
The entire optimization solving is GPU-based and runs at real-time frame rates for HD resolution.  
We showcase that our temporally consistent output is preferred over the output of competing methods by conducting a user study.
As part of future work, we would like to employ learning-based temporal denoising to further improve the quality of results.
Moreover, we would like to explore the usage of depth-based and saliency-based masks to spatially vary consistency parameters according to perceptual principles. 
We hope that our design paradigm of interactive consistency control will potentially make per-frame video stylization more user-friendly.

\section*{Acknowledgements}
We thank the anonymous reviewers for their valuable feedback.
We thank Jobin Idiculla Wattasseril for helping with the user study.
This work was funded by the German Federal Ministry of Education and Research (BMBF) (through grants 01IS15041 -- \enquote{mdViProject} and 01IS19006 -- \enquote{KI-Labor ITSE}) and the Research School on \enquote{Service-Oriented Systems Engineering} of the Hasso Plattner Institute.

\FloatBarrier

%

  \typeout{}
  \bibliographystyle{eg-alpha-doi} 
  \bibliography{egsr_bib}       


\FloatBarrier
\clearpage

\newcommand{\mytitle}[1]{%
    \twocolumn[%
    \centering
    {\Large\bfseries #1\vspace{1em}} 
    ]%
}

\mytitle{Supplementary Material}
  
\setcounter{section}{0}

\noindent Some details had to be omitted from the main paper due to the page limit; we present those details here. 
In the following, we report on ablation experiments in \Cref{Sup:SubSec:Ablation}, which were carried out to determine the best performing fast optical-flow network, and also expand on quantization and pruning which were employed for our mobile-optimized network. In \Cref{Sup:subsec:implementation} we expand on training and implementation details. In \Cref{Sup:sec:warperror} we provide detailed numbers for the warping error. 
In \Cref{Sup:sec:ex_user_study} we compare our method subjectively against Shekhar~\etal~\cite{Shekhar_Consistent2019} and Thimonier~\etal~\cite{Thimonier_learning2021} through a user study.
Finally, in \Cref{Sup:Sec:MoreFlowResults}, we present further visual results of our optical-flow network.


\section{Fast Optical Flow Network} 
\subsection{Ablation Study}
\label{Sup:SubSec:Ablation}

To analyze our optimization steps, we compare different variants of our \ac{CNN}.
All variants are trained on the full dataset schedule unless stated otherwise.
We make use of Sintel Final Train dataset~\cite{SintelDataset} as a benchmark and measure accuracy (in terms of \ac{EPE}), number of parameters, and run-time for different variants.
Due to different desktop and mobile \ac{GPU} hardware, the run-time performance can vary between platforms, thus we measure them separately.

\begin{table}[h]
	\centering
	\caption{Comparison of DenseNet~\cite{DenseNet} and light~\cite{ARFlow} connections on a desktop system and mobile devices on Sintel Final Train.}
	\label{Sup:Tab:AblationDense}
	\begin{footnotesize}
		\begin{tabular}{l r r r r r}
			\toprule 
			&			& \shst{Params}	& \shst{Desktop} & \shst{iPad Air} 	& \shst{iPad Pro} \\
			Variant						& \ac{EPE} ($\downarrow$)	& \ac{M} ($\downarrow$)	& \ac{FPS} ($\uparrow$) &   \ac{FPS} ($\uparrow$) 		&  \ac{FPS} ($\uparrow$)  \\
			\midrule
			dense 						& 2.507						& 9.36					& 29.97  		& 1.53 			& 2.80	\\
			\abest{light}				& \abest{2.825}				& \abest{5.99}			& \abest{40.26}	& \abest{2.83} 	& \abest{5.09}		\\
			\bottomrule
		\end{tabular}
	\end{footnotesize}
\end{table}

\paragraph*{DenseNet Connection Replacement.}

As a first architectural improvement, we replace DenseNet~\cite{DenseNet} connections in the flow estimators with light connections~\cite{ARFlow}.
Replacing these results in a significant run-time improvement on both desktop systems and mobile devices, with a larger relative speed-up on mobile devices (\Cref{Sup:Tab:AblationDense}).
We conjecture that convolutions with a large number of channels (dense architecture uses up to 565 channels) might perform worse on mobile \acp{GPU} due to smaller memory and cache sizes.
Thus, reducing these high channel counts results in a larger speed-up on mobile devices.
The light connections result in a loss in accuracy (\Cref{Sup:Tab:AblationDense}), but due to the significant run-time improvements, we find it a reasonable trade-off and use light connections in the following experiments and in our proposed mobile architecture.

\begin{table}[h]
	\centering
	\caption{Ablation of different channel reduction on desktop system}
	\label{Sup:Tab:AblationChannelsDesktop}
	\begin{footnotesize}
		\begin{tabular}{lrrrrrrr}
			\toprule 
			&			& \shst{Params}	& \shst{Desktop} & \shst{iPad Air} 	& \shst{iPad Pro} \\
			Variant						& EPE ($\downarrow$)	& \ac{M} ($\downarrow$)	& \ac{FPS} ($\uparrow$) &   \ac{FPS} ($\uparrow$) 		&  \ac{FPS} ($\uparrow$)  \\
			\midrule
			\abest{5 estimators}			& \abest{2.825}	 & \abest{5.99}			& \abest{40.26} & \abest{2.83} 	& \abest{5.09}	\\
			-25\% channels					& 3.659			 & 3.55					& 46.86		& 3.95 & 6.65 \\
			-50\% channels					& 4.236			 & 1.73		 			& 56.07		& 6.33 & 10.92	\\
			\bottomrule
		\end{tabular}
	\end{footnotesize}
\end{table}

\paragraph*{Channel Reduction.}

We reduce the number of channels throughout the \ac{CNN}~\cite{MobileNets}.
In this case, the loss in accuracy and achieved trade-off is not beneficial (\Cref{Sup:Tab:AblationChannelsDesktop}).
We hypothesize that channel reduction is potentially better for high-level \ac{CV} tasks,
where high-dimensional convolution features are mapped to very low-dimensional results~\cite{MobileNets}. 
Optical flow, however, requires pixel-precise predictions of continuous values (motion vectors) and thus requires a much higher spatial fidelity.
Furthermore, we observe that the relative speed-up on mobile devices is again higher than on desktop systems which supports our previous belief that larger convolutions are more difficult for mobile \acp{GPU} with smaller memory and cache sizes.

\begin{table*}[p]
	\centering
	\caption{Ablation of different light flow-estimator configurations.}
	\label{Sup:Tab:AblationFlowestimators}
	\begin{footnotesize}
		\begin{tabular}{l r r r r r r r r r r r r}
			\toprule
			& \twoc{\shst{Sintel Final\\Train}}			& \twoc{\shst{Number of\\parameters}}	& \twoc{\shst{Desktop\\run-time}} & \twoc{\shst{iPad Air\\run-time}} 	& \twoc{\shst{iPad Pro\\run-time}}  \\
			\cmidrule(lr){2-3}
			\cmidrule(lr){4-5}
			\cmidrule(lr){6-7}
			\cmidrule(lr){8-9}
			\cmidrule(lr){10-11}
			Variant									& EPE ($\downarrow$)	&					& \ac{M} ($\downarrow$)	&		& \ac{FPS} ($\uparrow$) &		& \ac{FPS} ($\uparrow$) 	& 				& \ac{FPS} ($\uparrow$) 	& 		   \\
			\midrule
			5 estimators, none separated			& 2.825			&								& 5.99		& 					& 40.26		& 				 & 2.83		& 						& 5.09			&   \\
			5 estimators, all separated				& 3.813			& \abad{+34.9\%}				& 2.66		& \agood{-55.6\%}	& 43.49		& \agood{+8.0\%} & 4.52		& \agood{+59.7\%}		& 7.76			& \agood{+52.4\%}  \\
			5 estimators, last two separated		& 3.104			& \abad{+9.8\%}					& 4.77		& \agood{-20.3\%}	& 44.06		& \agood{+9.4\%} & 4.22		& \agood{+49.1\%}		& 7.28			& \agood{+43.0\%}  \\
			\abest{4 estimators, none separated}	& \abest{3.229}	& \abad{\abest{+14.3\%}}& \abest{5.31} 		& \agood{\abest{-11.3\%}} & \abest{79.12} & \agood{\abest{+96.5\%}} & \abest{6.17}		& \agood{\abest{+118.0\%}}		& \abest{10.41}			& \agood{\abest{+104.5\%}}  \\
			4 estimators, last separated			& 3.460			& \abad{+22.4\%}				& 4.68		& \agood{-21.8\%}	& 81.28		&\agood{+101.7\%}& 7.35		& \agood{+159.7\%}		& 12.80			& \agood{+151.4\%}  \\
			\bottomrule
		\end{tabular}
	\end{footnotesize}
\end{table*}

\begin{table*}[p]
	\centering
	\caption{Ablation of different refinement configurations.}
	\label{Sup:Tab:AblationRefinementAll}
	\begin{footnotesize}
		\begin{tabular}{l r r r r r r r r r r r r}
			\toprule
			& \twoc{\shst{Sintel Final\\Train}}			& \twoc{\shst{Number of\\parameters}}	& \twoc{\shst{Desktop\\run-time}} & \twoc{\shst{iPad Air\\run-time}} 	& \twoc{\shst{iPad Pro\\run-time}}  \\
			\cmidrule(lr){2-3}
			\cmidrule(lr){4-5}
			\cmidrule(lr){6-7}
			\cmidrule(lr){8-9}
			\cmidrule(lr){10-11}
			Variant									& EPE ($\downarrow$)	&					& \ac{M} ($\downarrow$)	&		& \ac{FPS} ($\uparrow$) &		& \ac{FPS} ($\uparrow$) 	& 				& \ac{FPS} ($\uparrow$) 	& 		   \\
			\midrule
			4 estimators, default refinement					& 3.229		&								& 5.31		& 					& 79.12		& & 6.17		& 					& 10.41			&   \\
			4 estimators, no refinement							& 3.258		& \abad{+0.9\%}					& 4.79		& \agood{-9.8\%}	& 86.72		& \agood{+9.6\%}  & 7.44		& \agood{+20.5\%}		& 13.22			& \agood{+27.0\%}  \\
			\abest{4 estimators, separated refinement}		& \abest{3.169}		& \agood{\abest{-1.8\%}}	& \abest{4.85}		& \agood{\abest{-8.6\%}}	& \abest{83.53}		& \agood{\abest{+5.5\%}}  & \abest{7.23}		& \agood{\abest{+17.1\%}}		& \abest{12.87}			& \agood{\abest{+23.6\%}}  \\
			\midrule
			5 estimators, default refinement					& 2.825		&									& 5.99		& 					& 40.26			& & 2.83		& 						& 5.06			&   \\
			5 estimators, no refinement							& 3.248		& \abad{+14.97\%}					& 5.47		& \agood{-8.6\%}	& 51.80			& \agood{+28.6\%}  & 4.22		& \agood{+49.1\%}		& 7.31			& \agood{+44.4\%}  \\
			5 estimators, separated refinement					& 2.922		& \abad{+3.43\%}					& 5.53		& \agood{-7.6\%}	& 47.43			& \agood{+17.8\%}  & 3.47		& \agood{+22.6\%}		& 6.00			& \agood{+18.5\%}  \\
			\bottomrule
		\end{tabular}
	\end{footnotesize}
\end{table*}

\begin{table*}[p]
\centering
\caption[Ablation of different pruning amounts (desktop)]{Ablation of different pruning amounts on desktop system.}
\label{Sup:Tab:AblationPruning}
\begin{footnotesize}
	\begin{tabular}{l r r r r r r r r r r r r}
		\toprule
		& \twoc{\shst{Sintel Final\\Train}}			& \twoc{\shst{Number of\\parameters}}	& \twoc{\shst{Desktop\\run-time}} & \twoc{\shst{iPad Air\\run-time}} 	& \twoc{\shst{iPad Pro\\run-time}}  \\
		\cmidrule(lr){2-3}
		\cmidrule(lr){4-5}
		\cmidrule(lr){6-7}
		\cmidrule(lr){8-9}
		\cmidrule(lr){10-11}
		Variant									& \ac{EPE} ($\downarrow$)	&								& \ac{M} ($\downarrow$)	&							& \ac{FPS} ($\uparrow$)		& \\
		\midrule
		4 estimators, separated refinement				& 3.169		&						& 4.85			& 						& 79.12		& & 7.23		& 					& 12.87			&   \\
		30\% pruned				& 3.275		& \abad{+3.3\%}		& 3.38			& \agood{-30.3\%}		& 86.47		& \agood{-9.3\%} & 8.79			& \agood{+21.5\%}		& 16.45				& \agood{+27.8\%}  \\
		\abest{40\% pruned}	& \abest{3.443} 	& \abad{\abest{+8.6\%}}		& \abest{2.83}			& \agood{\abest{-41.6\%}}		& \abest{86.61}		& \agood{\abest{-9.4\%}} & \abest{10.00}			& \agood{\abest{+38.3\%}}		& \abest{18.94}				& \agood{\abest{+47.1\%}}  \\
		50\% pruned				& 4.036 	& \abad{+27.3\%}	& 2.24			& \agood{-53.8\%}		& 88.94		& \agood{-12.4\%} & 13.76			& \agood{+90.3\%}		& 27.50				& \agood{+113.6\%}  \\
		\midrule
		5 estimators				& 2.825		&						& 5.99			& 						& 40.26		& & 2.83			& 						& 5.06				&   \\
		30\% pruned				& 2.900		& \abad{+2.6\%}			& 4.06			& \agood{-32.2\%}		& 46.79 	& \agood{-16.2\%} & 3.89			& \agood{+37.4\%}		& 6.95				& \agood{+37.3\%}  \\
		40\% pruned				& 3.022		& \abad{+6.9\%}			& 3.35			& \agood{-44.0\%}		& 50.68		& \agood{-25.8\%} & \abest{4.63}			& \agood{\abest{+63.6\%}}		& \abest{8.17}				& \agood{\abest{+61.4\%}}  \\
		50\% pruned				& 3.106		& \abad{+9.9\%}			& 3.01			& \agood{-49.7\%}		& 51.51		& \agood{-27.9\%} & 4.86			& \agood{+71.7\%}		& 8.47				& \agood{+57.4\%}  \\
		\bottomrule
	\end{tabular}
\end{footnotesize}
\end{table*}

\begin{table*}[p]
	\centering
	\caption[Ablation of different quantization options (mobile)]{Ablation of different quantization options on mobile devices using our lite-flow CNN. 
	}
	\label{tab:AblationQuantizationIPad}
	\begin{footnotesize}
	\begin{tabular}{r c r r r r r r r r}
		\toprule
												 \twoc{Quantization settings}	& \twoc{\shst{\ac{CNN}\\file size}} & \twoc{\shst{Sintel Final\\Train}}	& \twoc{\shst{iPad Air\\run-time}} 	& \twoc{\shst{iPad Pro\\run-time}} \\
		\cmidrule(lr){1-2}
		\cmidrule(lr){3-4}
		\cmidrule(lr){5-6}
		\cmidrule(lr){7-8}
		\cmidrule(lr){9-10}
		\shst{Number of\\bits} & \shst{low-prec.\\ acc.}				& \ac{MB} ($\downarrow$)		& 			& \ac{EPE} ($\downarrow$)		&			& \ac{FPS} ($\uparrow$) 	& 				& \ac{FPS} ($\uparrow$)		&   \\
		\midrule
		32-bit &								& 16.3		& 						& 3.577				& 					& 9.95		& 					& 18.94			&   \\
		32-bit & \checkmark						& 16.3		& 						& 3.584				& \abad{+0.2\%}		& 12.73		& \agood{+27.9\%}		& 24.04			& \agood{+26.9\%}   \\
		16-bit & 								& 8.2		& \agood{-49.6\%}		& 3.577				& \agood{$\pm$0.0\%}		& 10.00		& \agood{+0.5\%}		& 18.64			& \abad{-1.5\%}  \\
		16-bit & \checkmark						& 8.2		& \agood{-49.6\%}		& 3.584				& \abad{+0.2\%}		& 12.68		& \agood{+27.4\%}		& 23.96			& \agood{+21.9\%}  \\
		8-bit &									& 4.1		& \agood{-74.8\%}		& 3.609				& \abad{+0.9\%}		& 9.69		& \abad{-2.6\%}			& 18.74			& \abad{-1.0\%} \\
		\abest{8-bit} & \abest{\checkmark}		& \abest{4.1}	& \agood{\abest{-74.8\%}}		& \abest{3.621}				& \abad{\abest{+1.2\%}}		& \abest{12.89}		& \agood{\abest{+29.5\%}}	& \abest{23.84}			& \agood{\abest{+27.2\%}}  \\
		4-bit & 								& 2.1		& \agood{-87.1\%}		& 6.665				& \abad{+86.3\%}		& 9.77		& \abad{-1.8\%}			& 18.75			& \abad{-1.0\%} \\
		4-bit & \checkmark						& 2.1		& \agood{-87.1\%}		& 6.665				& \abad{+86.3\%}		& 12.93		& \agood{+29.9\%}		& 23.81			& \agood{+27.0\%}  \\
		\bottomrule
	\end{tabular}
	\end{footnotesize}
\end{table*}


\paragraph*{Flow Estimators.}

We evaluate different configurations of separable convolutions for the five flow-estimator modules.
Replacing all convolutions in the flow estimators with separable convolutions leads to a significant loss in accuracy (\Cref{Sup:Tab:AblationFlowestimators}).
The last two flow estimators operate on the highest pyramid resolutions and have the largest impact on run-time performance.
Thus, loss of accuracy can be minimized by using separable convolutions only for the last two flow estimators. 
Moreover, we find that removing only the last flow estimator leads to a larger speed-up and overall better trade-off~\cite{LiteFlowNet2}, both on desktop systems and mobile devices (\Cref{Sup:Tab:AblationFlowestimators}).
The last flow estimator -- operating on quarter input resolution -- comprises only \SI{11.3}{\percent} of parameters, but removing it results in more than \SI{100}{\percent} speed-up on mobile devices. 

\paragraph*{Refinement.}

For the four previously chosen flow estimators, we find that dense refinement 
can be replaced by separable convolutions which even results in a slight increase in accuracy on both desktop and mobile devices (\Cref{Sup:Tab:AblationRefinementAll}).
For five flow estimators, we observe that dense refinement has a larger impact on accuracy.

\paragraph*{Pruning:} As an additional improvement for mobile deployment, we evaluate pruning as a post-training step. Convolution filter pruning is applied as proposed by Li~\etal\cite{PruningCNNFilters} and a $l_1$ strategy combined with automatic consistency checks~\cite{ImplFilterPruning} is used for selecting which filters to prune.
We apply it to each convolution layer that has more than two output channels.
To keep pruning simple, we prune the same percentage of filters from each layer and then perform a single re-training to account for the loss of accuracy.
We find that pruning \SI{40}{\percent} of filters achieves a good trade-off for the final architecture -- reducing accuracy by less than \SI{10}{\percent} ($<$ 0.5px \ac{EPE}) for more than \SI{40}{\percent} speed-up (\Cref{Sup:Tab:AblationPruning}).
We re-train the pruned \ac{CNN} with the same dataset schedule and settings as initial training, except for training on FlyingChairs~\cite{FlowNet} where we start with the lower learning rate of $\num{1e-5}$ and train for fewer iterations as training converges quickly, \eg a maximum of 15 epochs (1.5 hours).
 
We evaluate different options of pruning as post-training optimization.
As our initial training consists of multiple stages with different datasets,
we first evaluate after which stage to prune and at which stage to start re-training.
We observe that the best accuracy is obtained when pruning the fully trained \ac{CNN}
and re-training from the beginning of the dataset schedule (\Cref{tab:AblationPruningTraining}).
We believe this works best as learned representations after only training on Chairs~\cite{FlowNet} or Things3D~\cite{FlyingThings3D} are not as distinctive as after full training. 

Next, we evaluate which trade-offs result from different amounts of pruned channels.
For mobile devices and our final \ac{CNN}, we find that pruning up to 40\% of channels results in significant run-time improvement with plausible accuracy loss. 
Pruning 50\% of channels results in a substantially higher accuracy loss (\Cref{Sup:Tab:AblationPruning}).
For desktop, pruning -- similar to reducing the number of channels (\Cref{Sup:Tab:AblationChannelsDesktop}) -- results in a smaller speed-up than on mobile devices. 
Considering only a small improvement of the already high frame rate in exchange for a significant loss in quality, we do not recommend pruning for the desktop version.

\begin{table}[t]
	\centering
	\caption[Ablation of pruning and re-training at different stages]{Ablation of pruning and re-training at different training stages. The \textit{Initial} model is the one with 4 estimators and separated refinement and then we have its variations with respect to pruning at different stages. Note that increase in \acs{EPE} is the least for the \nth{3} variant.}
	\label{tab:AblationPruningTraining}
	\begin{footnotesize}
		\begin{tabular*}{.48\textwidth}{p{32mm} r r}
			\toprule
			\multicolumn{1}{r}{Training:}	& \twoc{Chairs $\rightarrow$ Things3D $\rightarrow$ Sintel} \\
			\cmidrule(lr){2-3}
			Variant				& \ac{EPE} ($\downarrow$) on Sintel Train &							\\
			\midrule
			Initial				& 3.169 			& 							\\
			Prune 30\% after Chairs, re-train from Chairs			& 3.505 			& \abad{+10.6\%}				\\
			Prune 30\% after Things3D, re-train from Things3D			& 3.290				& \abad{+3.8\%}			\\
			\abest{Prune 30\% after Sintel, re-train from Chairs}	& \abest{3.275} 	& \abad{\abest{+3.3\%}}			\\
			Prune 30\% after Sintel, re-train from Sintel			& 3.413				& \abad{+7.7\%}			\\
			\bottomrule
		\end{tabular*}
	\end{footnotesize}
\end{table}

\paragraph*{Quantization and Mobile Deployment:}
For mobile deployment, we make use of CoreML~\cite{CoreML} as the framework for executing our \acp{CNN} on Apple mobile devices.
We apply \SI{8}{\bit} linear weight-quantization and enable the accumulation of low-precision intermediate results (\Cref{tab:AblationQuantizationIPad}).
This minimizes the file size by \SI{75}{\percent} (compared to \SI{32}{\bit} weights) and further improves run-time performance by \SI{30}{\percent} (on mobile devices) with only negligible accuracy loss. 
Further analysis shows that our method does not profit from using the on-device \ac{NPU}. 


\subsection{Implementation Details}
\label{Sup:subsec:implementation}



\paragraph*{Pruning.}
For convolution filter pruning we use a PyTorch~\cite{PyTorch} implementation
by Gongfan Fang~\cite{ImplFilterPruning}.
We use $l_1$ strategy for selecting filters to prune. $l_2$ strategy is available too but Li~\etal\cite{PruningCNNFilters} show that these strategies perform comparable.
We round the number of resulting channels to a multiple of 8, as other filter counts result in a run-time overhead on mobile devices.
We re-train the pruned \ac{CNN} with the same dataset schedule and settings as initial training, except for training on FlyingChairs~\cite{FlowNet} where we start with the lower learning rate of $\num{1e-5}$ and train for fewer iterations as training converges quickly, \eg a maximum of 15 epochs (1.5 hours).

\paragraph*{Mobile execution.}
We use CoreML~\cite{CoreML} as the framework for executing our \acp{CNN} on Apple mobile devices.
We evaluate using iPad Pro (11-inch, \nth{2} gen 2020) and iPad Air (3rd gen, 2019).
CoreML efficiently implements standard \ac{CNN} operations, however, the two operations specific to optical flow, \ie correlation and warping, need to be implemented as custom layers using Metal \ac{GPU} shaders for parallel and efficient computation using the mobile \ac{GPU}.
After \ac{CNN} conversion to CoreML, we apply 8-bit linear weight quantization using coremltools
, low precision acculumation is enabled to enforce low precision accumulation in all operations. 

\begin{table*}[htpb]
    
    \begin{minipage}{\linewidth}
    	\centering
    	\caption[Our training settings for different datasets]{Our training settings for different datasets.
    		Training times are specified for one Nvidia V100 \ac{GPU}.}
    	\label{tab:DetailsDatasetsTraining}
    	\begin{footnotesize}
    		\begin{tabular}{l c c c}
    			\toprule
    			& FlyingChairs~\cite{FlowNet}	 & FlyingThings3D~\cite{FlyingThings3D}	& Sintel (Final)~\cite{SintelDataset} 		\\
    			\midrule
    			Batch size 				& 8 				& 4				& 4				\\
    			Dataset resolution		& 512 $\times$ 384	& 960 $\times$ 540 & 1024 $\times$ 436 \\
    			Training resolution		& 448 $\times$ 320	& 768 $\times$ 384 & 448 $\times$ 384				\\
    			Loss function			& \ac{EPE} (\Cref{eq:EPELoss}) & robust $l_1$  (\Cref{eq:RobustLoss})	& robust $l_1$ (\Cref{eq:RobustLoss})	\\
    			&					& $q=0.4$, $\epsilon=0.01$ & $q=0.4$, $\epsilon=0.01$ \\
    			Initial learning rate	& $\num{1e-4}$		& $\num{1e-5}$	& $5 \cdot \num{1e-5}$	\\
    			Learning rate schedule  & as $S_\text{long}$ in~\cite{FlowNet2}	& as $S_\text{fine}$ in~\cite{FlowNet2}& as in~\cite{PWCNet}\tablefootnote{Start at $5 \cdot \num{1e-5}$, divide by 2 at 45K, 65K, 85K, 95K and 98K iterations. At 100K, restart with $3 \cdot \num{1e-5}$ and follow same schedule for another 100K iterations.}				\\
    			Total epochs			& 60				& 10			& 200			\\
    			Total iterations		& 1200K		& 400K		& 200K		\\
    			Training time			& 6 hours			& 10 hours		& 4 hours		\\
    			\bottomrule
    		\end{tabular}
    	\end{footnotesize}
    \end{minipage}

\vspace{10pt}

    \begin{minipage}{.49\linewidth}
    \vspace{0pt}
      \centering
	\caption[Ablation of different hyperparameters for baseline training (desktop)]
	{Ablation of different hyperparameters for baseline training on desktop.
		Sintel Final \ac{EPE} is measured after training the original PWC-Net~\cite{PWCNet} architecture for 30 epochs on FlyingChairs~\cite{FlowNet}.
		}
	\label{tab:AblationHyperTrainingDesktop}
	\begin{footnotesize}
		\begin{tabular}{l c c c c r}
			\toprule
			& \multicolumn{4}{c}{Hyperparameters}					& \shst{Sintel Final\\Train} \\
			\cmidrule(lr){2-5}
			\cmidrule(lr){6-6}
			& Optimizer & Loss weights & \shst{Regular-\\ization}  & \shst{Gradient\\stopping}	& \ac{EPE} ($\downarrow$)			\\
			\midrule
			& Adam & exponential 		& 		&	 			& 5.561						\\
			& Adam & exponential		& $l_2$	& 				& 6.938						\\
			& AdamW & exponential 		& 		& 				& 5.153						\\
			& AdamW & exponential 		& $l_2$	& 				& 5.263						\\
			& \underline{AdamW} & \underline{equal}		 		& \underline{$l_2$}	& 				& \underline{5.086}						\\
			& AdamW & exponential		& $l_2$	& \checkmark			& 6.269						\\
			& AdamW & equal		 		& $l_2$	& \checkmark		& 5.465						\\

			\bottomrule
		\end{tabular}
	\end{footnotesize}
    \end{minipage} 
    \hfill
    \begin{minipage}{.49\linewidth}
    \vspace{0pt}
	\caption[Our augmentation settings]{Our augmentation settings. The same settings are applied to all training stages (\cref{tab:DetailsDatasetsTraining}),
		except when fine-tuning on Sintel~\cite{SintelDataset} where no Gaussian noise is added.}
	\label{tab:DetailsAugmentations}
	\begin{footnotesize}
		\begin{tabular}{l c c}
			\toprule
			Augmentation 		& Value range			& Probability \\
			\midrule
			Rotation			& uniform in [-17°; 17°]	& 0.2			\\
			Random crop			& training resolution, \cf \cref{tab:DetailsDatasetsTraining} & 1.0			\\
			Horizontal flip		& 							& 0.5			\\
			Vertical flip		& 							& 0.5			\\
			Additive brightness	& normal, $\sigma = 0.02$ 	& 1.0			\\
			RGB multiplier		& uniform in [0.9; 1.1]	& 1.0			\\
			Contrast multiplier	& uniform in [0.7; 1.3]	& 1.0			\\
			Gamma adjustment	& uniform in [0.7; 1.5]	& 1.0			\\
			RGB random order	&							& 1.0			\\
			Additive gaussian noise & $\sigma$ uniform from [0; 0.04] & 1.0	\\
			\bottomrule
		\end{tabular}
	\end{footnotesize}
    \end{minipage}%
\end{table*}

\subsection{Training Settings}
\label{sup:sec:training}

Similar to the original PWC-Net~\cite{PWCNet}, we train our mobile architecture on the training dataset schedule FlyingChairs~\cite{FlowNet} $\rightarrow$ FlyingThings3D $\rightarrow$ Sintel~\cite{SintelDataset}.
\Cref{tab:DetailsDatasetsTraining} lists training settings for respective stages.
We compute the multi-scale losses by taking the per-pixel difference between the output of each flow estimator and an accordingly downscaled ground-truth optical flow.
The pixel-level loss values are summed up to a final single value which is then used as a training objective by the optimizer.
Like PWC-Net~\cite{PWCNet}, we scale the ground-truth optical flow with a factor of 20 prior to calculating the loss and thus have to divide the flow estimate by 20 at test time. 
For training we use AdamW optimizer~\cite{AdamW} with $\beta_1 = 0.09$, $\beta_2 = 0.99$,
and $l2$ weight regularization with trade-off $\gamma = 0.0004$. 

\noindent 
Given predicted flow field $\text{uv}_{\text{Pred}}$ and ground-truth flow field $\text{uv}_{\text{GT}}$, \ac{EPE} loss (\Cref{eq:EPELoss}) and a robust $l_1$ loss (\Cref{eq:RobustLoss}) is defined as follows:

\begin{equation} \label{eq:EPELoss}
\mathcal{L}_\text{EPE}(\text{uv}_{\text{Pred}}, \text{uv}_{\text{GT}}) = \frac{1}{W \cdot H} \sum_{x, y} \norm{\text{uv}_{\text{Pred}}(x, y) - \text{uv}_{\text{GT}}(x, y)}_2
\end{equation}

\vspace{5px}

\begin{equation} \label{eq:RobustLoss}
\mathcal{L_\text{robust}}(\text{uv}_{\text{Pred}}, \text{uv}_{\text{GT}}) = \frac{1}{W \cdot H} \sum_{x, y} \left( \norm{\text{uv}_{\text{Pred}}(x, y) - \text{uv}_{\text{GT}}(x, y)}_1 + \epsilon \right) ^q
\end{equation}
with typical values of $\epsilon = 0.01$ and $q = 0.4$.
$q < 1$ results in less penalty to large error values and thus makes the loss more robust to outliers, which is necessary for fine-tuning on realistic, difficult datasets~\cite{PWCNet}.

\subsection{Hyperparameters Configuration}
\label{sup:sec:hyperparam}

We evaluate different hyperparameters while training the original PWC-Net~\cite{PWCNet} to determine a baseline that achieves the best possible accuracy.
We find that AdamW optimizer~\cite{AdamW} reaches a significantly better accuracy than Adam~\cite{Adam} -- with and without $l_2$ weight regularization (\Cref{tab:AblationHyperTrainingDesktop}).
Furthermore, we find that equally weighted multi-scale losses -- as opposed to commonly used exponential weighting~\cite{PWCNet, LiteFlowNet, LiteFlowNet2, VCN} -- result in slightly better accuracy.
Additionally, we consider gradient stopping as proposed by Hofinger~\etal\cite{ImprovingPyramidOF}, but observe that it does not result in any improvement.

 \begin{table*}[htb]
\begin{center}
  \begin{tabular}{l|cccc|cccc}
            &  \multicolumn{8}{c}{\text {Warping Error:  \textbf{L1} }}                \\  
            & \multicolumn{4}{c}{\text { DAVIS }} & \multicolumn{4}{c}{\text { VIDEVO }} \\
            Task &  $V_p$ & \cite{Bonneel_Blind2015} & \cite{Lai_Learning2018} &  Ours & $V_p$ & \cite{Bonneel_Blind2015} & \cite{Lai_Learning2018} &  Ours \\
            \hline
            CycleGAN/photo2ukiyoe & 0.037 & \textbf{0.028} & 0.028 & 0.033 & 0.036 & \textbf{0.026} & 0.028 & 0.032 \\
            CycleGAN/photo2vangogh & 0.049 & \textbf{0.033} & 0.037 & 0.042 & 0.047 & \textbf{0.032} & 0.036 & 0.041 \\
            fast-neural-style/rain-princess & 0.079 & \textbf{0.043} & 0.055 & 0.062 & 0.076 & \textbf{0.040} & 0.055 & 0.061 \\
            fast-neural-style/udnie & 0.044 & \textbf{0.025} & 0.031 & 0.036 & 0.038 & \textbf{0.021} & 0.026 & 0.031 \\
            WCT/antimonocromatismo & 0.052 & \textbf{0.030} & 0.036 & 0.041 & 0.046 & \textbf{0.023} & 0.030 & 0.035 \\
            WCT/asheville & 0.067 & \textbf{0.040} & 0.049 & 0.056 & 0.061 & \textbf{0.033} & 0.042 & 0.049 \\
            WCT/candy & 0.065 & \textbf{0.035} & 0.045 & 0.051 & 0.058 & \textbf{0.029} & 0.038 & 0.045 \\
            WCT/feathers & 0.058 & \textbf{0.035} & 0.040 & 0.049 & 0.054 & \textbf{0.030} & 0.037 & 0.045 \\
            WCT/sketch & 0.054 & \textbf{0.037} & 0.038 & 0.046 & 0.050 & \textbf{0.032} & 0.035 & 0.041 \\
            WCT/wave & 0.052 & \textbf{0.033} & 0.037 & 0.044 & 0.048 & \textbf{0.029} & 0.034 & 0.040 \\ 
            \hline
            Average & 0.056 & \textbf{0.034} & 0.040 & 0.046 & 0.051 & \textbf{0.036} & 0.036 & 0.042
        \end{tabular}
        \end{center}
        \caption{Flow Warping Error over stylization tasks. The optical flow evaluation is computed using GMA~\cite{jiang2021learning}}
        \label{table:warpingerror_full}
\end{table*}

\subsection{Data Augmentation}
\label{sup:sec:augmentations}

Dosovitskiy~\etal\cite{FlowNet} found that augmentations are important for learning-based optical flow methods to prevent overfitting on synthetic training data and to ensure generalization for real-world data.
Similarly, we apply geometric and color transformations to input frames and corresponding flow fields,
as listed in \cref{tab:DetailsAugmentations}. 
Geometric transformations are applied equally to both frames of an input pair, and must be reflected accordingly in the flow field. 
For example, a translation applied to the frames requires a translation applied to the flow field; a rotation of frames requires a rotation of the flow field and its motion vectors.
Color transformations need to be applied only to the frames, not the flow field.
While it would be possible to apply different transformations (geometric, colors) per frame of an input pair~\cite{Teed_RAFT2020} -- to further increase robustness against illumination changes for example --
we find that transformations applied equally per frame pair are sufficient augmentations that prevent overfitting, while not making training of small \ac{CNN} variants too difficult~\cite{MobileNets}.

\section{Warping Error}
\label{Sup:sec:warperror}

In \Cref{table:warpingerror_full} we show the warping error (using $\ell_1$ metric, as defined in the main paper) over the stylization tasks following Lai \etal \cite{Lai_Learning2018}.

\section{Extended User Study}
\label{Sup:sec:ex_user_study}

\begin{figure}[htb]
	\centering
	\includegraphics[trim={8cm 4.5cm 12cm 5cm},clip,width=1.0\columnwidth]{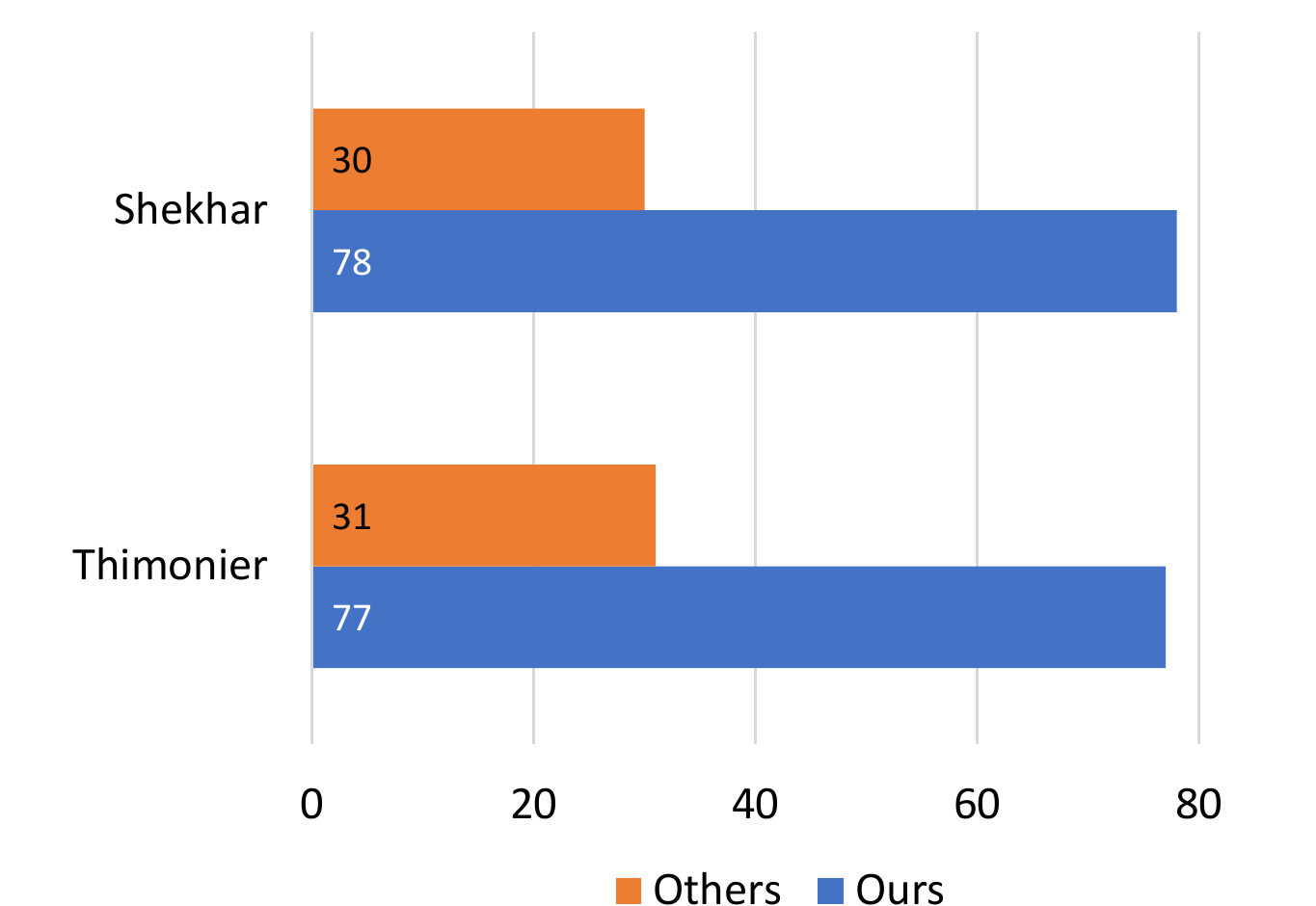}
	\caption{Statistics of the user study results on the removal of temporal flickering from per-frame stylized videos. For $12$ participants and $9$ different videos, we compare our method against Shekhar~\etal and Thimonier~\etal through a total of $108$ randomized A/B tests.}
	\label{fig:ex_user_study}
\end{figure}

To also compare with the methods of Shekhar~\etal~\cite{Shekhar_Consistent2019} and Thimonier~\etal~\cite{Thimonier_learning2021} we conducted another user study, involving only these two methods. 
The setup is similar to the one described in the main paper except that it was performed by a different group of participants to avoid bias.  
In total, $12$ persons ($3$ female, $8$ male, and $1$ did not specify) between the ages of $25$ to $40$ years participated in the study. 
\Cref{fig:ex_user_study} shows that our method surpasses the other methods by a large margin. 

\section{More Optical-Flow Results}
\label{Sup:Sec:MoreFlowResults}

In \Cref{Sup:Fig:QualityFlowSintelCoarse} and \Cref{Sup:Fig:QualityFlowDavisCoarse} we show further results for our lite optical flow network (configured as presented in the main paper) compared to other methods on Sintel and DAVIS.

\FloatBarrier
\clearpage

\begin{figure*}[tb]
\centering
\input{graphics/evaluation/compare_flow_sintelfinal_coarse_full.tex}
\caption{Optical flow estimated using the synthetic Sintel dataset\cite{SintelDataset}.}
\label{Sup:Fig:QualityFlowSintelCoarse}
\end{figure*} 

\begin{figure*}[tb]
\centering
\input{graphics/evaluation/compare_flow_davis_coarse_full.tex}
\caption{Optical flow estimated for the real-world dataset DAVIS~\cite{Davis2017Dataset}.}
\label{Sup:Fig:QualityFlowDavisCoarse}
\end{figure*}


\end{document}